\documentclass{aa}  
\usepackage{graphicx}
\usepackage{txfonts}
\usepackage{listings}
\usepackage{hyperref}
\hypersetup{colorlinks=true,linkcolor=blue,citecolor=blue,filecolor=magenta,urlcolor=blue}

\begin{document}

   \title{The 800pc long tidal tails of the Hyades star cluster}
   \titlerunning{The 800pc long tidal tails of the Hyades star cluster}
   \subtitle{Possible discovery of candidate epicyclic overdensities from an open star cluster}

  \author{
          Tereza Jerabkova \inst{1} 
           \and 
          Henri M.J. Boffin   \inst{2} 
          \and 
          Giacomo Beccari  \inst{2}
          \and 
          Guido de Marchi \inst{1}
          \and
          Jos H.~J. de Bruijne \inst{1}
          \and
          Timo Prusti \inst{1}
          }

  \institute{
  European Space Agency (ESA), European Space Research and Technology Centre (ESTEC), Keplerlaan 1, 2201 AZ Noordwijk, The Netherlands,\\ Emails: Tereza.Jerabkova@esa.int
        \and 
European Southern Observatory, Karl-Schwarzschild-Strasse 2, 85748 Garching bei M\"unchen\\
Emails: hboffin@eso.org, gbeccari@eso.org
             }
   \date{Received ; accepted }

% \abstract{}{}{}{}{} 
% 5 {} token are mandatory
 
  \abstract
{
The tidal tails of stellar clusters provide an important tool for studying the birth conditions of the clusters and their evolution, coupling, and interaction with the Galactic potential. 
The \textit{Gaia} satellite, with its high-quality astrometric data, opened this field of study, allowing us to observe large-scale tidal tails. Theoretical models of tidal-tail formation and evolution are available. However, the exact appearance of tidal features as seen in the \textit{Gaia} catalogue has not yet been studied. Here we present the $N-$body evolution of a Hyades-like stellar cluster with backward-integrated initial conditions on a realistic 3D orbit in the Milky Way (MW) galaxy computed within the AMUSE framework. For the first time, we explore the effect of the initial cluster rotation and the presence of lumps in the Galactic potential on the formation and evolution of tidal tails. For all of our simulations we present \textit{Gaia} observables and derived parameters in the convergent point (CP) diagram. 
We show that the tidal tails are not naturally clustered in any coordinate system and that they can span up to 40 km/s relative to the cluster centre in proper motions for a cluster age of 600-700~Myr. Models with initial rotation result in significant differences in the cluster mass loss and follow different angular momentum time evolution. Thus the orientation of the tidal tails relative to the motion vector of the cluster and the current cluster angular momentum constrain the initial rotation of the cluster.  
We highlight the use of the standard CP method in searches for co-moving groups and introduce a new compact CP (CCP) method that accounts for internal kinematics based on an assumed model. Using the CCP method, we are able to recover candidate members of the Hyades tidal tails in the \textit{Gaia} Data Release 2 (DR2) and early Data Release 3 (eDR3) reaching a total extent of almost 1kpc. 
We confirm the previously noted asymmetry in the detected tidal tails.
In the eDR3 data we recovered spatial overdensities in the leading and trailing tails that are kinematically consistent with being epicyclic overdensities and thus would present candidates for the first such detection in an open star cluster. We show that the epicyclic overdensities are able to provide constraints not only on the cluster properties, but also on the Galactic potential.
Finally, based on $N-$body simulations, a close encounter with a massive Galactic lump can explain the observed asymmetry in the tidal tails of the Hyades. 
}


   \maketitle
%
%-------------------------------------------------------------------

\section{Introduction}
Stars and stellar clusters form in sub-parsec (pc) dense filamentary regions of
molecular clouds \citep{Kroupa95a, Kroupa95b, LL03, MarksKroupa12,Andre2014}. The detailed small-scale physics of star formation and its initial conditions coupled to the galaxy-wide scales through gravitational forces and stellar populations still remains to be fully understood.
 Tidal tails of star cluster and large-scale
 relics of star formation in filaments \citep{Jerabkova2019, Beccari2020}  are unique structures providing a physical link between the initial conditions of cluster formation and the Galactic potential, its tides, and shears. These structures are in transition,  becoming part of the galactic field \citep{BM03, Kroupa05,DK20a,DK20b, Meingast+20,DW20} and creating an additional link to galactic stellar population signatures \citep{MarksKroupa11, Jerabkova+18}.

Embedded star clusters that survive the process of the expulsion of their residual gas re-virialise to open clusters (OCs) with half-mass radii of about~$3\,$pc \citep{Kroupa+01, BK17} and as OCs evolve through mass loss from their dying stars and through loss of members driven by the energy equipartition or two-body relaxation processes. 
This process causes a small increase in binding energy per dynamical time for
a small fraction of stars, such that the stars leave the cluster and orbit the Galaxy \citep{BM03}.  If the cluster is on a circular orbit, stars that leave with a lower velocity will fall towards the Galactic centre and will thus speed ahead as the angular velocity increases on their eccentric orbit with apogalacticon at the cluster orbit. A star that leaves with a higher velocity will fall behind it as its orbital angular velocity decreases on its eccentric orbit with perigalacticon at the cluster orbit. The stars drifting from the cluster form a symmetrical S-shaped stellar distribution in the immediate surroundings of the cluster.

Because the large majority of stars leave an OC through the energy equipartition process \citep[rather than being ejected through three-body  or binary-binary encounters,][]{Oh2015,Oh2016}, their velocities relative to the cluster are within a few km/s. The lost stars slowly drift from the cluster following very similar orbits. The cluster thus develops a thin leading tidal tail closer to the Galactic centre and a thin trailing tidal tail farther away. The tails are expected to compose a symmetric structure around the cluster \citep{Thomas+18}. While the stars initially drift away from the cluster potential, they later partially and repeatedly return as tidal-tail stars, and their cluster of origin moves on orbits with slightly different eccentricities and consequently with different orbital frequencies. This drifting-away and returning motion leads to overdensities along the tidal tails that are known as epicyclic overdensities \citep{Kuepper+08, Just+09, Kuepper+12}. The tidal tails of clusters on eccentric orbits have more complicated structures~\citep{Kuepper+10}, but they still display regularly spaced epicyclic overdensities, the spacing of which depends on the Galactic potential. The distance between the cluster and the  K\"upper overdensities decreases as the cluster looses mass \citep{Kuepper+10}, allowing a very precise measurement of the Galactic potential \citep{Kuepper+15}. 

Much progress has been made on the detection and study of tidal tails around globular clusters (GC) in the halo of the MW 
(e.g. \citealt{Carballo+20, PiattiCarballo20} or the identification of the long stellar stream of the massive globular cluster $\omega$ Centauri by \citealt{Ibata2019}). The reason is mostly that because the typical orbit of a GC is very high above the Galactic plane, contamination by field stars is removed. Moreover, GCs are massive ($> 10^4\, M_\odot$) and have long lifetimes ($\approx 13$ Gyr), allowing the tidal tails to be well populated by the cluster stars and to display large extensions on the sky. A landmark discovery, for example, is given by the long tidal tails of the globular cluster Palomar~5 \citep{Odenkirchen+03}.  

Detecting the tidal tails around OCs is more challenging because of their young ages (from a few hundred million years up to a few billion years at most) and their low mass ($\approx 10^2-10^3$ M$_{\odot}$). Moreover, OCs are mostly confined to the Galactic disc, which makes identifying their tidal tails challenging because they contain few stars and are heavily contaminated by field stars. However, in the same way as for GC, it is expected that tidal tails around OCs display an S-shape and epicyclic overdensities, the existence of which is related to the orbit of the cluster, the initial conditions at formation, and the Galactic potential. Complications arise because OCs may have orbits that take them somewhat out of the Galactic mid-plane, and the bar and spiral patterns of the Galaxy may lead to non-axisymmetric perturbations, which may affect the properties of the tails.

 The advent of the ESA \textit{Gaia} satellite for the first time allows the study of young stellar populations on large scales (>100pc) 
 in six-dimensional (6D) position and velocity space. For example, using \textit{Gaia} DR2 \citep{GaiaDR2_2018}, \cite{Kounkel+18} and \cite{Zari+19} 
demonstrated that the Orion star-forming region is composed of a variety of populations with different spatial and kinematic properties each that are all likely generated in multiple events instead of in a progressive star formation history \citep[see also][discussing three bursts of star formation in the Orion Nebula cluster]{Beccari2017,Jerabkova2019_ONC,Kroupa2018}.
 \cite{Zari+18} described the 6D properties and age structures in which young stars are found within a region of~500~pc around the Sun, and a number of other studies focusing on individual star-forming regions have been reported. One example is the large-scale picture of the Gamma Velorum region  \citep{Beccari2018, Cantat-Gaudin2019}. 
 The \textit{Gaia} DR2 catalogue also led to the discovery of stellar relic filaments. These are spatial structures of a few~pc that are more than 90~pc long and consist of stars of equal ages that are younger than a few hundred~million years \citep{Jerabkova2019,Beccari2020}. A Galaxy-wide survey using the \textit{Gaia} DR2 catalogue reveals many elongated structures made of coeval stars, many of which are tidal tails of dissolving star clusters \citep{KounkelCovey19}.  \cite{Meingast+20} described the nearby extended spatial distribution of stars that co-move with their clusters but are not bound to them. 
After the release of the \textit{Gaia} DR2 catalogue, tidal tails have been found around four nearby ($<300\,$pc) OCs, namely Blanco~1 ($\approx 100\,$ Myr, \citealt{Zhang+20}), the Hyades \citep[$\approx 600-700\,$Myr;][]{MeingastAlves19, Roeser+19, Hy_age1,Hy_age2, Hy_age3, Hy_age4, GDR2_Hyades}, Coma~Berenices \citep[$\approx 750\,$Myr;][]{Tang+19, Fuernkranz+19}, and Praesepe~\citep[$\approx 800\,$Myr;][]{RS19}.

The Hyades cluster has attracted much attention because it is only about $45\,$pc distant from the Sun (e.g. see \citealt{Chumak+05} for a visualisation), such that its stellar content and its tidal tails can most likely be mapped out with high precision. Using~6D phase-space constraints \citep{roeser2010}, \citet{Roeser+11} found the half-mass radius of the Hyades to be $4.1\,$pc and its tidal radius to be $9\,$pc with a bound mass of $275\,M_\odot$. The total mass of  Hyades stars found so far is about $435\,M_\odot$. Interestingly, the measured bulk (3D) velocity dispersion within $9\,$pc is $\approx 0.8\pm0.1\,$km/s, but the theoretical value for Hyades stellar mass is expected to be $0.36\,$km/s (table~3 in \citealt{Roeser+11}). The Hyades therefore show a possible mass discrepancy of a factor of four; part of this may be due to unresolved multiple stars \citep{Dabringhausen2016} and/or to the fact that the cluster is not  in dynamical equilibrium \citep{Oh2020}. The cluster contains stars with masses between about $0.1\, M_\odot$ and $2.6\,M_\odot$, distributed according to a present-day mass function (PDMF) that is broadly consistent with evaporation from the cluster, a canonical initial mass function (IMF), and a majority of stars that have been born in binaries \citep{Kroupa95c, Kroupa11, Ernst+11}. 

The astrophysical evolution of this cluster has been studied in depth using $N-$body methods linked with stellar evolution codes \citep{Portegies+01, Madsen03}, confirming the expected theoretical velocity dispersion. \cite{Chumak+05} in addition provided the first detailed $N-$body study of the expected tidal-tail structure of the Hyades. They showed that each of the two symmetrical tails should be about $500\,$pc long and also documented the degeneracy between the initial mass and the initial radius of the cluster for the models to fit its present-day bulk properties. The symmetrical tidal tails contain very pronounced  K\"upper overdensities about $200\,$pc from the cluster. \cite{Ernst+11} based their models on the \cite{Roeser+11} Hyades survey constraints, thereby verifying and significantly improving the details of the cluster evolution. The resultant best-fitting initial mass of a Plummer model is $1230\,M_\odot$ with an initial half-mass radius of $2.62\,$pc, which leads to an average mass-loss rate of $1.4\,M_\odot$/Myr. The present-day tidal tails reach out to a length of $\approx800\,$pc according to this. These derived quantities depend on the choice of the Galactic potential, however, whereby the current position and velocity vectors of the cluster are well constrained. 

In this paper we perform $N-$body simulations of the Hyades cluster in a realistic Galactic potential with the aim to investigate the expected extent of the tidal tails. With this in mind, we develop a new recipe to find these stellar structures using the Gaia DR2 and eDR3 data with the final goal of reporting evidence for or against the expected $\approx 800\,$ pc long tidal tails. This work thus concentrates on the problem of developing a  new method 
(called the compact convergent point (CCP) method) for finding the leading and trailing tails in their full expected length.

This work is structured as follows: We introduce Hyades-like cluster evolution on a realistic orbit and reveal the structure of its tidal tails in various \textit{Gaia} spaces in Sec.~\ref{sec:simul}. 
In the same section we also introduce the new 
CCP method in order to project the tidal tails that are otherwise spread in spatial and velocity coordinates into a compact configuration. In  Sec.~\ref{sec:search} we use the information gained in  Sec.~\ref{sec:simul} and identify candidate Hyades tidal-tail members in the \textit{Gaia} data spanning a length of 800 pc. In Sec.~\ref{sec:lumps} we study the effect of an initial angular moment and  of a lumpy Galactic potential on the formation and evolution of tidal tails for the first time. The last two sections, Sec.~\ref{sec:disc} and Sec.~\ref{sec:concl}, present a discussion and some conclusions.

\section{Simulated \textit{Gaia} view of tidal tails}
\label{sec:simul}
The main aim of this section is to present for the first time  the results of a simulation in the observable (and derived from observable) parameters of the \textit{Gaia} catalogue of the tidal tails of this star cluster. We then determine whether these simulations can be used to detect the full or at least a larger extent of the tidal tails of  the Hyades star cluster.

We used the code called astronomical multipurpose software environment,  \texttt{AMUSE} \citep{amuse_book,amuse2013,amuse2013b,amuse2009}. It allowed us to couple the evolution of a star cluster and the description of a Galactic potential to simulate the evolution of a Hyades-like star cluster on a realistic orbit.

\subsection{Present-day parameters of the Hyades and coordinate setup }
\label{sec:coords}

We summarise in Tab.~\ref{table:Hyad} some of the present-day properties of the Hyades as derived in~\cite{Roeser+11}, \cite{Goldman2013}, and \cite{Ernst+11}. In particular, \cite{Roeser+11} and \cite{Goldman2013} provided a catalogue of Hyades members down to  0.116 $M_{\odot}$ and 0.1 $M_{\odot}$, respectively. The reported parameters are still discrepant because the method that are used in the different studies differ. As an example, the age of the Hyades is estimated to be $625\pm 50$ Myr in \cite{Perryman1998}, $648\pm 50$ Myr in \cite{DeGennaro2009}, 500-650 Myr in \cite{Lebreton2001}, 675-700 Myr \citep{Hy_age1}, 680 Myr \citep{Hy_age2}, 675-700 Myr \citep{Hy_age3}, 650 $\pm$ 70 Myr \citep{Hy_age4}, and 795 Myr \citep{GDR2_Hyades}. For the purpose of this study, we adopted a nominal age value of 655 Myr and considered an age spread of 620 Myr to 695 Myr (see Tab.~\ref{table:Hyad}) for the models we present.
The velocity dispersion of the Hyades star cluster has been estimated in a number of studies. \cite{Madsen03} and \cite{Makarov2000} reported a velocity dispersion of about $0.3$ km/s based on the cluster virial mass. 
\cite{Hy_age1} concluded that the upper limit on the velocity dispersion is about $0.5$ km/s. 
\cite{Roeser+11} and \cite{Oh2015} derived an even higher value for the velocity dispersion of about 0.8 km/s. When we discuss our result, we consider all these values.

\begin{table}
\caption{Present-day properties of the Hyades star cluster based on \cite{Roeser+11}, \cite{Ernst+11}, \cite{Goldman2013}, \cite{GDR2_Hyades}, \cite{Hy_age1}, \cite{Hy_age2}, \cite{Hy_age3}, and \cite{Hy_age4}. ** See the text for a discussion of the velocity dispersion.  }
\label{table:Hyad}      
\centering                                
\begin{tabular}{l l }          
\hline\hline                       
parameter & value \\    
\hline     
age & $\approx$ 600-700 Myr \\
number of stars ($r<30$ pc) & 724 \\
stellar mass ($r< 30$ pc) & 469 $M_{\odot}$\\
3D velocity dispersion ($r< 9$ pc) & $\approx 0.3-0.8$ km/s ** \\
\hline
%inserts single line
\end{tabular}
\end{table}
%Jacobi radius rJ 8.6 pc (9.0 pc)
%Number of stars N(rJ) 354 (359)

We used the \texttt{Astropy SkyCoord} package to compute all coordinate transformations in this work \citep{astropy}. Galactic and equatorial coordinates are centred on the barycenter of the Solar System. For the galactocentric coordinates we used the \texttt{'pre-v4.0'} \texttt{Astropy} frame default setup, which adopts the coordinates of the Sun in the galactocentric Cartesian system as
\begin{align*}
[X_{\odot},Y_{\odot},Z_{\odot}] = [8300.0 \, \mathrm{pc},  0.0 \, \mathrm{pc}, 27.0 \, \mathrm{pc}],
\end{align*}
and the velocities as  
\begin{align*}
[V_{X,\odot},V_{Y,\odot}, V_{Z,\odot} ] = 
[11.1,\, 232.24,\, 7.25]\, \mathrm{km/s}.
\end{align*}
This framework was chosen because it is one of the default settings in the  widely used tool \texttt{AstroPy} \citep{astropy2013,astropy}. The circular velocity at the distance of the Sun agrees well with observations \citep[e.g.][]{Reid2014}. The adopted distance of the Sun from the Galactic mid-plane is 10 pc larger than the distance found in recent studies \citep{Yao2017,Karim2017}. This difference does not 
significantly affect the interpretation 
 of the computed star cluster, in addition, \cite{Bland2016} estimated the distance of the Sun from the Galactic plane to be $25\pm5$pc. Thus the default setting from \texttt{AstroPy} was used here. 

When we refer to the Hyades star cluster, we use the cluster parameters published by the \citet[][their Table A.3]{GDR2_Hyades}. That is, the velocity of the cluster in equatorial barycentric spherical coordinates is
\begin{align*}
[\mu^{*}_{\alpha},\mu_{\delta}, V_r] = 
[101.005 \, \mathrm{mas/yr}, -28.490\,  \mathrm{mas/yr}, 39.96 \, \mathrm{km/s} ]. 
\end{align*}

In their Table A.3, \citet[][]{GDR2_Hyades} published the values of the equatorial barycentric Cartesian velocities ($U'$, $V'$, $W'$) and parallax, $\varpi$, which allowed us to derive the cluster R.A. and Dec coordinates by transforming between the spherical and Cartesian coordinates, \\
\begin{align*}
[\mathrm{R.A.}, \mathrm{Dec}, \varpi] = [67.985 \, \mathrm{deg}, 17.012 \, \mathrm{deg}, 21.052 \,  \mathrm{mas}].
\end{align*}

\subsection{$N$-body setup}
\label{sec:nbody}

The formation and evolution of tidal tails mostly depend on the interplay between the dynamical and stellar evolution of the host star cluster and the interaction with the Galactic potential. The complexity of the study is therefore mostly driven by the complexity of the physical processes at play, the uncertainties of the initial conditions of a star cluster, and the parameters describing the Galaxy. 
The computations we present here were all performed with \texttt{AMUSE} \citep{amuse_book,amuse2013,amuse2013b,amuse2009}, which allowed us to couple the star cluster evolution and the description of a Galactic potential in a convenient way. 

\subsubsection{Star cluster setup and evolution}\label{sec:int}
To integrate the internal dynamical evolution of star clusters,  we used 
the \texttt{Huayno} code \citep{Huayno2012,Huayno2014}, 
which is a class of $N$-body integrators that implements 
a variety of kick-drift-kick algorithms through the Hamiltonian
splitting strategy of adjustable order \citep[see][for detailed descriptions]{amuse_book,Huayno2012,Huayno2014}. 
To integrate the evolution of a Hyades-type star cluster, the \texttt{Huayno} code was implemented within the \texttt{AMUSE} environment, which allowed us to couple the direct N-body integration with the stellar evolution code \texttt{SeBa} \citep{Zwart1996,Toonen2012} and the integration along the orbit (see Sec.~\ref{sec:gal} for the details). 
We used the optimal softening length,
\begin{equation}\label{eq:eps}
\varepsilon =  
2GM_{\mathrm{cl}}^2/(N |U|),
\end{equation}
where $M_{\mathrm{cl}}$ is the stellar mass of the cluster, $G$ is the gravitational constant, $N$ is the number of stars, and $U$ is the potential energy of the cluster. For our simulations the value of the softening length was $\varepsilon = 0.004 \, \mathrm{pc}$ for the initial cluster parameters. The mean stellar separation was $\approx 0.2$ pc, which is significantly larger than the softening length and ensured  a sufficiently correct evaluation of two-body relaxation.
The physical origin of the optimal softening length described by Eq. (\ref{eq:eps}) is $\varepsilon = 4 R_{\mathrm{virial}}/N$.
For the other parameters, we used the default \texttt{Huayno} values such as the type~8 integrator HOLD\_DKD  \citep[for details, see][]{Huayno2012,Huayno2014, amuse_book}.
The parameters we adopted to define the initial setup for the cluster are given in Tab.~\ref{table:CLP}. In particular, the stellar initial mass function  (IMF) $\xi_{\star}(m)$ was implemented as a 
two-part power-law function based on \cite{Kroupa+01},\begin{equation}
\xi_{\star} (m) =    \left\{ \begin{array}{ll}
\,\,\, \,\,\,\,\,\,m^{-1.3} \hspace{1.65cm} 0.08\leq m/M_{\odot}<0.50 \,, \\
0.5\, m^{-2.3} \hspace{1.65cm} 0.50\leq m/M_{\odot}<120 \,, \\
\end{array} \right.
\label{eq:IMF}
\end{equation}
where $\mathrm{d} N_{\star} = \xi_{\star}(m) \, \mathrm{d} m$
is the number of stars in the mass interval $m$ to $m+\mathrm{d} m$.  We did not include primordial binaries in the initial cluster setup.

Essentially, the initial model can be seen as constituting a post-gas-expulsion re-virialised model \citep{Kroupa+01}. 
We note that it is interesting and important to study the effect of the exact initial conditions in detail (e.g. relation of initial mass to radius that matches the observed present-day properties of the Hyades, the gas expulsion phase, the stellar IMF, and the initial binary fraction). In this respect,  \cite{DK20a, DK20b} have recently studied the effects of gas expulsion on the early evolution stage ($100\,$Myr) of a tidal tail. With our setup we are limited to study the overall morphology of tidal tails; see Sec.~\ref{sec:nbody_dis}, where the limitations of our method are discussed at length. 
%It is chosen to be suitable for our aim of addressing the detectability of tidal tails with \textit{Gaia} data. 

To ensure that our setup provides a reliable output comparable with the Hyades and the morphology of its tidal tails, we proceeded as follows: 1)  For a number of simulations we ran the same setup with the \texttt{PhiGRAPE} code \citep{grape2007, grape2009} that is available in AMUSE and was also used by \cite{Kharchenko2009} to study open clusters and their tidal tails. We confirmed that these simulations give comparable results. 
2) We used the same initial setup as \cite{Ernst+11}, who used the Aarseth {\tt Nbody6} code to simulate the tidal tails of the Hyades, to allow a comparison. 

%In Sec.~\ref{sec:nbody_dis} we discuss the limitation of our $N$-body method/integrator used. In a nutshell, the aim of our study is the formation and evolution of tidal tails which is not sensitive to the cluster's strong encounters that are not accurately described by our method which is using softening. 

\begin{table}
\caption{Initial parameters of the Hyades star cluster based on \cite{Roeser+11} and \cite{Ernst+11}. \textbf{\textbf{\textbf{M1, M2, and M3}}} are the names of our models} 
\label{table:CLP}      
\centering                                
\begin{tabular}{l l }          
\hline\hline                       
parameter & value \\    
\hline     
Cluster profile & Plummer mode \\
initial mass $M_{cl,0}$ & $1230\,M_{\odot}$\\
half-mass radius, $r_h$ & 2.62 pc\\
IMF & canonical IMF, Eq. (\ref{eq:IMF})\\
angular momentum, $L_z$ & \textbf{M2} $L_z$<0, \textbf{M1} $L_z$=0, \textbf{M3} $L_z$>0\\ 
\hline       
initial cluster coordinates & (galactocentric Cartesian f.)\\
\hline 
$X,Y,Z$ & 5.605, 7.382, -0.045 kpc \\
$v_X, v_Y, v_Z$ & 163.8, -114.0, 3.4 km/s\\
\hline
%inserts single line
\end{tabular}
\end{table}

\subsubsection{Galaxy parameters and computation setup} \label{sec:gal}
To follow the orbital evolution of stellar clusters in the Galaxy, we 
used a detailed model of the Galactic potential from the \texttt{Galaxia} module \citep{Martinez2016, Martinez2017} available in \texttt{AMUSE}. The \texttt{Galaxia} module consists of a semi-analytical axisymmetric potential given by a Galactic bulge, disc, and halo \citep{Allen1991} and allows additional non-axisymetric components, such as a bar and spiral arms. We did not implement any non-axisymetric components. The adopted physical values of the MW are summarised in Tab.~\ref{table:MWp}. The parameters that are not listed are the same as the \texttt{Galaxia} module default values. 
For the MW potential we considered two sets of parameters from \cite{Allen1991} and \cite{Irrgang2013}. 
\cite{Allen1991} has been used in recent studies \citep[e.g.][]{DK20b} and therefore serves as a good comparison benchmark. \cite{Irrgang2013} presented updated values of the MW potential, see Tab.~\ref{table:MWp} for comparison.
We used the sixth-order rotating \texttt{BRIDGE} (to bridge integration within the cluster to orbit integration of stars in the Galaxy) by setting the option \texttt{SPLIT\_6TH\_SS\_M13} 
with~13 symmetric evaluations of the force per time step, $\Delta t = 0.25$ Myr. We ensured that the relative energy error per time step was always below $10^{-7}$ \citep{Martinez2016,Martinez2017,amuse_book}.

\begin{table}
\caption{MW parameters. Values are taken from \cite{Allen1991} for \textbf{M1-M4} and from \cite{Irrgang2013} for \textbf{M5}. }
\label{table:MWp}      
\centering                                
\begin{tabular}{l l l}          
\hline\hline                       
parameter & value {\textbf{M1-4}} & value {\textbf{M5}}\\    
\hline                                   
Mass of the bulge ($M_b$) & $1.41 \cdot 10^{10}\, M_{\odot}$ & $0.95 \cdot 10^{10}\, M_{\odot}$ \\
Bulge scale-length ($b_1$) & 0.03873 kpc & 0.23 kpc\\
Disc mass ($M_d$) & $8.56 \cdot 10^{10}\, M_{\odot}$ & $6.6 \cdot 10^{10}\, M_{\odot}$\\
Disc scale-length 1 ($a_1$) & 5.31 kpc & 4.22 kpc \\
Disc scale-length 2 ($a_2$) & 0.25 kpc & 0.292 kpc\\
Halo mass ($M_h$) & $1.07 \cdot 10^{11}   \, M_{\odot}$ & $0.24 \cdot 10^{11}   \, M_{\odot}$\\
Halo scale-length ($a_3$) & 12 kpc  & 2.565 kpc\\
lumpy galaxy & NO (\textbf{M1,2,3}) & NO\\
 & YES (\textbf{M4}) & \\
-- lump mass function & power-law (slope -1.6),\\
-- minimum lump mass & $M_{min} \in (10,10^3)\,M_{\odot}$\\
-- maximum lump mass & $M_{max} \in (10^4,10^8)\,M_{\odot}$ \\
-- lump size &  10 --100 pc \\
\hline                                             %inserts single line
\end{tabular}
\end{table}

\begin{figure}
        \centering
        \includegraphics[width=\hsize]{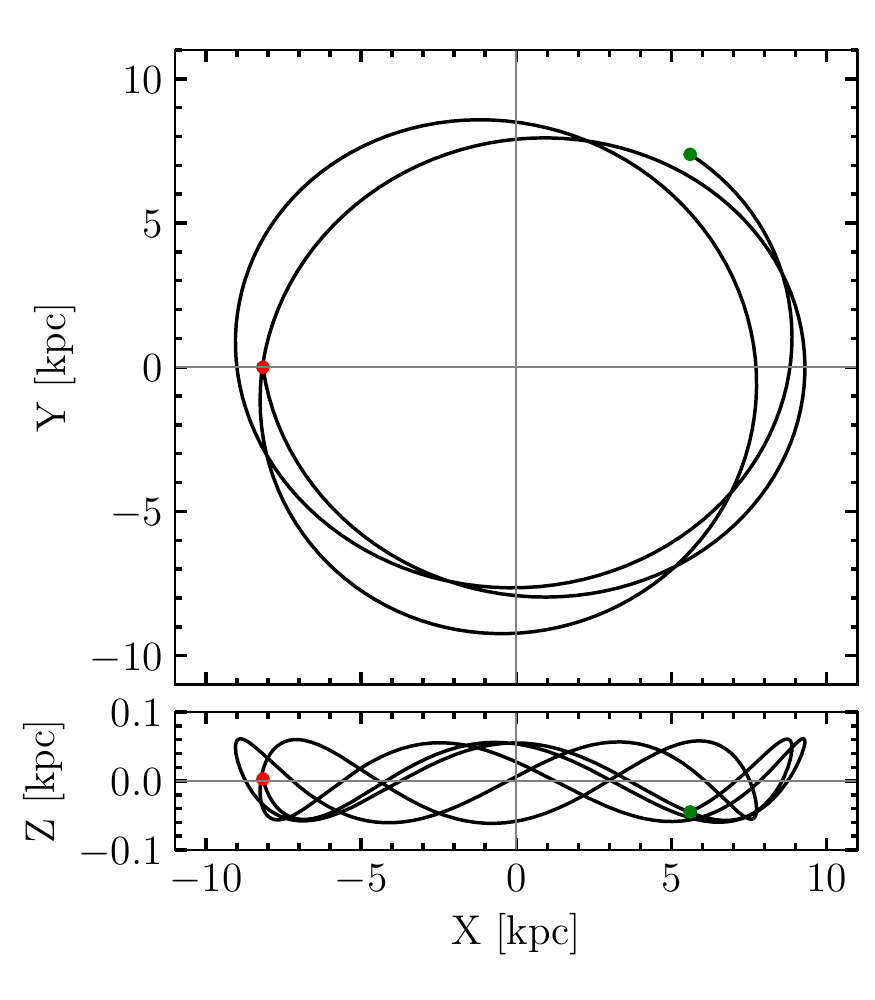}
        \caption{Backward-integrated orbit in galactocentric Cartesian coordinates from the present-day coordinates for the Hyades (red circle) to its likely birth location (green circle). The backward-integration was performed using a point mass in the same Galactic potential as the simulated Galactic cluster. }
        \label{fig:orbit}
    \end{figure}

We estimated the initial position of the Hyades in the Galaxy using its current position as described in Sec.~\ref{sec:coords} and by integrating backwards (using the integration method described above, i.e. the same setup as for the forward cluster integration the orbit of a point mass for the nominal age of the Hyades of 655 Myr; see Tab.~\ref{table:Hyad}). This was done for the two potential MW parametrisations we considered. 
The derived trajectory is shown in galactocentric Cartesian coordinates in Fig.~\ref{fig:orbit} (see Tab.~\ref{table:CLP} for the initial coordinates for the Hyaes).

With Galactic parameters we computed the tidal radius of the star cluster. 
The tidal radius, $R_t$ is here defined as (see \citealt{BinneyTremaine87}) 
\begin{equation}
    R_t = \frac{M(R<R_t)}{3 M_{\mathrm{Gal}}} \cdot R_{\mathrm{Gal}} \,
,\end{equation}
where $M(R<R_t)$ is the stellar mass within the cluster tidal radius, $M_{\mathrm{Gal}}$ is the Galactic mass within the cluster orbit, and $R_{\mathrm{Gal}}$ is the distance of the star cluster from the Galactic centre. 
We used an iterative procedure to estimate $R_t$ and $M(R<R_t)$ for a given position of the cluster in the Galactic potential.

To facilitate comparison of our simulation results with the observed Hyades star cluster, we followed several steps: 
1) The backward-integrated initial conditions ensured that the simulated star cluster was almost exactly at the position of the present-day Hyades. However, because the cluster is losing stars, its orbit is slightly different from the orbit of a single particle. To correct for this difference, we shifted each simulation snapshot to the present-day position of the Hyades star cluster and aligned their velocity vectors,  which necessitated a rotation of the system. This operation thus makes the necessary corrections in projection on the sky without changing internal kinematics.
This is standard practice, but previous papers have usually shifted the star cluster to the centre (zero) position and aligned the velocity vector with one of the  Cartesian axes such that no rotation was applied. The advantage of our method is that our results are directly comparable with the \textit{Gaia} data in all parameter spaces. 
2) The age of the cluster is uncertain. This presents a clear obstacle because, at a given age, the cluster might have a certain special feature in one of the projections that shifts slightly with age. To account for this uncertainty and ensure that we did not discard valuable stars from the \textit{Gaia} catalogue, we stacked several snapshots spanning over 75 Myr from the simulation. 
3) We computed the cluster and its tails in the Galactic (spherical and Cartesian) and galactocentric (spherical, cylindrical, and Cartesian) coordinate systems and representations (as CP projected velocities), which are directly comparable with \textit{Gaia} data. We note that to the best of our knowledge, this is the first time that simulations are reported that are fully comparable with \textit{Gaia} measurements. 
We emphasize that the internal kinematics of a star cluster and of its tail in general depends on the acceleration that is generated along the orbit. For an orbit with  a large variation of the acceleration, that is, for highly eccentric orbits, the simulation snapshots would therefore differ greatly for different spatial positions. In the case of the Hyades star cluster, the orbit is close to circular. The stacking of snapshots along the orbit is therefore not significantly affected by changes in acceleration along the orbit. 

\begin{figure}
        \centering
        \includegraphics[width=\hsize]{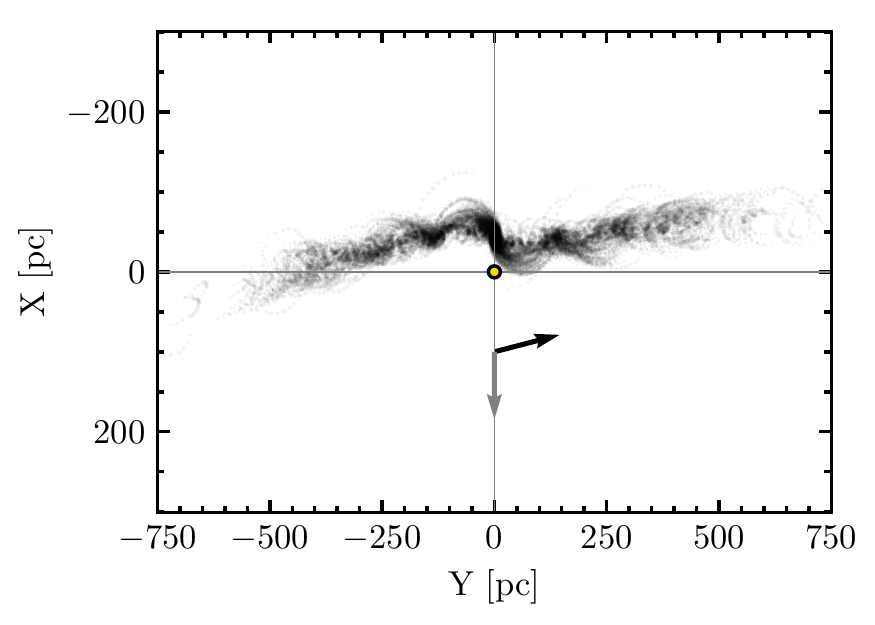}
        \caption{Hyades-like simulated star cluster at the series of snapshots at 620, 625, 630, 635, 640, 645, 650, 655, 660, 665, 670, 675, 680, 685, 690, and 695 Myr in Galactic Cartesian coordinates. The Sun is marked as a yellow point. The grey arrow points to the Galactic centre, and the black arrow is the cluster velocity vector in the corresponding coordinates. The time stacking of snapshots shows the movement of individual stars to and from the epicyclic overdensities.}
        \label{fig:YX_M}
    \end{figure}

We now discuss  the resulting star cluster and its tidal tails for combined snapshots from 620 Myr to 695 Myr with a time step of 5 Myr time-step for the non-rotating cluster in the smooth Galactic potential
(model {\bf M1}). This is plotted for several parameters measured by the \textit{Gaia} satellite. 
In Fig.~\ref{fig:YX_M} the star cluster and its tidal tails are plotted. The time-stacking of several snapshots shows the epicyclic motions of individual stars.
Otherwise, the extent and structure of the tails are comparable to previous simulations \citep{Ernst+11,Chumak+05}. 
We note that the realistic star cluster trajectory with excursions out of the Galactic plane does not have a large effect on the physical appearance of the tails. However, it is very important
to consider the projection effects caused by the present-day position of the cluster and its tail on the sky. That is, whether  the tails are in or out of the Galactic plane changes their appearance on the sky and in other parameter spaces significantly (proper motions and their derived projections).

\begin{figure}
        \centering
        \includegraphics[scale=1.0]{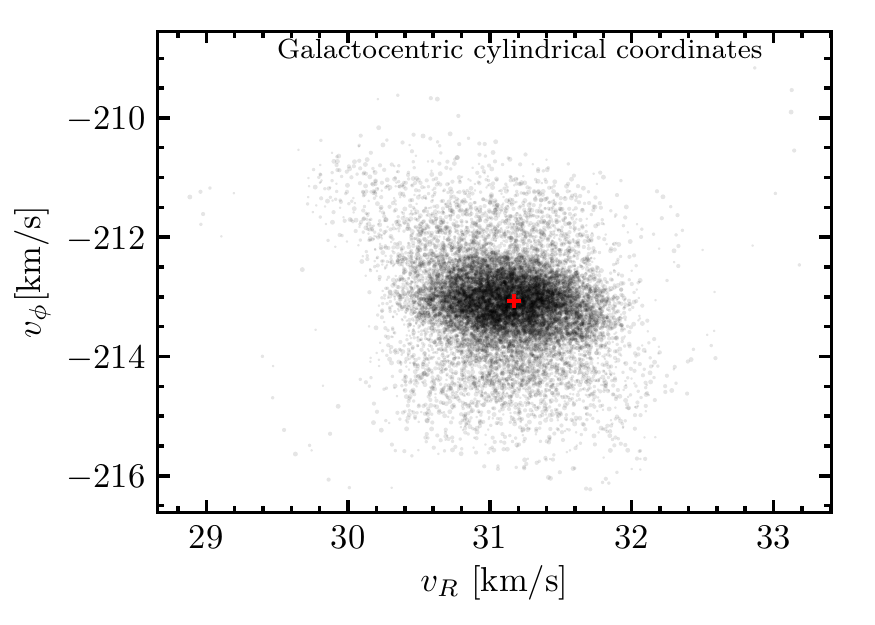}
        \includegraphics[scale=1.0]{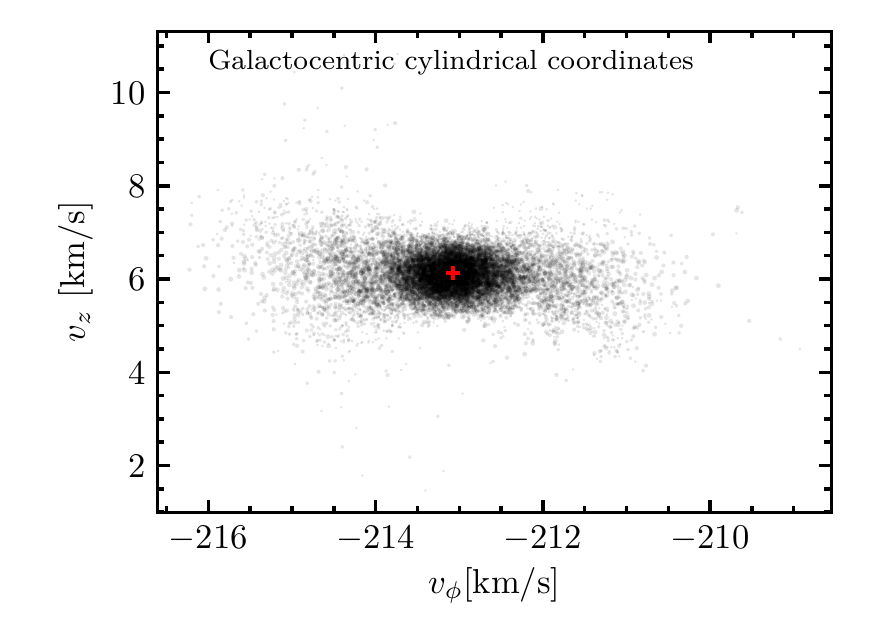}
        \caption{Simulated star cluster and its tidal tails (\textbf{M1}) plotted in velocity space in galactocentric cylindrical coordinates. The red cross marks the central velocity value of the Hyades.
        }
        \label{fig:cylin}
    \end{figure}

\cite{MeingastAlves19} used galactocentric cylindrical coordinates centred on the Hyades cluster with a search radius of 1.5 km/s in order to search for its tidal tails. 
Fig.~\ref{fig:cylin} shows that in galactocentric cylindrical coordinates, the two tidal tails in their entirety indeed extend for only a few km/s. This is because galactocentric cylindrical coordinates eliminate projection effects introduced by the extended nature of the tails. 
This coordinate system presents the most compact representation of the members of the tail and is independent of the model. 

However, \cite{MeingastAlves19}  recovered only a small part of the tidal tails in comparison to \cite{Roeser+19}, who used the CP method (discussed in more detail below).
One reason for this is that radial velocity measurements are required to compute velocities in galactocentric coordinates from the \textit{Gaia} astrometric data. This presents a significant limitation because only bright stars have radial velocity measurements in the \textit{Gaia} catalogue ($G < 15$ mag),  and thus they constitute a very incomplete data sample to work with.  
The other aspect is the extension of the tail in velocity space. As shown in Fig.~\ref{fig:cylin}, the central region is dense and thus can be recovered by a standard clustering algorithm such as DBSCAN \citep[e.g.][]{Jerabkova2019}, but the outer parts drop rapidly in density while the contribution from contamination grows (because the extended tails are farther from the Sun), making it difficult to distinguish the tails from the field contamination.  
 
\begin{figure}
        \centering
        \includegraphics[width=\hsize]{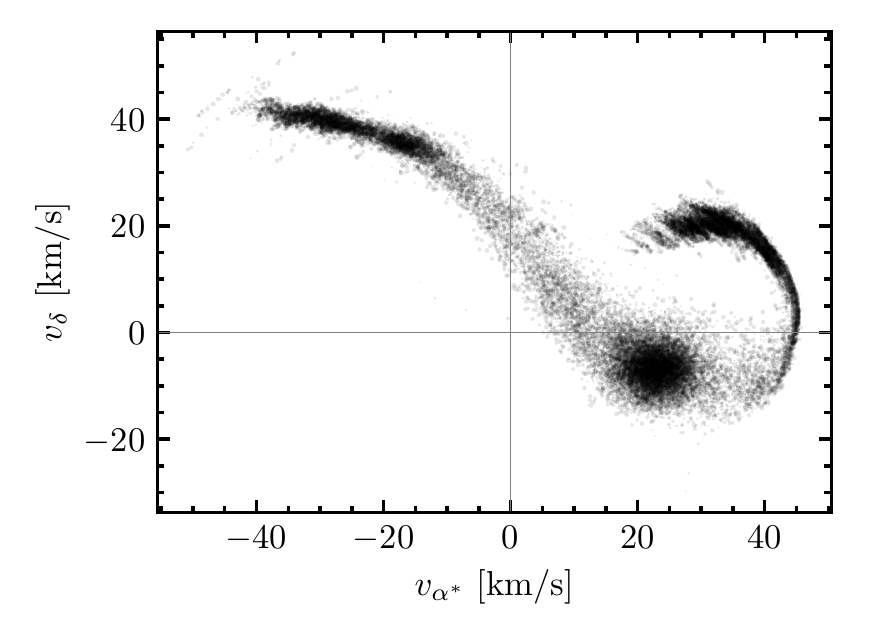}
        \includegraphics[width=\hsize]{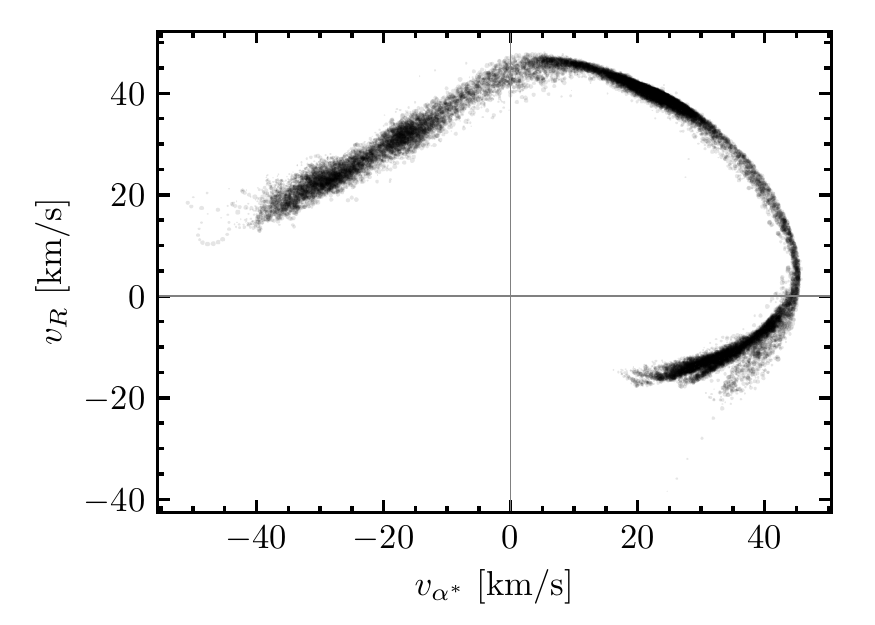}
        \includegraphics[width=\hsize]{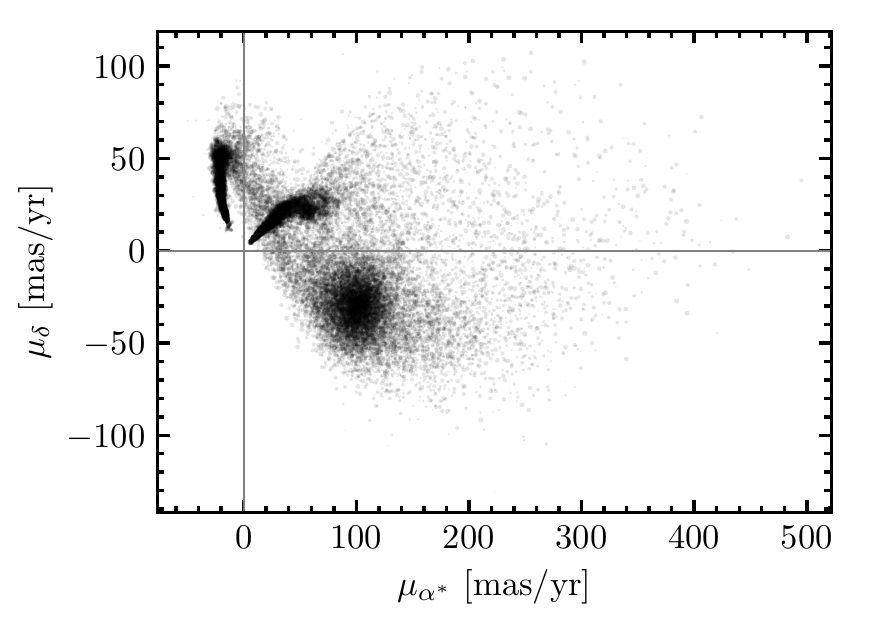}
        \caption{Modeled (\textbf{M1}) Hyades star cluster and its tail at time-stacked snapshots from 620 Myr to 695~Myr. \textbf{Top panel:} Proper motion distribution in physical km/s units. The central clump represents the star cluster, and the extended distribution (up to 80 km/s) consists of members of the tidal tail. \textbf{Middle panel:} Proper motion in R.A. and radial velocity distribution showing the same 80 km/s span for the tidal tails members. \textbf{Bottom panel:} On-the-sky distribution of proper motions of the modeled cluster and its tails. Because of the proximity of Hyades to the Sun, the spread in proper motion is large (see the loose lump representing the cluster members), and the two overdensities in the shape of tentacles caution against using a blind clustering algorithm in order to search for tidal tails in the data. }
        \label{fig:proj}
    \end{figure}

Fig.~\ref{fig:proj} shows the projection of the model tidal tails into  space defined by  some selected combinations of \textit{Gaia} parameters. The core of the cluster is represented by a dense clump in proper motions (on the sky $\mu^*_{\alpha}$, $\mu_{\delta}$, or distance-corrected $v^*_{\alpha}$, $v_{\delta}$) and in radial velocities. However, the tidal tail is then projected in various shapes with several clumps or overdensities. 
Without the knowledge given to us by the model, it would be very difficult if not impossible to recover these structures by searching for overdensities in proper motions and radial velocities in the real dataset.
 The shape of a cluster and its tails in proper motions and radial velocities depends on the physical properties of the tails and on the position in the galaxy. This means that the very same tidal tails would be represented by different structures in proper motions in the sky projections if they were located elsewhere. 
 
\begin{figure*}
    \centering
    \includegraphics[scale=1]{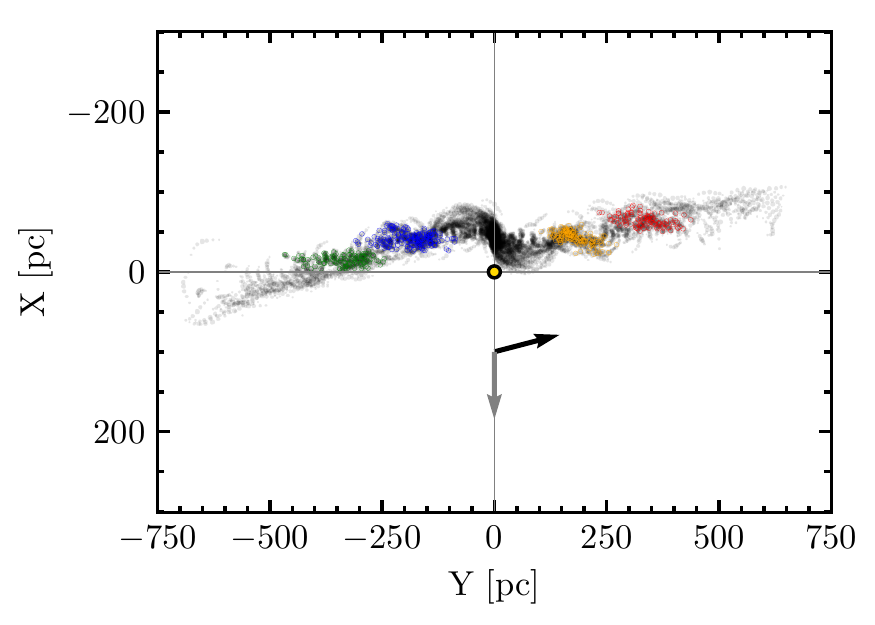}
    \includegraphics[scale=1]{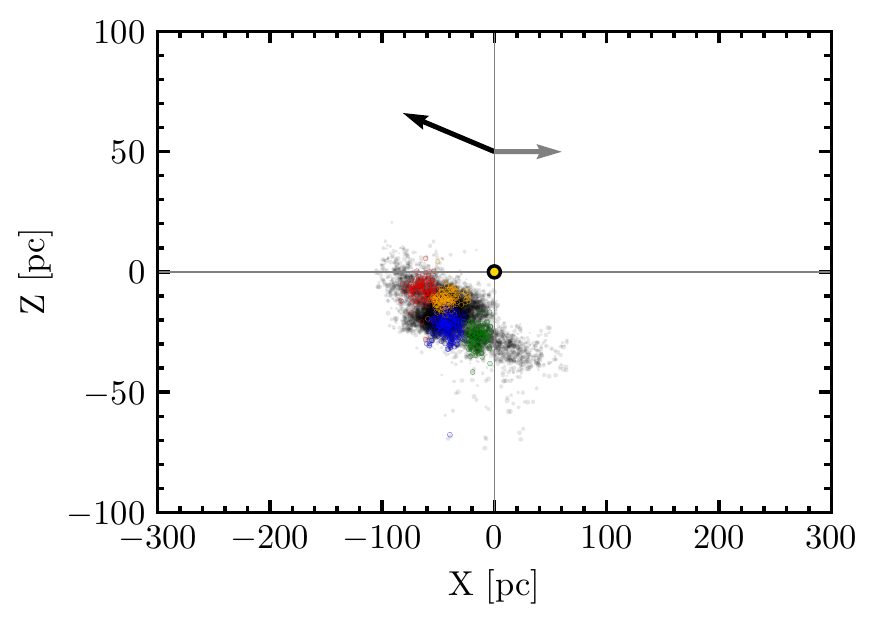}
    \includegraphics[scale=1]{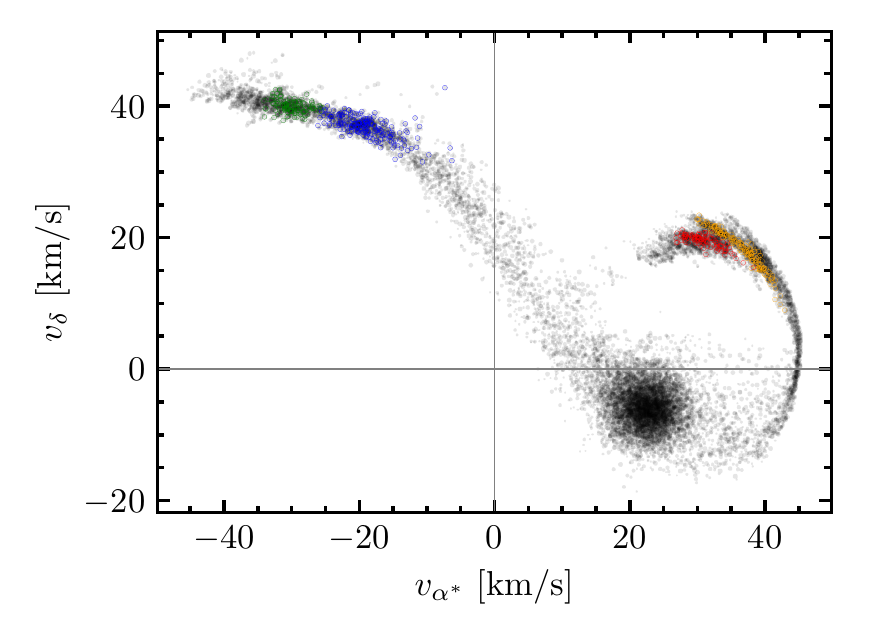}
    \includegraphics[scale=1]{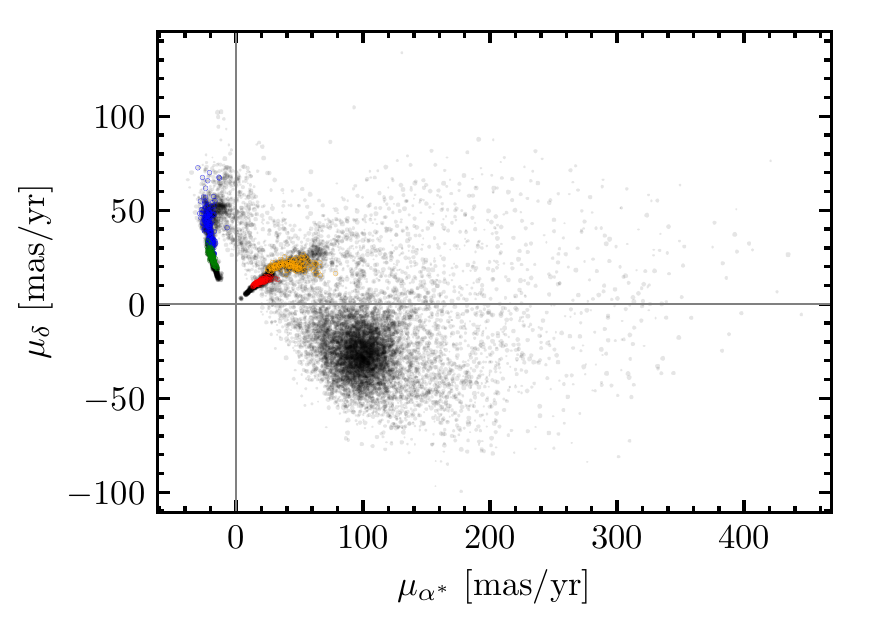}
     \includegraphics[scale=1]{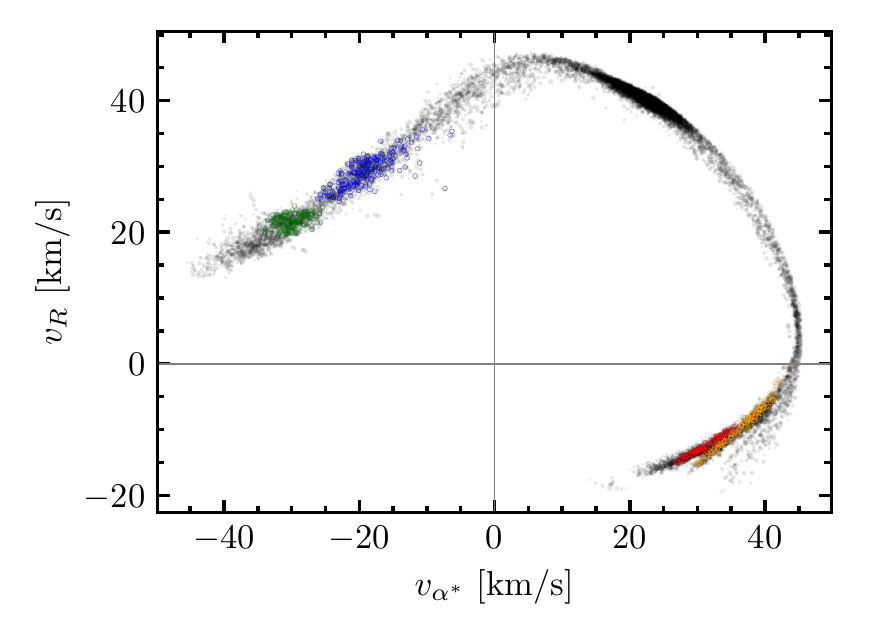}
     \includegraphics[scale=1]{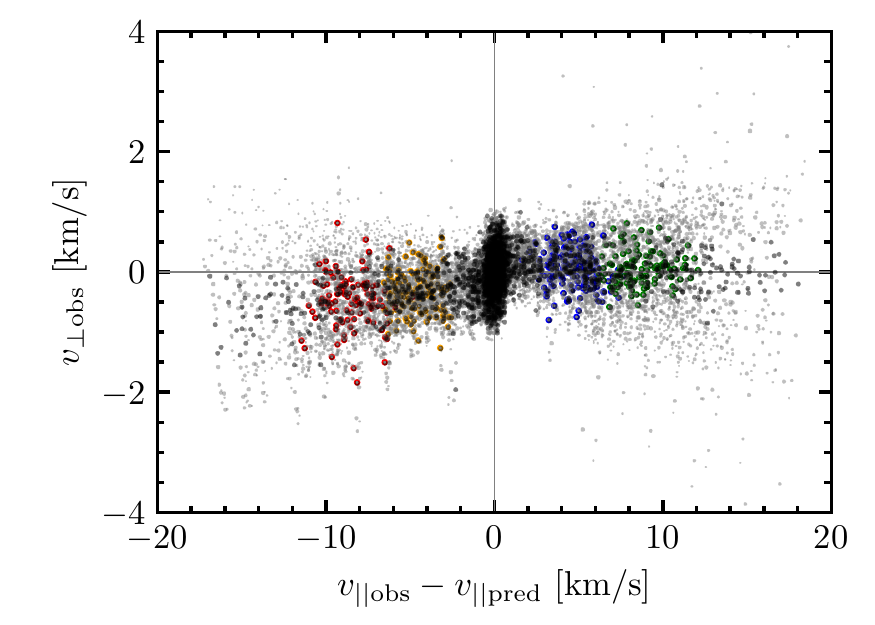}
    \caption{Epicyclic overdensities depicted in colour in different plots showing various phase-space coordinates for model \textbf{M1}. We used the $X-Y$ phase-space to identify epicyclic overdensities. Stars belonging to the epyciclic overdensities are plotted in colour and appear in various \textit{Gaia} parameter spaces.}
    \label{fig:epyc}
\end{figure*}

Models are expected to be able to easily test for the epicyclic overdensities, which may also be used to identify tail members in the \textit{Gaia} data, as has been pointed out in a recent study of \cite{Oh2020}. In order to understand the position of the epicyclic overdensities in different spatial and velocity space, we used the simulated \textbf{M1} model and identified the overdensities in $X-Y$ Galactic Cartesian coordinates, the see upper left panel of Fig.~\ref{fig:epyc}. In the other panels of the same figure, the epicyclic overdensities are plotted with the same colour-coding.

\subsection{Compact convergent point method: Recovery of the full extent of tidal tails} 
\label{sec:ccp}
The conclusion so far is that tidal tails are not in general represented by a simple overdensity in velocity space. While plots showing modelled proper motions and radial velocities of an open cluster and its tail have been documented here for the first time, the difficulties of detecting extended co-moving structures have been known and explored in detail for quite a while \citep[e.g.][]{vanLeeuwen09}.
Two effects contribute to the appearance of tidal tails in velocity space. We describe them in detail below.  

\textbf{The first} effect is a projection effect. 
In an imagined set of perfectly co-moving stars on a circular orbit in the Galaxy, these points will have the same Cartesian velocities in the Galactic coordinates. However, the on-the-sky projection effects mean that stars with different positions on the sky will have different proper motion values. A good solution to this approach is the well-established CP method \citep[][and references therein]{Smart1939,Jos1999,vanLeeuwen09}. The CP method is based on correcting proper motions for projection effects of velocities on the celestial sphere that are caused by the spatial extent of a stellar structure. 
The proper motions (in km/s taking individual parallaxes into account) of the constituent stars in a spatially co-moving group point to a so-called CP on the sky. It is then possible to compute the expected or predicted parallel and perpendicular velocity components that point to the CP, $v_{||\mathrm{pred}},v_{\bot\mathrm{pred}}$, for a star with a given position in space or on the sky by defining the CP, $v_{\bot\mathrm{pred}}=0$. These values are then compared with the projected values of measured proper motions (in km/s), $v_{||\mathrm{obs}},v_{\bot\mathrm{obs}}$. For stars that are co-moving, the differences $v_{||\mathrm{pred}} -  v_{||\mathrm{obs}}$ and $v_{\bot\mathrm{obs}}$ will be close to zero. 

\begin{figure}
        \centering
        \includegraphics[scale=1.0]{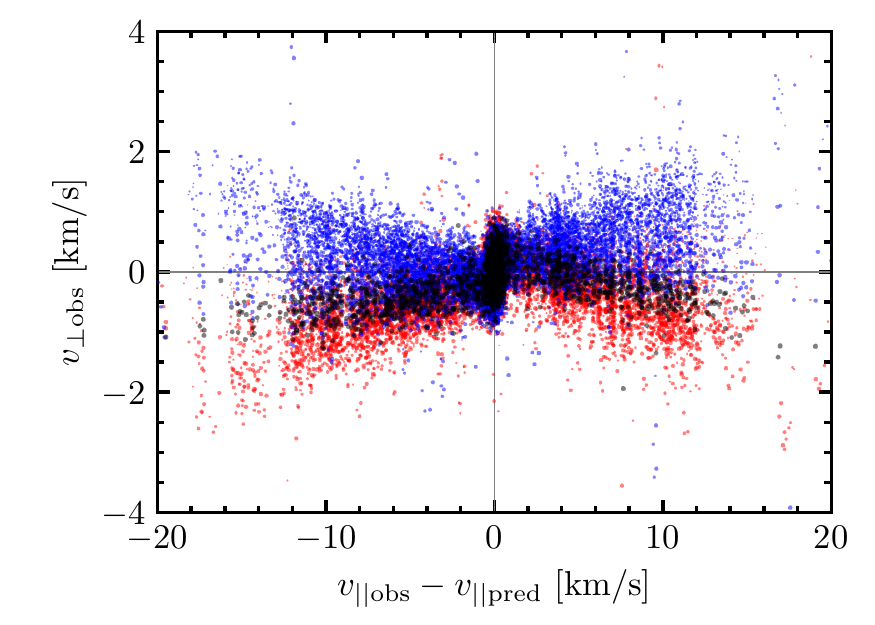}
        \includegraphics[scale=1.0]{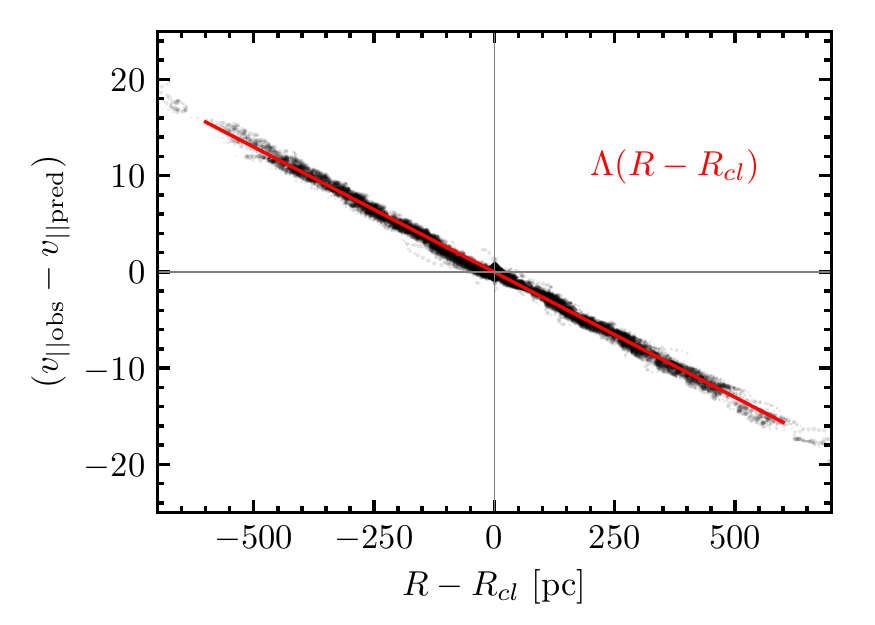}
        \includegraphics[scale=1.0]{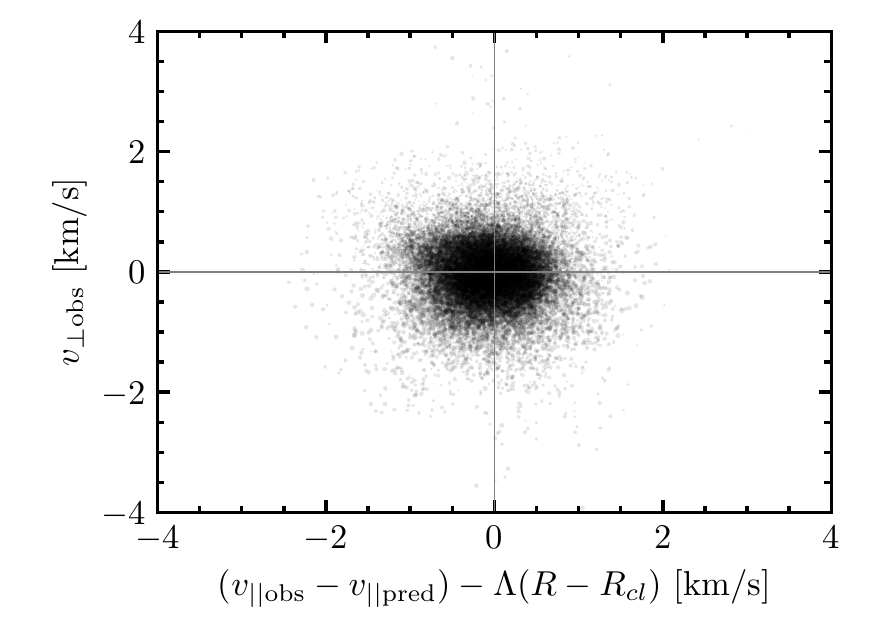}
        \caption{Hyades cluster and its tidal tails used to explain the transition from the classical CP metod to newly introduced CCP method. 
        \textbf{Top panel:} Classical CP method diagram showing the velocity distribution of \textbf{M1}-simulated Hyades and its tidal tails projected onto the parallel (horizontal axes) and perpendicular direction (vertical axes) to the CP. The predicted velocity component is removed based on the on-the-sky position of individual objects (see text for more details). The snapshots beginning at 620~Myr to 695Myr are plotted, each placed at the Hyades position in spatial and velocity space. In red we show snapshots at 620, 625, 630, 635, 640, 645, and 650~Myr, in black those at 655~Myr, and in blue the snapshots at 660, 665, 670, 675, 680, 685, and 690~Myr. This shows the scatter in the CP diagram that is caused by the time evolution of the tidal tail. The extent of the points on the horizontal axes  reaches 40 km/s.  \textbf{Middle panel:} Dependence of $v_{||\mathrm{pred}} -  v_{||\mathrm{obs}}$ as used in the CP method (horizontal axes in the top panel) as a function of the distance from the cluster centre ($R_{cl}$). The objects farther from the cluster centre move relatively faster. The red line shows the approximately linear relation denoted as the function $\Lambda(R-R_{cl})$. \textbf{Bottom panel:} The modelled points in the CP diagram (top panel) without the velocity trend from the middle panel described by the function $\Lambda(R-R_{cl}).$  This operation compacts the whole structure (star cluster plus the tidal tail) within a region spanning a few km/s. We call this method the CCP method and use it in Sec.~\ref{sec:gaia_res} to search for the full extent of the Hyades tidal tails in the \textit{Gaia} data.}
        \label{fig:CCP}
    \end{figure}

\textbf{The second} effect on the appearance of the tidal tails in velocity space consist of the actual velocity differences between the members of the tail that are in general caused by their drifting apart because they are on slightly different orbits in the Galaxy. Individual stars in tidal tails are indeed not on exactly the same orbits and have non-zero relative velocities with respect to the cluster. These effects are amplified with time as the members drift apart from each other.  
These differences are not corrected for by the CP method and complicate the detection of the outer parts of tidal tails and other extended structures. This is shown in the top panel in Fig.~\ref{fig:CCP}, which  plots the projected modelled star cluster and its tail (\textbf{M1}) using the the CP method (in black and for different ages in colour). 

The power of the standard CP method has been demonstrated by \cite{Roeser+19}. These authors used the \textit{Gaia} DR2 and found that the tidal tails of the Hyades cluster extend up to 200 pc from the cluster. 
The top panel of Fig.~\ref{fig:CCP} can be compared with Fig.~1 of \cite{Roeser+19}. We used the age stacking method again and plot several time-snapshots from our simulation in one plot. The CP plot shows the central dense part (the stellar cluster), and that the tidal tails extend up to a distance of 20~km/s in velocity space. \cite{Roeser+19} were only able to detect the overdensity above the background to an extent of up to 5~km/s, and their detected structure is not symmetrical. This is again a reminder that it is difficult to find tidal tails because they are most likely not dense enough to be distinguished from the background contamination. The structure of the tidal tails is a weak function of age in this diagram. 

The middle panel of Fig.~\ref{fig:CCP} shows the dependence of the velocity difference,  $v_{||\mathrm{pred}} -  v_{||\mathrm{obs}}$, in the CP method (i.e. the horizontal axis of the top panel in Fig.~\ref{fig:CCP}) on the spatial distance from the cluster centre. We demonstrate  that it is possible to approximate this relation by the linear function $\Lambda(R-R_{cl})$ for distances in Galactocentric coordinates, $R_{cl}-R$ denotes distance difference of individual stars from clusters centre. 
For the \textbf{M1} model at a simulation time of $625$ Myr, the relation has the form 
\begin{equation}\label{eq:lam}
    \Lambda(R-R_{cl}) = - 0.026 \, (\pm 0.005) \, (R-R_{cl}) - 0.06 \, (\pm 0.01) \, .
\end{equation}
While there is some variation for different time snapshots (see the uncertainty spread in the brackets above in eq.\ref{eq:lam}), the variations
are small enough for this method to be robust when the tail is developed (i.e. the age of the cluster exceeds 100~Myr). 
This means that the farther away the location of the two tidal tails from the star cluster centre ($R_{cl}$), the larger the velocity difference. We verified that this correlation is robust and not sensitive to the type of orbit and age (after detectable tidal tails have developed) for a cluster at a fixed position and orientation. This is shown in the bottom panel of Fig.~\ref{fig:CCP}: Even when single relation is used for a number of  snapshots that span almost 100 Myr, the final distribution transformed with eq.\ref{eq:lam} remains compact. 
This trend notably remains robust, and Eq.~\ref{eq:lam} is also valid for model \textbf{M5} when  the updated MW potential from \cite{Irrgang2013} is used.

In order to make the most compact representation of the  tidal tails, we removed the $\Lambda(R-R_{cl})$ correlation from the CP method velocity difference 
$v_{||\mathrm{pred}} -  v_{||\mathrm{obs}}$. The bottom panel of  Fig.~\ref{fig:CCP} thus shows the newly introduced compact CP method (CCP). Because \cite{Roeser+19} and other studies have reported that the models (comparable with our \textbf{M1} model) agree well with the data, we use the CCP method in order to try to recover the full extent of the Hyades tidal tails in Sec.~\ref{sec:gaia_res}.

We emphasize that the results above (mainly the appearance of tidal tails in velocity space) demonstrate that tidal tails of open clusters are not in general represented by any model-independent overdensity in proper motions or derived quantities (e.g. using the CCP method). This needs to be kept in mind when searching for tidal tails in \textit{Gaia} data. A numerical model might even be necessary in order to be able to find the extended parts and interpret them correctly. 

\section{Searching for the tidal tails in the \textit{Gaia} data} \label{sec:search}
With the modelling described above, the newly gained knowledge, and the developed CCP method,  we now attempt to recover the full extent of the theoretically expected tidal tails around the Hyades in the \textit{Gaia} data. 
To do so, we downloaded \textit{Gaia}  DR2 \citep{GaiaDR2_2018}  and the eDR3 data \citep{eDR3} from the ESA archive\footnote{https://gea.esac.esa.int/archive/; see the \texttt{ADQL} set of commands in Appendix \ref{sec:gaia}.}. 
The chosen range of parallaxes has the potential that the Hyades tidal tail members can be identified up to a distance of $500\,$pc from the Sun, that is, the full length of the tail up to 1000~pc. 
We queried only targets with high-quality parallax measurement, $\texttt{paralax\_over\_error > 10}$, allowing us to compute the distance $d$ as 
\begin{equation*}
    d/[\mathrm{pc}] = \frac{1000}{\varpi / [\mathrm{mas}]}\,.
\end{equation*}
Moreover, we used the \texttt{RUWE} quality criterion to filter out objects with spurious astrometric solutions (Lindegren, document \texttt{GAIA-C3-TN-LU-LL-124-01}). 
We note that previous studies such as \cite{Roeser+19} applied a smaller range of parallaxes,  limiting the search volume in the \textit{Gaia} catalogue and thus the detectable extent of the tail. The smaller volume, however, also  means lower contamination. The models computed above show that the tail does not produce a simple overdensity in velocity space, but that the extended parts of the tails are much harder to recover because the contamination increases with cubic distance from the Sun. Our search for the full extent of the Hyades tidal tails, which are comprised of a few hundred stars in the catalogue of about $10^8$ targets is thus very challenging. 

 The new approach used here therefore is to employ the $N-$body simulations described above in order to separate the expected members of the Hyades star cluster and its tails from the Galactic field. The standard CP method already uses model expectations to define the tidal tails. By using the $N-$body models to formulate the CCP method, we therefore merely generalise this approach. For the purpose of this initial study, we used the benchmark \textbf{M1} model. See also \cite{Roser2020} who extended CP method based on empirical correlation in a similar way as the CCP method applied here. 

  \cite{Roeser+19} reported that the detected parts of the Hyades tidal tails agree very well with previously published $N-$body models that are basically identical to our \textbf{M1} model. This means that we should at least be able to detect the parts of the tail reported by \cite{Roeser+19} with the M1 model, and that any deviations from the model in the data will indicate the need for more realistic models.

\subsection{Hyades and tidal tail selection method}\label{sec:gaia_res}

In order to recover the Hyades cluster and its tidal tails from our initial catalogue, we followed a multi-step procedure, comprising steps S1, S2, and S3. We describe the steps in detail below.

\begin{figure*}
    \centering
    \includegraphics{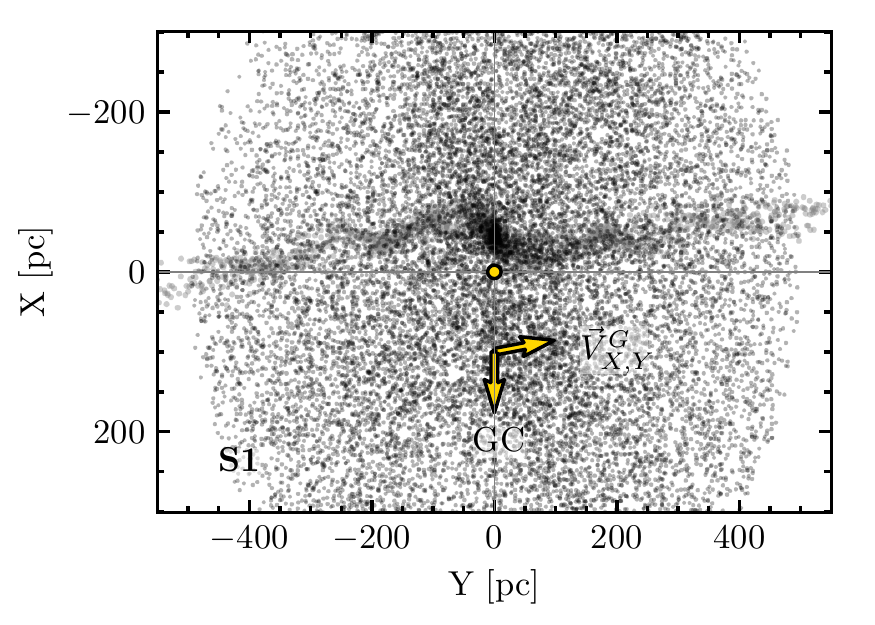}
    \includegraphics{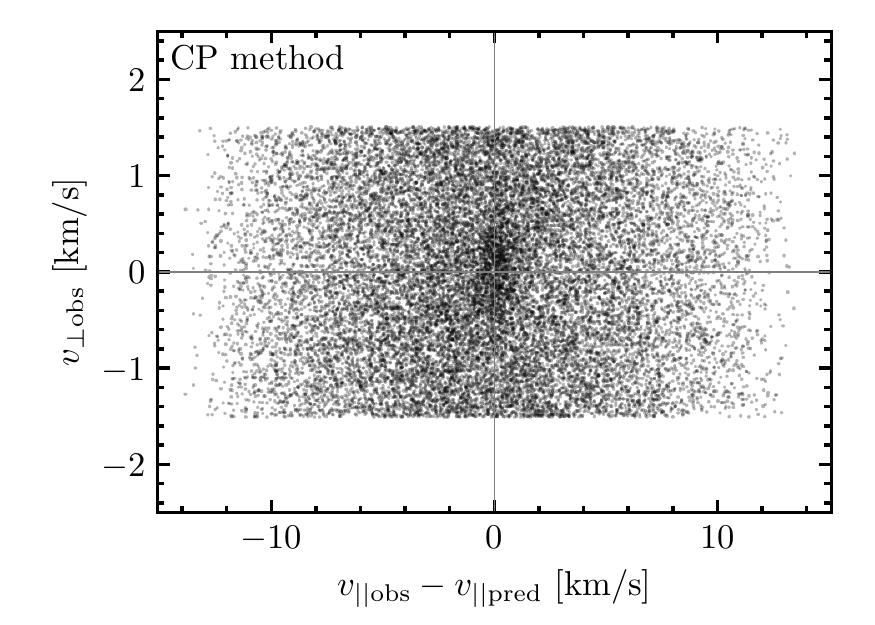}
    \caption{\textit{Gaia} data sample after the \textbf{S1} CCP criterion was applied, which filters out $\approx$ 22 000 stars from the initial  $\approx$ 11~000~000. \textbf{Left panel:} Galactic Cartesian coordinates. The yellow point shows the (0,0) position of the Sun, and the arrows show the velocity of the Hyades in the $X-Y$ plane and the direction to the GC. The black \textit{Gaia} data points are highlighted with the grey model to guide the eye. This suggests an overdensity to the eye. \textbf{Right panel:} Classical CP method plot. The overdensity at (0,0) is already visible after the \textbf{S1} selection. }
    \label{fig:S1sel}
\end{figure*}

\textbf{S1} We used the CCP method as described in Sec.~\ref{sec:ccp}. 
In the CCP method, the stellar cluster and its extended tidal tails are mapped to a clump with a radius of  $1.5$~km/s; see the bottom panel in Fig.~\ref{fig:CCP}. The \textbf{S1} criterion selects objects from the initial catalogue that have $-1.5 < (v_{\mathrm{|| obs}} - v_{\mathrm{|| pred}})- \Lambda(R-R_{cl}), v_{\perp \mathrm{obs}} < 1.5$ km/s. This procedure reduces our initial DR2 catalogue  of~11,223,898~sources to 22,909~objects and our initial eDR3 catalogue  of~13,896,271~sources to 28,218~objects. In spatial coordinates, the leading tail of the Hyades cluster can be distinguished from the contamination, see Fig.~\ref{fig:S1sel}. This shows that while contamination is still present after the \textbf{S1} step, the method is successful because it allows the cluster and the tail to emerge from it. In the initial catalogue, the overdensity visible in the standard CP diagram is not evident, as is the case for the smaller data set shown in Fig. 1 in \cite{Roeser+19}.

\begin{figure}
        \centering
        \includegraphics[width=\hsize]{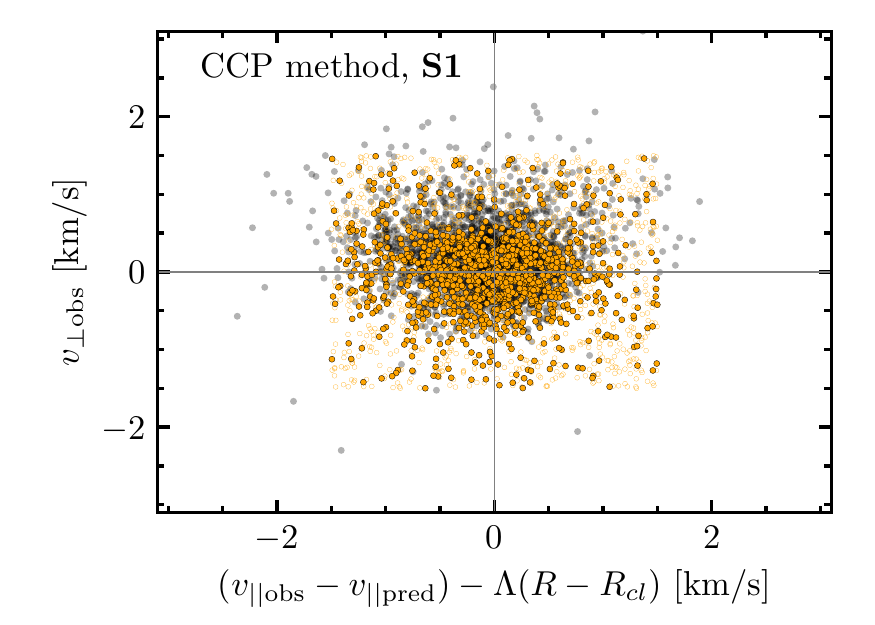}
        \includegraphics[width=\hsize]{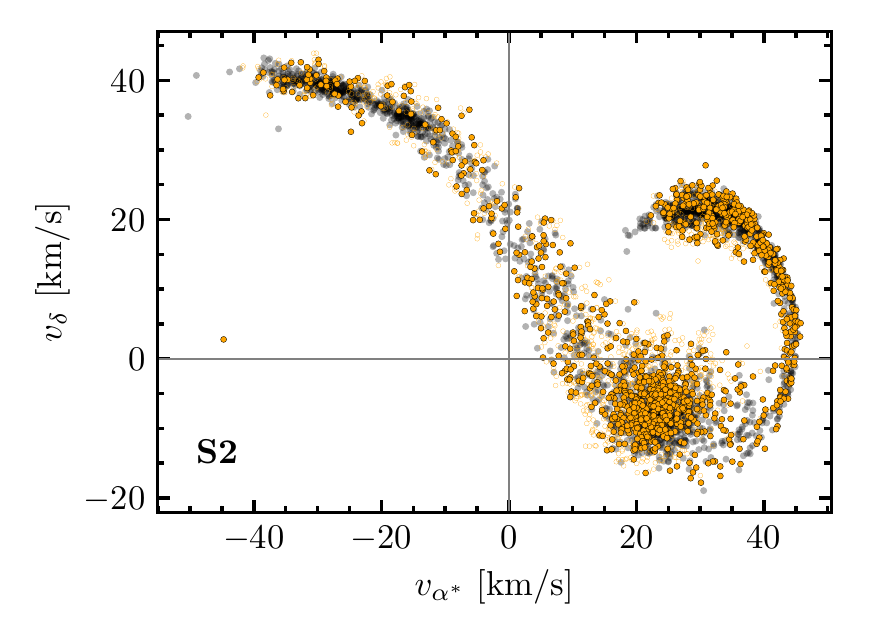}
        \caption{Gaia data showing selection criteria \textbf{S1} and \textbf{S2}.  The black background points show the \textbf{M1} model, the open yellow circles are \textit{Gaia} objects that are consistent with the CCP modelled distribution (i.e. after the \textbf{S1} selection was applied) and with the  modelled proper motion values (\textbf{S2} selection). The full orange points show the \textit{Gaia} objects that in addition to this, are also consistent with the spatial distribution of the \textbf{M1} model (\textbf{S2} selection). 
        \textbf{Top panel:} Diagram of the CCP method. \textbf{Bottom panel:} Proper motion distribution in equatorial spherical coordinates, R.A. and Dec, in physical~km/s units. }
        \label{fig:S2}
    \end{figure}

\begin{figure}
        \centering
        \includegraphics[scale=1]{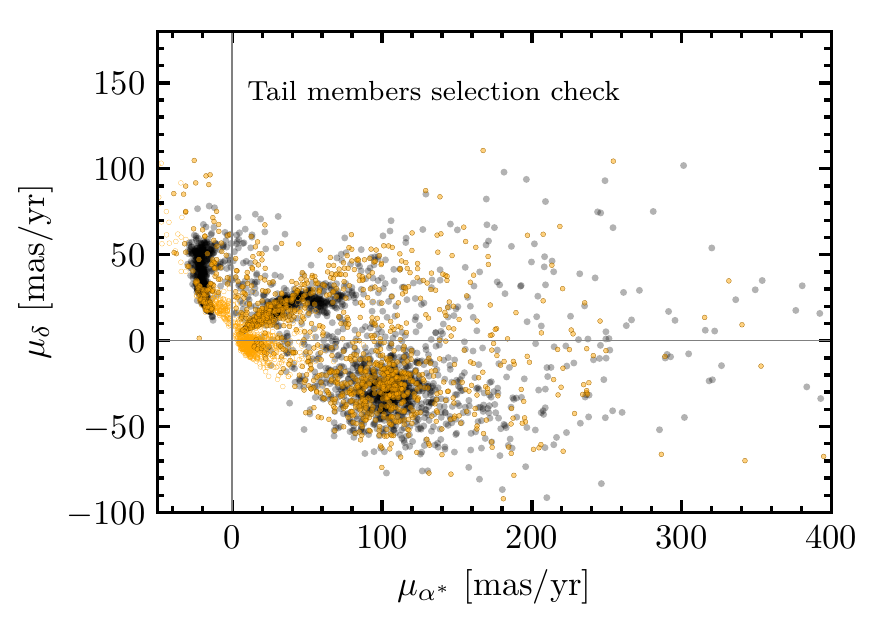}
        \caption{On-the-sky proper motions in equatorial spherical coordinates R.A. and Dec in mas/yr. See Fig.~\ref{fig:S2} and text for the detailed description. The black points are the \textbf{M1} model points, the open orange circles show the remaining objects after the \textbf{S1+S2} selection was applied, and the filled orange circles show the objects after the \textbf{S1+S2+S3} selection. }
        \label{fig:sky}
    \end{figure}

\begin{figure*}
        \centering
        \includegraphics[scale=1]{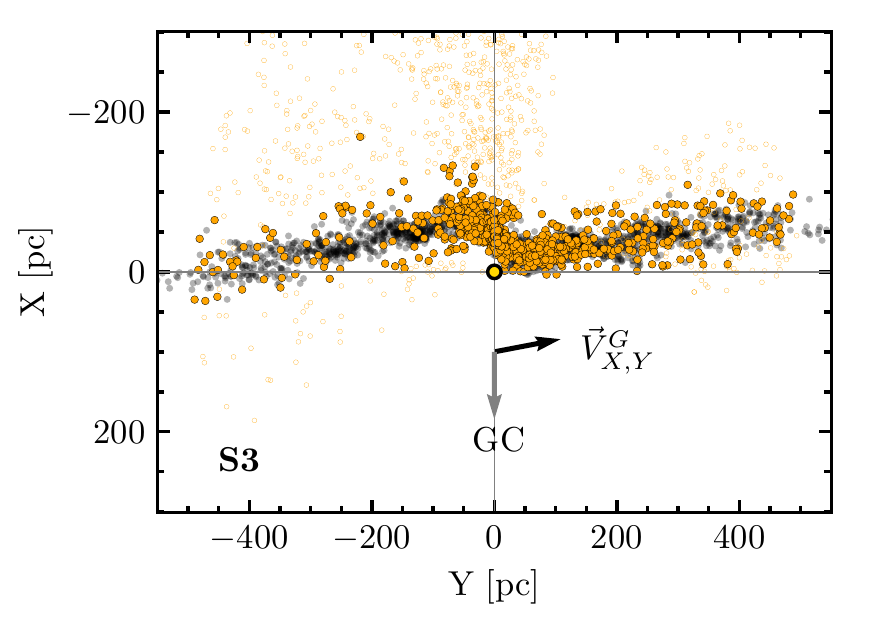}
         \includegraphics[scale=1]{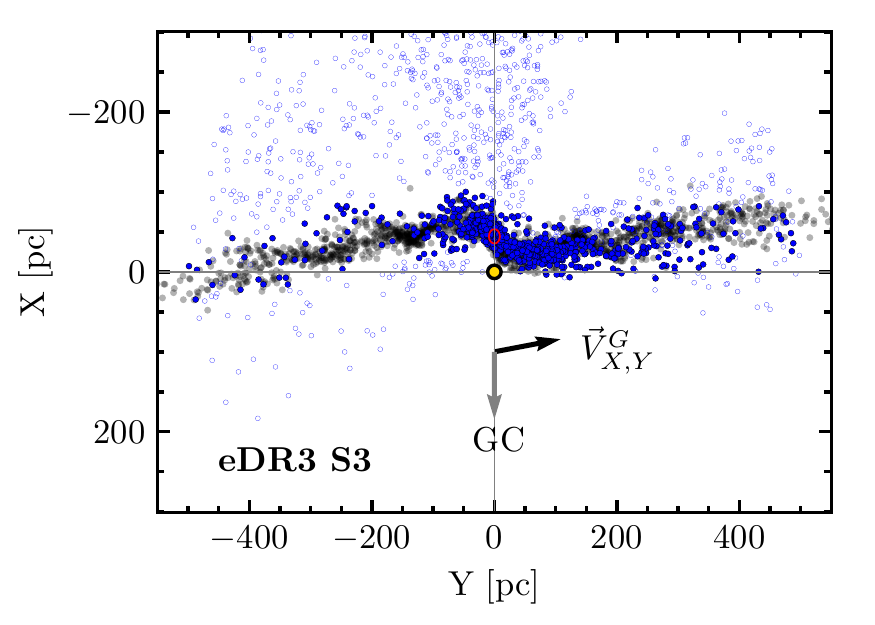}
        \includegraphics[scale=1]{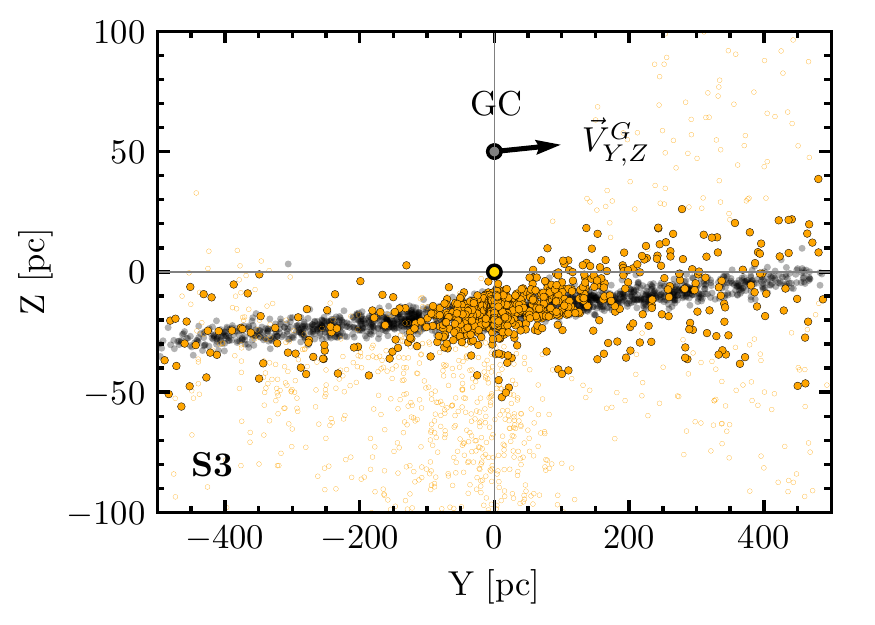}
        \includegraphics[scale=1]{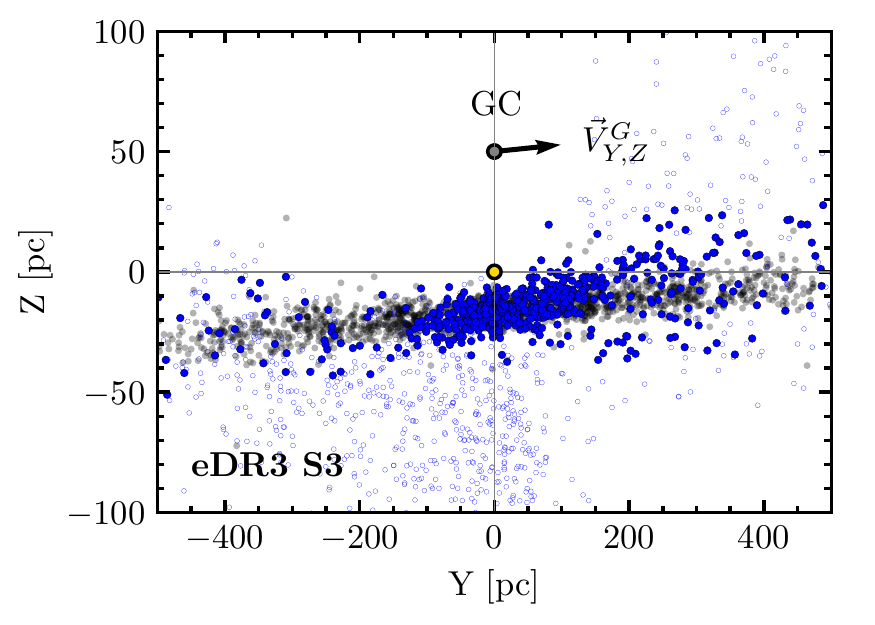}
        \includegraphics[scale=1]{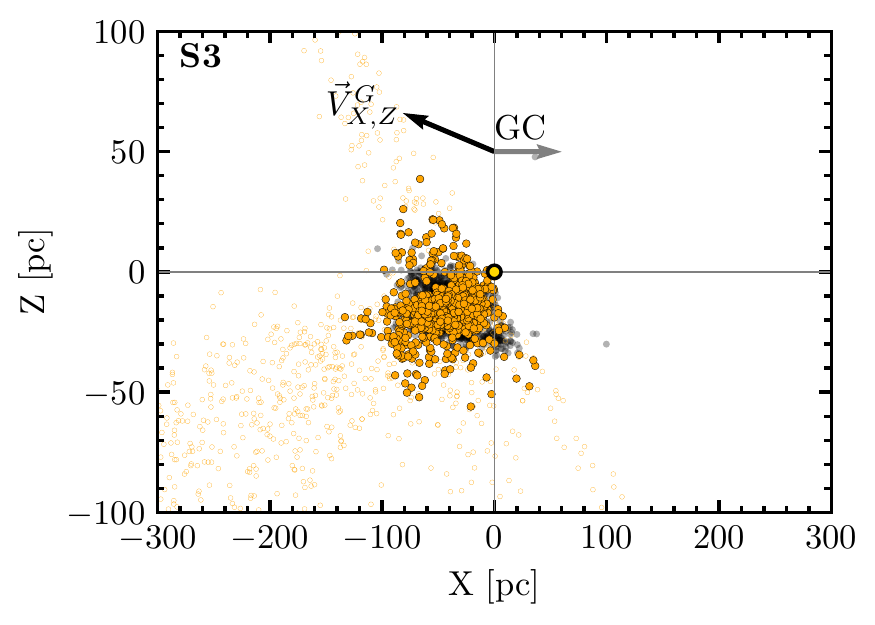}
        \includegraphics[scale=1]{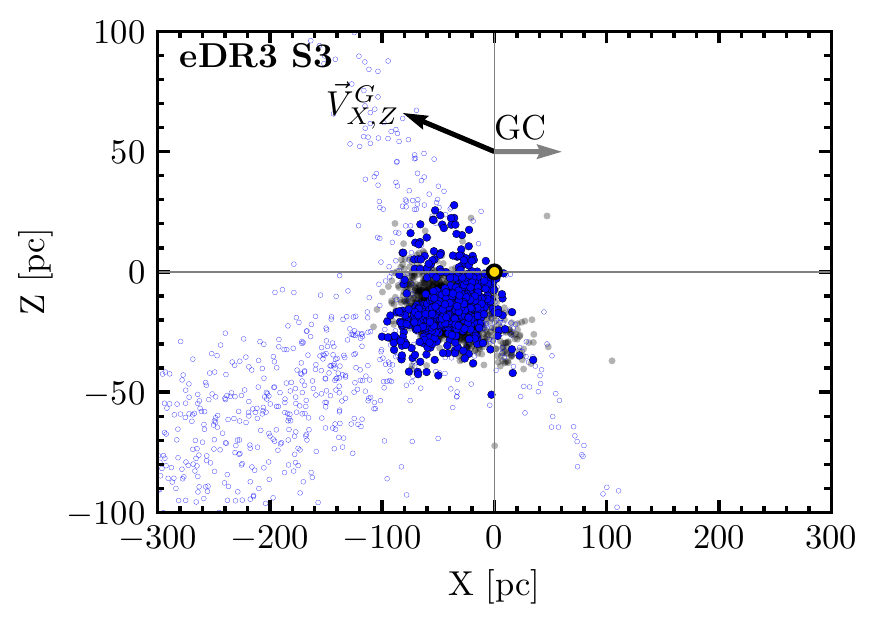}    \caption{Tidal tails in Galactic Cartesian coordinates. The large yellow point at (0,0) in all panels marks the Sun. The black arrow shows the velocity vector in the given coordinate system, and the grey arrow points to the GC.  See Fig.~\ref{fig:S2} and text for the detailed description. The black points are the \textbf{M1} model points, the open orange circles show the remaining points after the \textbf{S1+S2} selection was applied, and the filled orange circles show the objects after the \textbf{S1+S2+S3} selection.}
        \label{fig:space}
    \end{figure*}

\textbf{S2} We used the model-predicted proper motions (top panel of Fig.~\ref{fig:proj}) and selected only objects with a distance from any modelled proper motion within 1.5~km/s. We used tangential velocities in~km/s. The coordinate system can be either equatorial or galactic. 
For the purpose of this study, we chose to make the cut at 1.5~km/s so that targets with an uncertainty of 0.5~km/s on their proper motion were still accounted for within $3 \sigma$. The inclusiveness or completeness of the selected sample changes when this value is changed, and it will be interesting to explore this in more detail in a specific future study. This selection further reduces the number of objects from~22,909 to~2102. We plot these targets as open orange circles in Fig.~\ref{fig:S2}. They perfectly overlap (by construction) with the model predictions in Fig.~\ref{fig:proj}. However, some of these points are still contamination, which is shown in the proper motion plot (the lump of points near (0,0) mas/yr in Fig.~\ref{fig:sky}). For the eDR3 we used the benchmark model M5 and the corresponding tangential velocities. The number of objects from the S1 selection value of 28,218 is further reduced to 1,774. They are plotted as open blue circles in  Fig.~\ref{fig:space} and Fig.~\ref{fig:epy-data}.

\textbf{S3} To clean the sample from the contaminants that were still present after \textbf{S2} was applied, we used the modelled spatial coordinates and chose only the targets that overlap with the 3D tidal tail shape within their 3$\sigma$ uncertainty. The finally selected DR2 data sample is plotted as full orange circles (Fig.~\ref{fig:S2},\ref{fig:sky},\ref{fig:space}) and contains 1109~objects,~411 of which belong to a leading tail and 331 to a trailing tail (assuming the tidal radius is 9~pc). 
For the eDR3 data sample, for which we used the \textbf{M5} model, the final catalogue contains 862 objects, 293 of which belong to the leading tail and 166 to the trailing tail.  The data are plotted as filled blue circles in Fig.~\ref{fig:space} and Fig.~\ref{fig:epy-data}.
The final data sets can be acquired through the CDS\footnote{Centre de Données astronomiques de Strasbourg, \url{https://cds.u-strasbg.fr/}} service and the tables showing the data format are available in the appendix, see Tab.~\ref{tab:DR2} and Tab.~\ref{tab:eDR3}.

\subsection{Members of the tidal tail: RV and CaMD check.}
\begin{figure}
        \centering
        \includegraphics[scale=1]{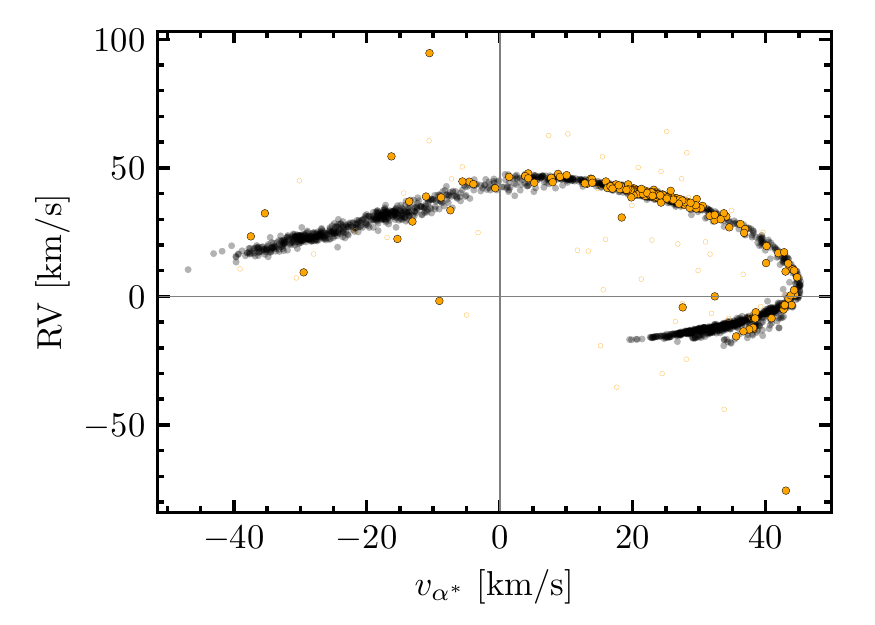}
        \caption{Tangential velocity in R.A. in km/s on the vertical axis and the radial velocities, RV, on the horizontal axis. See Fig.~\ref{fig:S2} and text for the detailed description. The black points are the \textbf{M1} model points, the open orange circles are obtained after the \textbf{S1+S2} selection was applied, and the filled orange circles remain after the \textbf{S1+S2+S3} selection criterion was applied. The radial velocity information is only available for a small number of objects.}
        \label{fig:RV}
    \end{figure}

We present a catalogue of candidate members of the extended tidal tails of the Hyades. The selection after the CCP method was applied is based on model~M1, as described above. We did not use radial velocities (RV) in the process of tail selection, however. Because we used the underlying $N-$body model, the RV values of the selected candidate members are expected to follow the measurements. For the few stars that have \textit{Gaia} RV measurements, this is indeed the case, as shown in Fig.~\ref{fig:RV}. 

\begin{figure}
        \centering
        \includegraphics[scale=1]{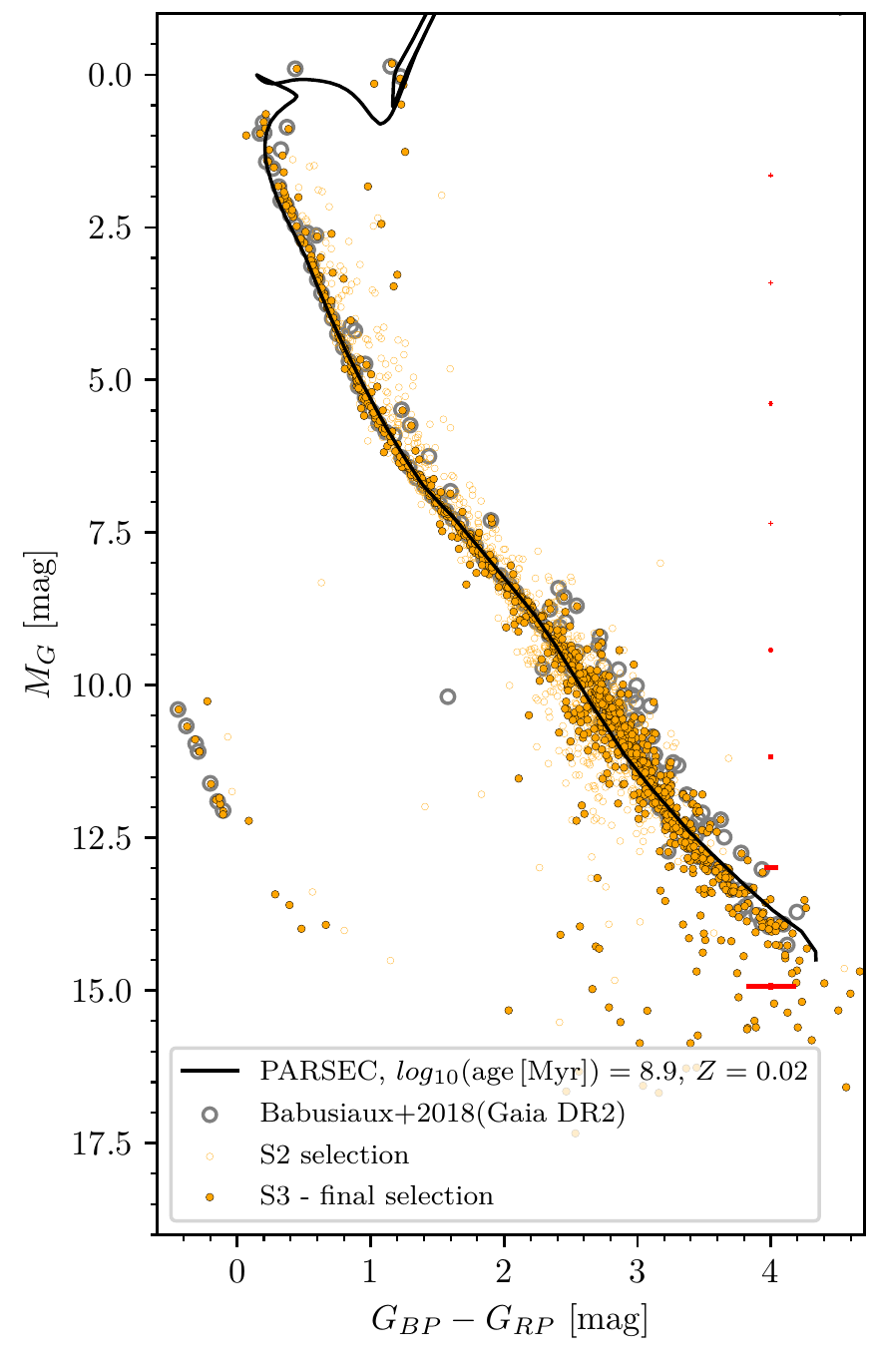}
        \caption{Colour-absolute magnitude diagram for \textbf{S1+S2} -selected objects (open orange circles), \textbf{S1+S2+S3}-selected objects (full orange circle),  and the \textit{Gaia} Hyades CaMD from \cite{GDR2_Hyades} as open grey circles. For the selection criteria \textbf{S1, S2, and S3}, see Sec.\ref{sec:gaia_res}. The PARSEC isochrone is plotted as a black line. The missing binary sequence in our selected points is most likely due to the \texttt{RUWE} filtering, which partially filters out unresolved binary stars. Mean error bars computed in $M_G$ bins computed based on Gaia photometric fluxes (see Gaia DR2 primer, issue 1.5) are shown in red at the right. }
        \label{fig:CMD}
    \end{figure}

The extended tidal tails can also be detected using the colour-absolute magnitude diagram (CaMD). This procedure is completely independent from the method used above.
Fig.~\ref{fig:CMD} shows the CaMD for the \textbf{S1+S2}-selected, \textbf{S1+S2+S3}-selected and the \textit{Gaia} Hyades CaMD from \cite{GDR2_Hyades}. 
The agreement of the Hyades \textit{Gaia} DR2 sample \citep{GDR2_Hyades} and our selected objects is encouraging.

In the upper part of the CaMD (for targets brighter than $M_G = 10$ mag) that shows the overall more precise photometric values, the scatter in the \textbf{S3}-final selected Hyades and the corresponding data sample for the tidal tails is clearly smaller. 
The \textbf{S3}-final data sample presents candidate members whose membership of the Hyades cluster should be explored through detailed chemical tagging. However, the decrease in scatter from selections \textbf{S2} to \textbf{S3} provides supportive evidence that they are associated with the Hyades star cluster. 

\subsection{Detection of epyciclic overdensities}

\begin{figure*}
        \centering
         \includegraphics[scale=0.9]{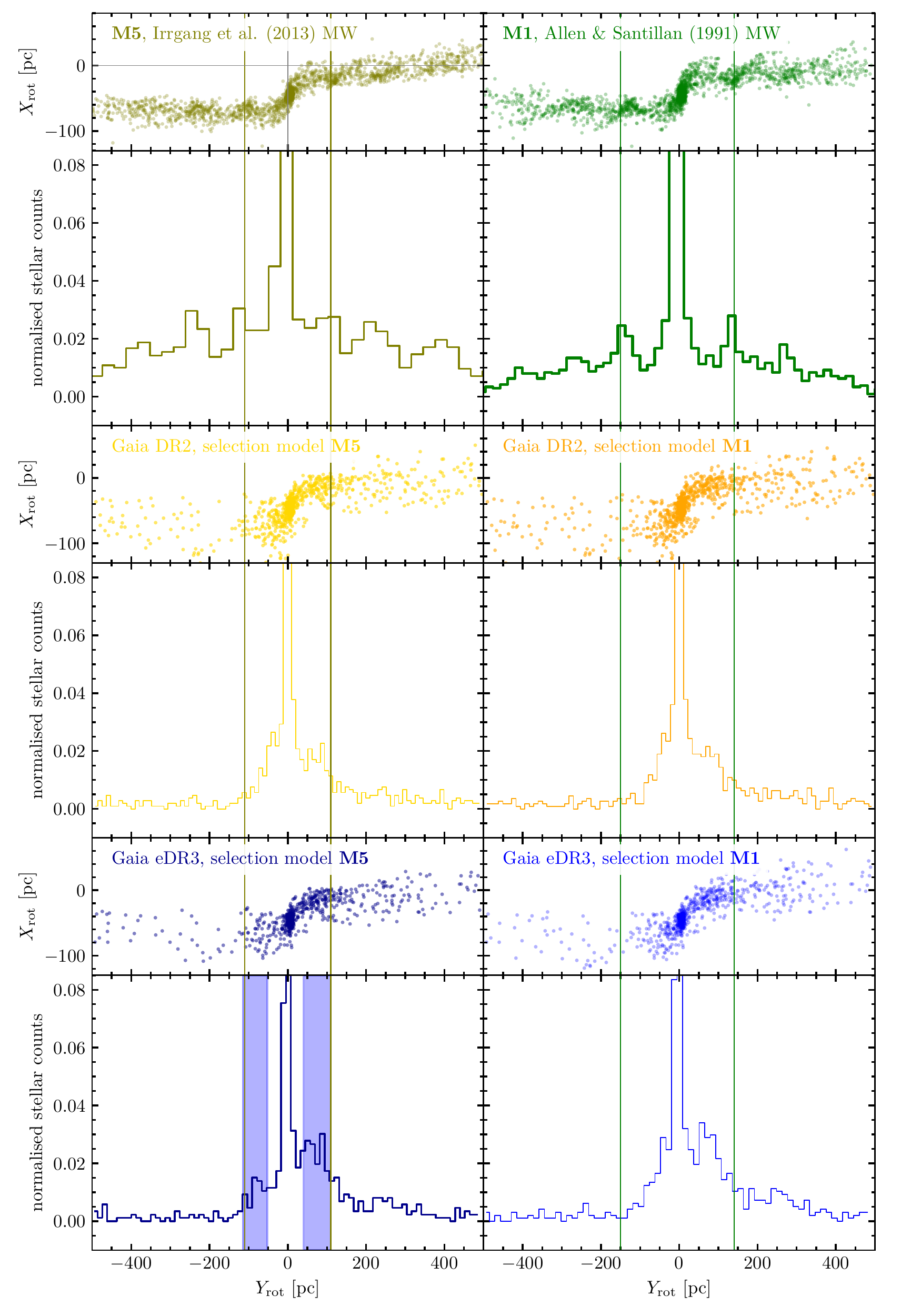}
        \caption{Each pair of panels (from top to bottom) shows the cluster and its tidal tails in $X-Y$ Galactic coordinates rotated so that the $V_Y$ component is horizontal. This allows us to plot the histogram along the tail and to reduce projection effects caused by the spatial alignment of the tail. The upper two double panels show models $M1$ and $M5$ at the age of 655 Myr. The lower two double panels show the data selected based on the $M1$ and $M5$ models.}
        \label{fig:epy-hiss}
    \end{figure*}

 Fig.~\ref{fig:epyc} shows the positions of epicyclic overdensities in various spaces. Because the tails contain a relatively small number of stars, the overdensities, if present,  might be difficult to spot in real data. In order to quantify the density structure along the tail in a robust way, we used the spatial distribution in $X-Y$ space. The histograms along the tail in $X-Y$ space are shown in Fig.~\ref{fig:epy-hiss}, accompanied by models M1 and M5. We used the \texttt{AstroPy} histogram function with the implemented Freedman-Diaconis rule to choose the optimal bin widths.

The simulation results of M1 and M5 follow qualitatively similar evolution paths. This results in three clear, symmetric, epicyclic overdensities in the leading and trailing tail. The detailed tail structure and exact position of the epicyclic overdensities differ in the models.

In order to recover the Hyades cluster and its tail in the Gaia data, we followed  a multi-step procedure.
We discussed above that the CCP method that implements  the velocity gradient along the tail is robust and does not change from model M1 to M5. The subsequent steps that employ the detailed shape of the distribution of proper motions then differ among the models, however. Thus we used models M1 and M5 to search for Hyades tidal tails in the data. The results are compared in the bottom four panels in Fig.~\ref{fig:epy-hiss}. For the case of DR2, some substructures in the tails are suggested by the data. These become much clearer for the eDR3 data. Because the position of the overdensities is better matched by the M5 model using the MW potential parameters from \cite{Irrgang2013}, the signal becomes clearer for this selection.

\begin{figure}
        \centering
        \includegraphics[scale=1]{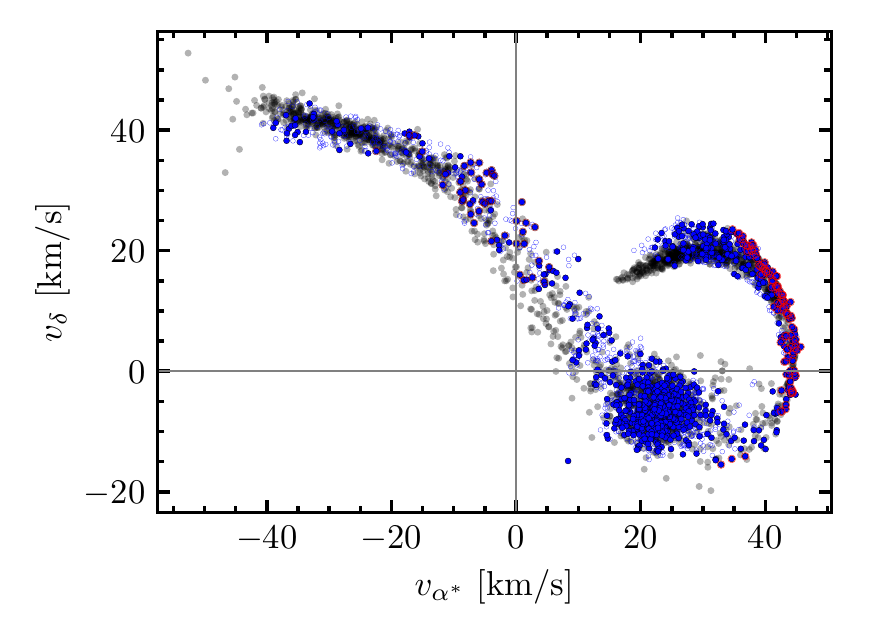}
         \includegraphics[scale=1]{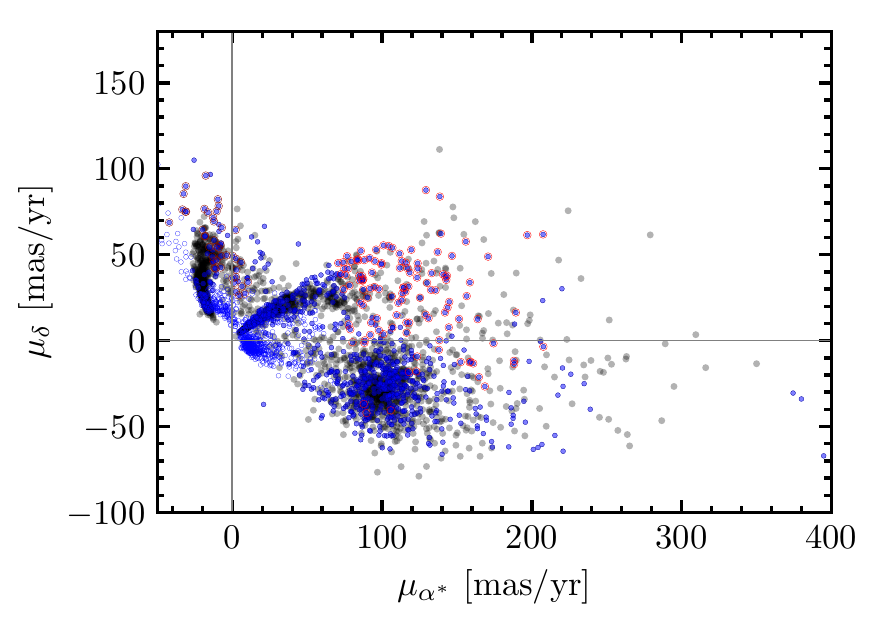}
        \caption{Candidate epicyclic overdensities plotted in proper motions and tangential velocities together with the eDR3 final data selection.  The top panel shows tangential velocities, and the bottom panel shows proper motions. The underlying black points belong to model {\bf M5} at an age of 655 Myr. Open blue circles show the data after the S2 selection was applied based on model {\bf M5}. The filled blue points  are the final data after the S3 selection. In Fig.~\ref{fig:epy-hiss} two spatial overdensities are clearly visible (highlighted by semi-transparent vertical blue bars), and stars that belong to them are plotted as filled red circles (for the overdensity in the leading part of the tail) and as dark red circles (for the overdensity in the trailing part of the tail). }
        \label{fig:epy-data}
    \end{figure}

We used the M5 model to select tidal-tail candidates in the eDR3 data that belong to the detected overdensities. 
Their stellar kinematics are shown in Fig.~\ref{fig:epy-data} as open red circles. 
The position of the stars that belong to the overdensities is as expected from the epicyclic overdensities. 
Further analysis achieved with a large modelling effort would provide more insights, but the detected substructures in the tidal tails of the Hyades appear to represent the first observed epicyclic overdensities in an open star cluster based on the spatial and kinematical properties.

\section{Effect of the initial angular momentum and lumpy Galactic potential}\label{sec:lumps}
We have discussed the importance of using detailed kinematic modelling in order to collect and interpret the most useful \textit{Gaia} data. With our novel model-based method, we are able to identify candidate members of the Hyades tidal tails up to the largest distances of the full extent of 800 pc and can confirm the detected asymmetry already reported by \cite{Roeser+19}. The $N$-body model and the observations differ in several instances, however. We appear to be unable to detect the epicyclic overdensities that are a firm prediction in the evolution of a tail in the Galactic potential. Furthermore, \cite{Roeser+11} pointed out that the cluster has an elevated (super-virial) velocity dispersion.
This has been confirmed by \cite{Oh2020},
but see the discussion of the value of the velocity dispersion in Sec.~\ref{sec:coords}.
The aim of this section is to explore whether the initial angular momentum and/or lumpy Galactic potential could remove some of the discrepancies between model and observation.

\subsection{Initial angular momentum: Cluster rotation} \label{sec:rot}
In addition to the initial setup, we also considered simulations with a non-zero $Z$-component of angular momentum, $L_z$, such that the star cluster initially rotates around its $Z$-axis perpendicular to the Galactic plane. To initialise the rotating cluster, we followed several steps: 1) We set up a Plummer model identical to the description in Tab.~\ref{table:CLP} with average zero angular momentum. 2) The sign of $L_z$ was chosen: when $L_z < 0,$ the cluster rotation sense is counter-orbit, and when $L_z > 0,$ it is co-rotating with the cluster orbit. 3) We verified the value of the angular momentum for each star, $L_z^{\star}$. When the sign of  $L_z^{\star}$ is opposite to the chosen direction of rotation in step~2, the velocity signs are changed to the opposite values. This procedure ensures that the total energy of the star cluster remains unchanged while allowing it to have a non-zero total angular momentum. The aim here is to assess for the first time how and whether an initially rotating cluster may affect the properties of its tidal tails. 

\subsection{Lumpy Galactic disc} \label{sec:lump}
In addition to the smooth semi-analytical Galactic potential, we also considered local potential fluctuations that might be caused by molecular clouds or other star clusters, for example. The question we would like to address with this setup has been illustrated by \cite{Roeser+19}. These authors pointed out that the detected tails of the Hyades cluster are asymmetric: the trailing tail is shorter and less populated. This might be caused by the interactions with potential perturbations, such as those generated by giant molecular clouds or star clusters. 

Introducing lumps to the smooth Galactic potential described above adds a number of additional parameters to be considered, such as their number, density distribution, their orbits, mass, and size distribution. 
This in-depth study is beyond the scope of this paper, in which we primarily seek a novel method to extract likely members of non-compact 6D structures.
The effect of a clumpy Galaxy on these structures deserves an individual consideration based on the exact setup and the interactions the cluster encounters during its evolution. For the purpose of assessing the possible effect of lumps on the tidal tails of the Hyades, we considered the following model. 
The smooth Galactic potential and initial stellar cluster setup was identical to the simulations above, see Tab.~\ref{table:CLP} and Tab~\ref{table:MWp}.  The basic model of the lumpy Galaxy presented in the \texttt{AMUSE} book \citep{amuse_book} was used for guidance. We initialised lumps with a power-law number distribution slope of $-1.6$ (the Salpeter value would be $-2.35$), sampling between 4,000 to 10,000 lumps that are randomly distributed in a region in the Galaxy from 3.5 kpc to 8.5 kpc distance from the GC on corresponding circular orbits. 
We varied the minimum, $M_{min} \in (10,10^3) \, M_{\odot}$, and maximum, $M_{max} \in (10^4,10^7)\, M_{\odot}$, masses of the distribution. Because we set up the lumps as point masses, their physical size is represented by a softening length, which  we varied from 10 pc up to 100 pc. 

These parameters allowed us to study various interactions of tidal tails with lumps,  from close encounters of massive lumps to a large number of small perturbations. We emphasize that the aim of this study is not to simulate a realistic distribution of GMCs in the MW, but to understand whether gravitational interactions with lumps can produce a non-symmetric tidal tail reminiscent of the observed one.

\subsection{Initial spin angular momentum of the cluster: Simulation results} 

\begin{figure*}
        \centering
        \includegraphics[scale=1.0]{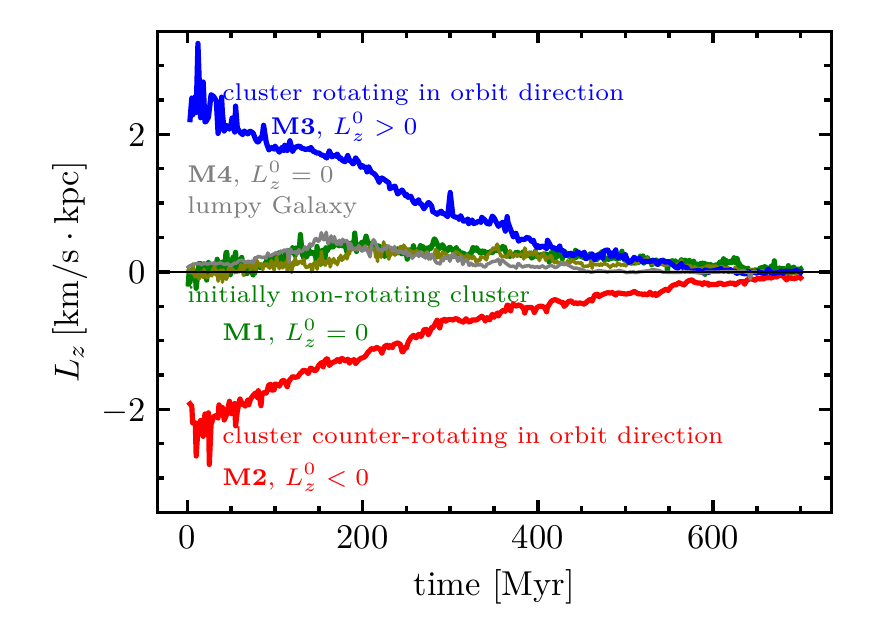}
        \includegraphics[scale=1.0]{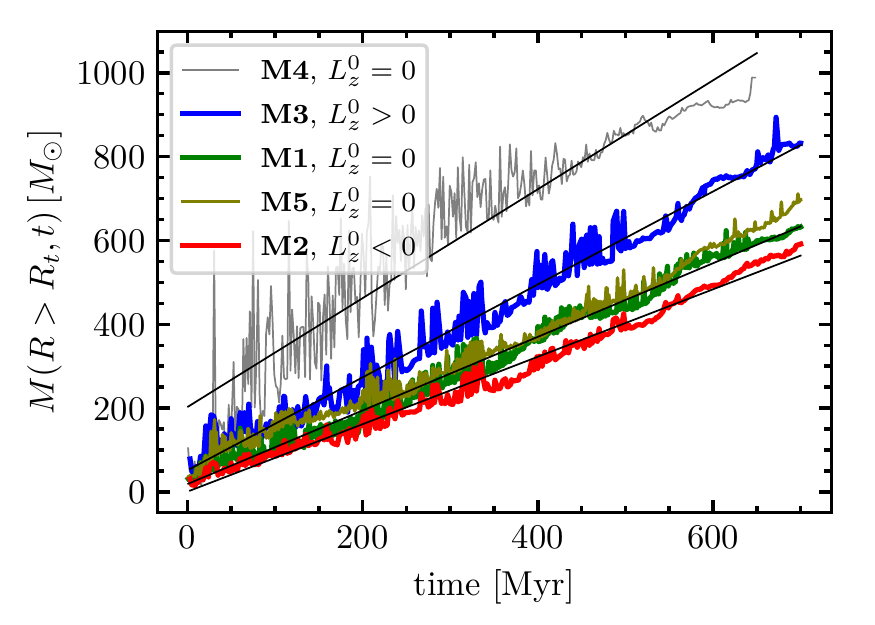}
        \caption{Angular momentum and stellar mass of the tidal tails as a function of time for simulated star clusters. \textbf{The left panel} shows the evolution of the $z$ component of the spin angular momentum, $L_z$, of the star cluster with time. We plot models \textbf{M1, M2,} and \textbf{M3,} which were evolved from the same initial conditions (see Tab.~\ref{table:CLP}) but with different initial spin angular momenta (see Sec.~\ref{sec:nbody}). Model M4 considers the same MW potential as M1-M3 with the addition of lumps, and model M5 uses the MW potential from \cite{Irrgang2013}. The $L_z$ values are always computed within the tidal radius of the cluster. \textbf{The right panel} shows the mass outside of the tidal radius ($R_t$) of the cluster at a give time for the same models (\textbf{M1, M2,} and \textbf{M3}). The black lines are linear fits to the individual tracks and are used to compute the average mass-loss rate. The angular momentum of all our modelled clusters given by the motion in the Galaxy has a positive value.
        }
        \label{fig:LZ}
    \end{figure*}

The mass loss and thus also the mass and length of the tidal tails are higher for the cluster model that has an initial spin angular momentum vector  
in the same direction as its orbital angular momentum vector in comparison to the initially non-rotating cluster. 
On the other hand, the mass loss and the mass and length of the tidal tails are lower for the cluster that initially rotates in the opposite direction compared to its orbital angular momentum. The mass loss for the different models is shown in Fig.~\ref{fig:LZ}. 
The initial spin angular momentum in our models is small because we initialised the Plummer models, therefore the effects on the length and orientation of the leading and trailing tidal tails relative to the cluster centre are evident but barely statistically significant.

The initial loss of stars from the cluster carries away most of the initial spin angular momentum because the cluster immediately begins to de-spin, and thus the stars farthest from the cluster, those that were lost first, carry most of the information on the initial cluster spin. Steps towards understanding the depopulation of the phase-space distribution function of stars in a rotating satellite have been performed by \cite{PiatekPryor95} and \cite{Kroupa97}, but much more work on this physical process is needed to fully understand the configuration of initial spin and subsequent tidal tail. A large number of simulations needs to be analysed in order to quantify the correlation between the initial cluster spin and the orientation and extent on the tidal-tail tips. The results obtained here are thus clearly suggestive of a potentially powerful method for constraining the initial cluster spin by measuring the orientation of the tail tips relative to the cluster velocity vector and their distance from the cluster position, but this particular problem is beyond the aims of this work, which are to introduce the new CCP method. We emphasise that while there is an effect caused by cluster rotation on the formation of tidal tails, there is no indication of asymmetric tidal tails.

\subsection{Lumpy Galactic potential: Results} 

In order to assess whether encounters with lumps might in principle be able to account for the observed Hyades tidal-tail asymmetry and for its potentially high velocity dispersion, 
we performed~ten simulations with varied lump parameters.  The aim here was not to perform a detailed analysis of all possible cases, but to broadly assess the possibilities. In the range of parameters we explored (see Tab.~\ref{table:MWp}), the main results based on the evolution of the tidal tails can be divided into three statistical categories that we list below.

\textbf{1)} No discernible effect on the tidal tails is detected. This is the case for large effective sizes of the lumps, with only a few or no  encounters with massive ($>10^5\,M_{\odot}$) lumps.  

\textbf{2)} A discernible effect on the tidal tails becomes evident, that is, the  tidal tails show a wider and more dispersed distribution of their members than in the smooth Galactic potential model, with an effect on the cluster itself. This is the result in simulations with many encounters, for which the effective sizes is not very important, without any close interaction with a massive ($>10^7\,M_{\odot}$) lump.  No significant asymmetries between the leading and trailing tails are produced, and the velocity dispersion of the cluster is not significantly affected.

\textbf{3)} An essentially complete destruction of the star cluster created by a close encounter with a massive lump, which leads to a temporal tail asymmetry and an increased velocity dispersion of the cluster. In one of the simulations for model \textbf{M4}, a close encounter with a massive ($10^7\,M_{\odot}$) lump has been detected around the age of the cluster at 640~Myr.
The star cluster winds up around the lump and is destroyed. The nominal age of the Hyades is 625~Myr, therefore this case represents a good comparison, noting that it would mean that the cluster is in the process of disruption \citep[see as well ][]{Oh2020}. In the following paragraph we discuss this event in more detail and use model \textbf{M4} for comparison with the simulations without lumps discussed above.

\begin{figure}
    \centering
    \includegraphics{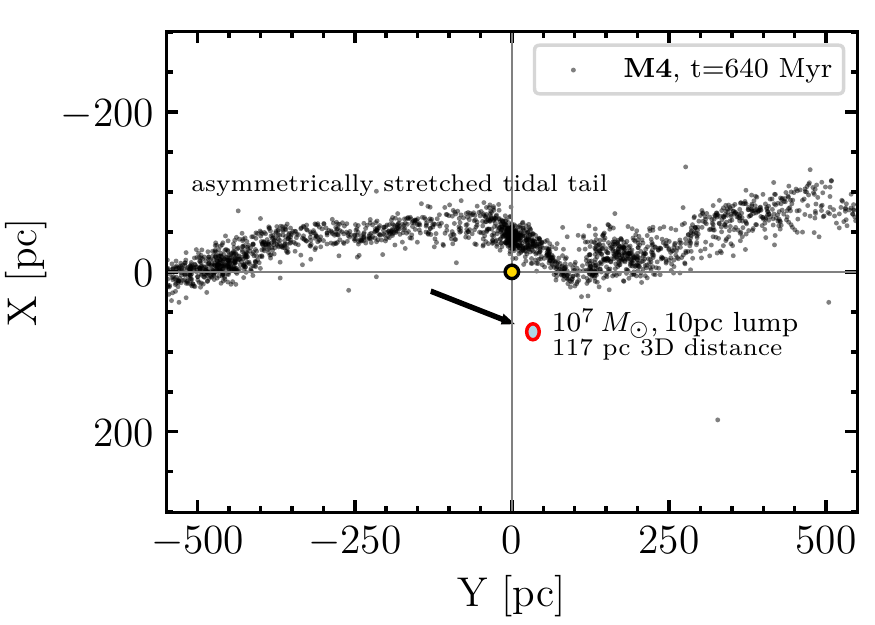}
\includegraphics{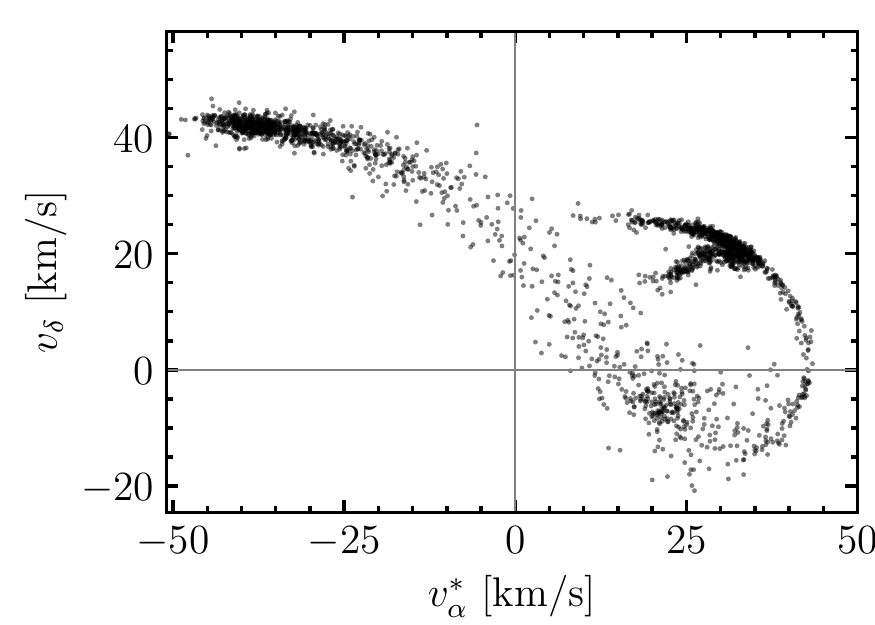}
\includegraphics{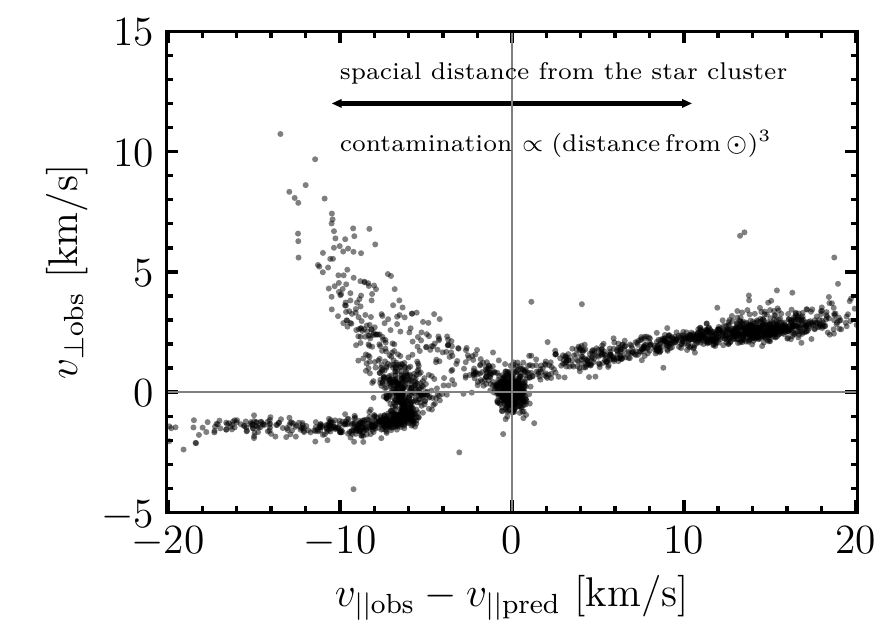}
    \caption{Encounter of the cluster and its tidal tails with a massive Galactic lump. This asymmetrically stretches the tail (note the displaced positions of the epicyclic overdensities; compare with Fig.~\ref{fig:YX_M}) and changes the velocity distribution in the star cluster and in its tails. As the star cluster is disrupted by the lump, its velocity distribution increases. The snapshot here combines lump distance and mass so that the velocity dispersion of the model cluster is increased to the  potentially higher observed value in the Hyades star cluster (see text for more details).   This affects the appearance of the star cluster and of its tail in different parameter spaces. The change when the CP method is applied in the bottom panel is clearly visible. The disruption process thus can not only produce a temporary asymmetry in the tidal tails (prior to complete disruption), but can affect the detection using the CP method, which can lead to an incomplete recovery of the tail.}
    \label{fig:comp_pmradec}
\end{figure}

The simulated cluster in the lumpy Galaxy model \textbf{M4} is destroyed at an age of 640 Myr by being gravitationally torn and sheared by a massive lump. However, the most interesting part of the simulation from the point of view of the Hyades stellar cluster is not the actual act of cluster destruction, but the moments before it, because this part can be compared with the observed Hyades regarding the reported potentially higher velocity dispersion and tidal-tail asymmetry. Because the simulated stellar cluster is close to being disrupted, its velocity dispersion increases with time as the lump approaches, as shown in the last two columns in Tab.~\ref{table:S_Hyad}. For the purpose of comparison with the Hyades, we therefore chose the snapshot at an age of 641~Myr for which the velocity dispersion 
within 9~pc of the cluster centre is 0.8~km/s (see Tab.~\ref{table:Hyad}). This is closest to the observed potentially higher value.  \cite{Oh2020} have concluded that based on the present-day velocity dispersion, the Hyades cluster is in a stage of disruption according to kinematic forward-modelling. 

This snapshot is plotted in Fig.~\ref{fig:comp_pmradec} showing the $X-Y$ positions in the Galaxy (Sun centred), tangential velocities in R.A. and Dec in km/s, and in the CP diagram. These plots can be compared with the same coordinate representation of simulation \textbf{M1} (see Fig.~\ref{fig:space} for the $X-Y$ positions of the cluster and tail stars in the Galaxy, Fig.~\ref{fig:proj} for the proper motion plot, and Fig.~\ref{fig:CCP} for the CP diagram). In addition to the elevated velocity dispersion, the close presence of the lump already started stretching the tidal tail. Fig.~\ref{fig:comp_pmradec} clearly shows an asymmetry such that i) the trailing tail is lagging behind more strongly, with the first epicyclic overdensity being stretched out almost completely. The proper motion distribution is also affected, and in combination,
an asymmetric CP velocity projection becomes evident. The bottom panel in Fig.~\ref{fig:comp_pmradec} shows the effect on the CP velocity distribution. The contamination from the background increases with the relative velocity difference from the cluster centre because this corresponds to regions that are more distant from the Sun because of the large extent of the tail. It is therefore straightforward to see that because of the asymmetry, the observed distribution would be reflected in the detected tidal tails when the CP method is used. 
Fig.~\ref{fig:epyc} shows that the leading and trailing tails are reversed in the CP distribution. This means that while the close-lump encounter in the simulation in model \textbf{M4} is consistent with the observation of a potentially higher velocity dispersion, it does not appear to explain the observed asymmetry. For reference, in Fig.~1 of \cite{Roeser+19} the first 200~pc of the Hyades leading tail is seen in the CP diagram between 0 and $-$5~km/s. Our diagram instead shows an under-density in the same region. Nevertheless, it is possible that if a wider parameter space or more cluster--lump encounters are explored in the future, the CP diagram of the Hyades might be matched. More precise astrometric parameters with future \textit{Gaia} data releases will provide firmer data constraints.

More research is clearly required to reach firm conclusions. Based on our results, however, we suggest that an encounter between the Hyades and a lump with a mass of about $10^7\,M_\odot$ appears to be a possible explanation not only for the measured potentially higher velocity dispersion of the Hyades \citep{Roeser+11, Oh2020}, but possibly also for the asymmetry of the detected tidal tails \citep[][and also this work]{Roeser+19}. This explanation may turn out to be implausible, however, because it would require a lump of this mass to be about 117~pc away from the real Hyades cluster. The Hyades are located close to the Sun and in an inter-arm region of the Galaxy that largely lacks massive molecular clouds. While the closest star-forming molecular clouds are about 100~pc away, their masses are about four orders of magnitude lower.

\subsection{Quantitative comparison to our simulations}
To make a quantitative comparison of the models \textbf{M1, M2, M3, M4} described above, we list in Tab.~\ref{table:S_Hyad} time-averaged quantities in a Hyades-compatible age span (620-690 Myr for \textbf{M1, M2, and M3} and 
prior disruption times 615-635 Myr for \textbf{M4}). We show the tidal radius ($R_t$), the 3D velocity dispersion ($\sigma$), the stellar mass of the cluster ($M$), and the number of stars and stellar remnants ($N$) within a specified radius. For the lumpy Galaxy model we also include the time that is closer to the disruption event caused by the close interaction with a massive lump. The increase in the velocity dispersion of the cluster is clearly visible.

\begin{table*}
\centering                                
\begin{tabular}{l l l l l l l}          
\hline\hline                       
 parameter & value -- \textbf{M1} & value -- \textbf{M2} & value -- \textbf{M3} & value -- \textbf{M4}$^{\star}$ & value -- \textbf{M4}$^\dagger$ & value -- \textbf{M5} \\    
  & $L_z^0 = 0$ & $L_z^0 < 0$ & $L_z^0 > 0$ & $L_z^0 = 0$, lumps & $L_z^0 = 0$, lumps  & $L_z^0 = 0$ \\
  & & & & & (during disruption)  & \\
\hline     
\hline 
$R_t$ [pc] & 9.2 $\pm$ 0.1 & 9.5 $\pm$ 0.1 & 7.9 $\pm$ 0.1 & 5.2 $\pm$ 0.2 & 4.9 $\pm$ 0.7 & 9.1 $\pm$ 0.1  \\
$\sigma(r<R_t)$ [km/s] & 0.55 $\pm$ 0.01 & 0.57 $\pm$ 0.02 & 0.53 $\pm$ 0.01 & 0.85 $\pm$ 0.07 & 0.95 $\pm$ 0.09 & 0.52 $\pm$ 0.01 \\
$\sigma(r<3\,\mathrm{pc})$ [km/s] & 0.62 $\pm$ 0.01 & 0.68 $\pm$ 0.02 & 0.66 $\pm$ 0.05 & 1.2 $\pm$ 0.1 & 1.17 $\pm$ 0.06 &  0.61 $\pm$ 0.01  \\
$\sigma(r<9\,\mathrm{pc})$ [km/s] & 0.55 $\pm$ 0.01 & 0.58 0.01 & 0.52 $\pm$ 0.01 & 0.68 $\pm$ 0.02  & 0.93 $\pm$ 0.27 & 0.53 $\pm$ 0.01 \\
$M(r<R_t)]$ & 394 $\pm$ 11 & 436 $\pm$ 6 & 249 $\pm$ 2 & 60 $\pm$ 5 & 56 $\pm$ 2 & 380 $\pm$ 7\\
$M(r<30\,\mathrm{pc})]$ &  536 $\pm$ 7 & 568 $\pm$ 4 &  423 $\pm$ 8 & 191 $\pm$ 10 &  175 $\pm$ 4 & 660 $\pm$ 5\\
$N(r<R_t)$ & 704 $\pm$ 26  & 845 $\pm$ 14 & 347 $\pm$ 8 & 70 $\pm$ 7 &  63 $\pm$ 3 & 690 $\pm$ 13 \\
$N(r<30\,\mathrm{pc})$ & 1092 $\pm$ 17 & 1220 $\pm$ 14 & 770 $\pm$ 15 & 363 $\pm$ 20 & 317 $\pm$ 12 & 980 $\pm$ 9\\
average mass loss [$M_{\odot}$/yr] $^{\star \star}$ & 0.9 $\pm$ 0.1  & 0.8 $\pm$  0.1 & 1.1 $\pm$ 0.1 &  1.3 $\pm$ 0.2  & ($=$) 1.3 $\pm$ 0.2 & 0.9 $\pm$ 0.1 \\
\hline
\end{tabular}
\caption{Simulated present-day properties (mean values between ages of 620 and 695 Myr for {\bf M1, M2} and {\bf M3} and 615 to 635 Myr for {\bf M4;} time before cluster disruption) of the Hyades star cluster.
$^{\star}$ The mass loss in model \textbf{M4} is more complex than in the other models, see the right panel of Fig.~\ref{fig:LZ}, thus the linear mass-loss average only serves as an orientation value.
$^{\star \star}$ The average mass loss of each of the models has been calculated by a linear fit to the cluster mass loss over the simulation time (see Fig.~\ref{fig:LZ}, left panel). 
 $\dagger$ As described in Sec.~\ref{sec:lumps}, the \textbf{M4} model is disrupted by a close encounter with a $10^7\, M_{\odot}$ lump. The disruption event occurs approximately 645~Myr after the start of the simulation, and we present the last column of values, which are averages of the values between 638~Myr and  643~Myr. This column shows the increase in velocity dispersion before the cluster is disrupted.  \label{table:S_Hyad}  }
\end{table*}

The left panel of Fig.~\ref{fig:LZ} shows the evolution in the angular momentum of the star cluster models within different setups (\textbf{M1, M2, M3, and M4}). The two initially non-rotating setups develop rotation in the direction of orbital motion (the cluster spin angular momentum points in the same direction as the orbital angular momentum). For the lumpy Galaxy model, the cluster has a higher mass loss and thus a lower cluster mass at the nominal age of the Hyades. This model also develops a non-zero spin angular momentum that faster returns to the zero value than for the \textbf{M1} model in conjunction with its significantly higher mass loss. 

The right panel shows the time evolution of the stellar mass that is outside of the tidal radius, $R_t$. 
As expected, the model in a lumpy Galaxy loses mass at the highest rate and with the highest dispersion. To compute the average mass loss, which is also provided in Tab.~\ref{table:S_Hyad}, we approximated each relation using a linear fit, and the average mass-loss rate is then represented by its slope.

\section{Discussion}
\label{sec:disc}

\subsection{Comparison with the study of \cite{Roeser+19} }
\cite{Roeser+19} used the CP method to recover a co-moving group of objects based on correcting their proper motions for on-the-sky projection. They considered the region of 200 pc around the Sun. In their sample, an overdensity was found. They used a clustering algorithm to recover objects belonging to this overdensity. These authors published their data set, which comprises the leading tail extending up to 170 pc from the centre of the Hyades with 292 stars, and a trailing tail up to 70~pc from the cluster, containing 237 stars. 

Based on the \textit{Gaia} IDs, we cross-matched our \textbf{S1}-selected data sample with the published data set of \cite{Roeser+19} and recovered 891 of 1316 targets (with our full initial catalogue, we recovered 1182 targets). 
The RUWE quality filtering removes about 100 targets from the data set of \cite{Roeser+19}. The additional missing targets are removed by the \textbf{S1} cut because they are the most kinematically incompatible with being tidal tails members based on our model. One simple example of this behaviour are targets with an opposite kinematic pattern based on their location in the leading or trailing tail.  

\begin{figure}
        \centering
        \includegraphics[scale=1.0]{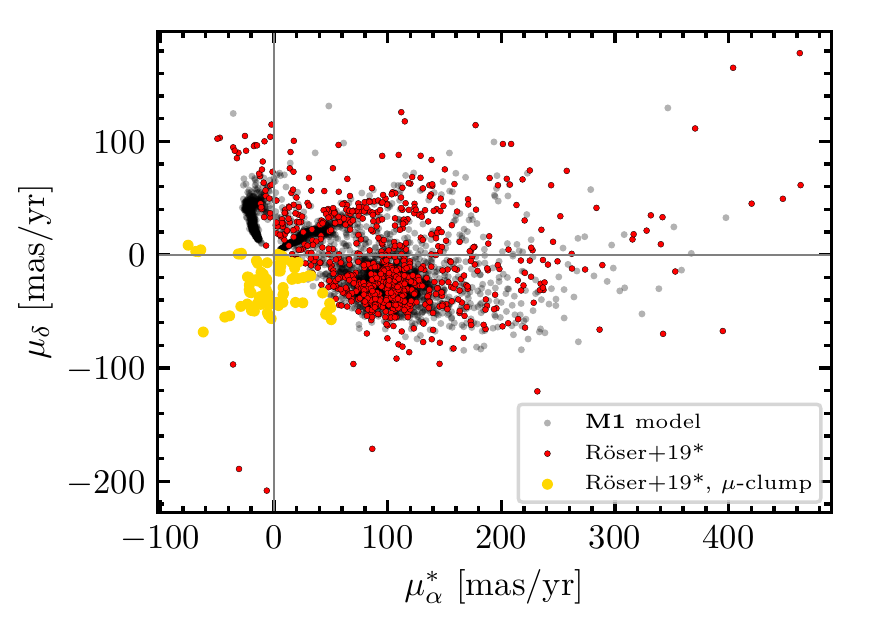}
        \includegraphics[scale=1.0]{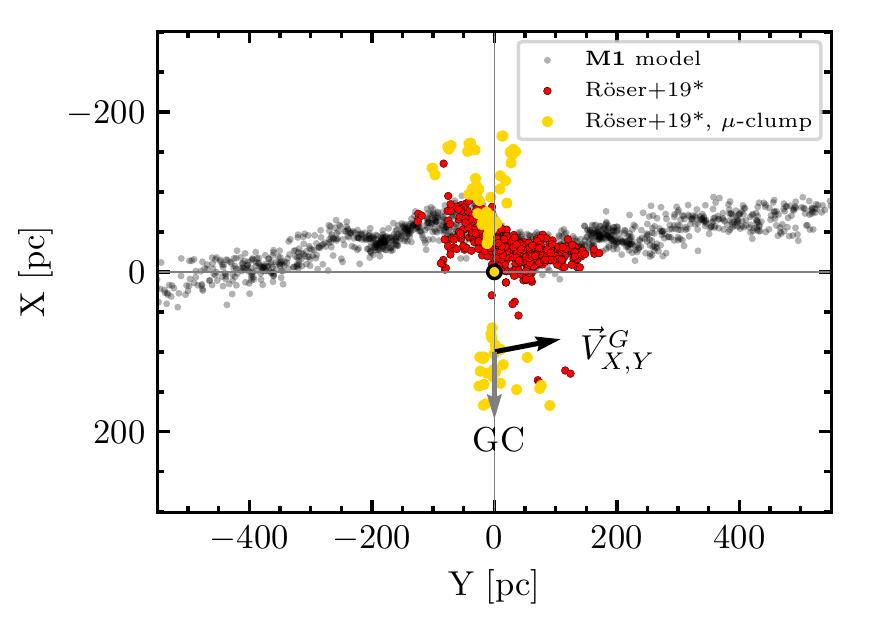}
                \caption{Comparison of models and data selection done by \cite{Roeser+19}. \textbf{Top panel:} Proper motion distribution in mas/yr plotted for the \textbf{M1} model as black background points. In red we show  targets identified as members of the Hyades and its tail by \cite{Roeser+19} in common with our S1-selected sample, and yellow points represent a selected subset of these that creates a clump in proper motion distribution that is offset from the model prediction. \textbf{Bottom panel:} Spatial distribution of the same targets as in the top panel, shown in Galactic Cartesian coordinates with its central sun plotted as a yellow point with a black line. The selected clump in proper motion (yellow points) is scattered over the tidal tails and partially overlaps it. The points are selected based on their position in measured proper motion without the need to introduce any model. }
        \label{fig:roeser}
\end{figure}

In Fig.~\ref{fig:roeser} we show our baseline model (as used in all plots so far) over-plotted with the cross-matched data set \citep{Roeser+19} and our \textbf{S1}-selected data sample. It is immediately visible that while there is general agreement, a number of points from \cite{Roeser+19} do not overlap with the model. As an example, we chose the clustering of points extending to (0,0) and plotted them in a different colour. This clustering is scattered around the Hyades and the tidal tail. 

\begin{figure}
        \includegraphics[scale=1.0]{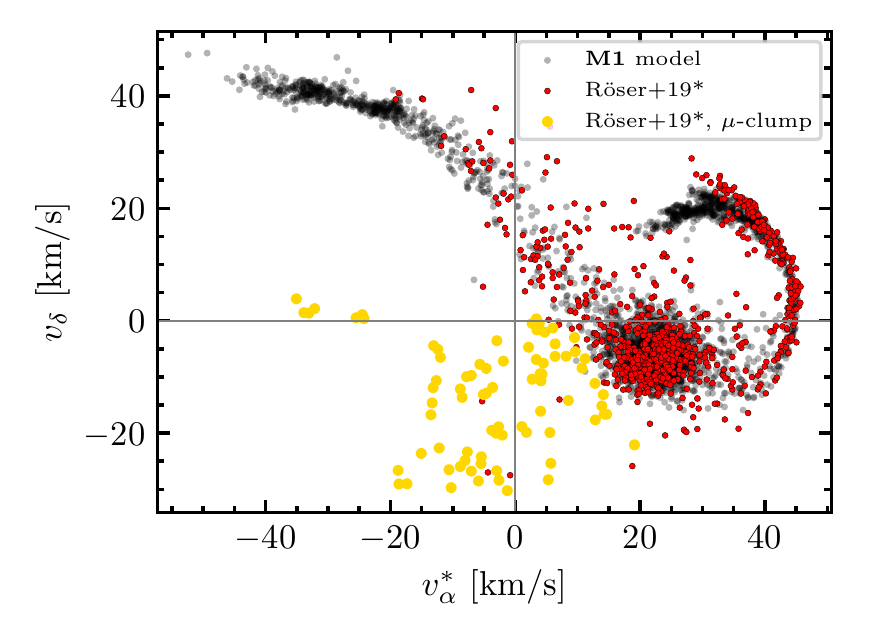}
        \includegraphics[scale=1.0]{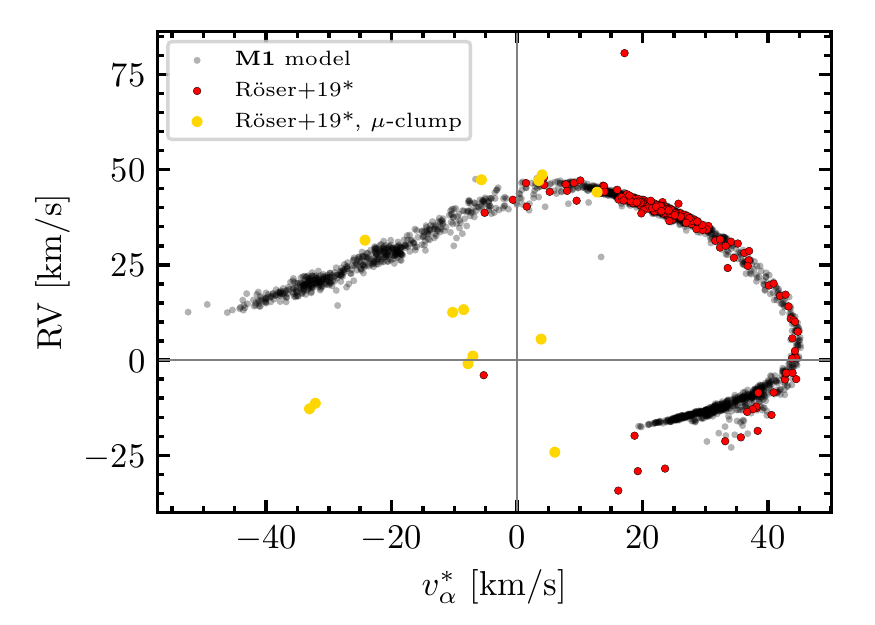}
        \caption{Same set of points to complement Fig.~\ref{fig:roeser} (black for \textbf{model M1} , red for \cite{Roeser+19} cross-matched with our S1-selected data sample, and yellow for the proper motion in mas/yr selected clump) plotted in tangential velocities in km/s (\textbf{top panel}) and proper motion in R.A. vs. radial velocities (\textbf{bottom panel}).  }
        \label{fig:roeserRV}
    \end{figure}

The CaMD does not provide a difference between the clump and the other data points from \cite{Roeser+19}.
More precise age and chemical composition estimates are required to investigate the association of these points with the Hyades. However, the spatial distribution, the proper motions (on-the-sky and in~km/s, and also the radial velocity measurements available for a data sub-sample) in Fig.\ref{fig:roeser} and Fig.\ref{fig:roeserRV} suggest that these are more likely contamination objects and not members of the Hyades cluster and its tail. 

\subsection{Limitation of the $N$-body integrator} \label{sec:nbody_dis}
%Nbody6/PeTar vs softening --> binaries, IMF, etc

The quantification of the tidal-tail models we performed relies on the $N-$body integrator \texttt{Huayno} \citep{Huayno2012,Huayno2014}, which uses an optimal softening length (as described in Sec.~\ref{sec:int}, see eq.~\ref{eq:eps}) to allow integrating close encounters between stars. This is a reasonable approximation to the overall energy-equipartition process coming from the many weak stellar encounters, but it fails to take the strong close gravitational interactions between stars into account. Because these are very rare and tend to cause stars to be ejected, the overall results as quantified in terms of the stellar mass loss from the cluster through evaporation are expected to be reasonably accurate over the time of the simulations. This evaporation, rather than the strong three-body encounters that eject stars, is the relevant physical process that populates the tidal tails. We tested whether our models are able to reproduce the tidal tails reasonably accurately by comparing the mass-loss rate from our cluster models with models that were computed with the very advanced and precise Aarseth $N-$body codes \citep[e.g.][]{Ernst+11}. The models presented here and those of \cite{Ernst+11} have a comparable mass loss, but the general trend of our models with softening is that the mass-loss values are slightly lower.  This is expected because softening makes the energy equipartition process less efficient (in the limit of a completely relaxation-free code, such as is used in galactic dynamics simulations, there would be no energy equipartition through two-body encounters and thus no mass loss through this process). The overall morphology of the tidal tails computed here with the K\"upper epicyclic overdensities and length indicates, however, that the energetics of the evaporating stars are calculated sufficiently accurately for our purpose in comparison with the Aarseth code. 

\subsection{Effect of the chosen MW potential parameters}
We have considered two different MW-like potentials, those by \cite{Allen1991} and \cite{Irrgang2013}. In general, a direct comparison of cluster evolution in different potentials is difficult because the clusters will have different trajectories and will be exposed to different tidal field. 
We have started with the present-day Hyades position and integrated back in a given potential for the same amount of time to obtain the initial conditions. We used the same cluster initial mass and radius. 
Fig.~\ref{fig:LZ} shows that the models \textbf{M1} and \textbf{M5} follow the same angular momentum evolution and have the same mass loss.

Interestingly, there is a difference in the structure of the tidal tails between model \textbf{M1} and \textbf{M5} for the same time snapshot, as shown in Fig.~\ref{fig:epy-hiss}. This demonstrates the great potential of epicyclic overdensities to constrain the Galactic potential, as described by \cite{Kuepper+08,Kuepper+15}. 
The updated model of the MW potential by \cite{Irrgang2013} reproduces the position of the detected overdensities in the eDR3 data better than the \cite{Allen1991} potential. 
This opens the possibility for future work to significantly constrain the Galactic potential.

\subsection{Limitations of the model-dependent selection}
The standard way of searching for star clusters is to use a clustering algorithm that identifies overdensities in velocity (often proper motion) space \citep{CantatGaudin2020}. This method not only allowed the recovery of stellar clusters, but also of some extended (100 pc long) stellar structures  \citep{Jerabkova2019,Kounkel2019, Beccari2020}. Removing projection effects in the standard CP method \citep{vanLeeuwen09} showed to be very promising in recovering up to 200 pc long  (in length extending from the cluster) tidal tails of open clusters \citep[e.g. Hyades,][]{Roeser+19}. 
The $N-$body modelling of Hyades-like star clusters on a realistic orbit in the Galactic potential here has demonstrated that the full extent of the tidal tails cannot be recovered with any of the methods mentioned above because it is represented by an extended (overall up to $\pm40\,$km/s) structure in different velocity spaces, and because the extent depends on the initial conditions such as the age and position of the cluster and the tail in the Galaxy relative to the Sun. 
We therefore developed the new CCP method that takes the modelled parameters into account  (see Sec.~\ref{sec:ccp}).

The model-dependent CCP search method comes with the same caveat as the standard CP method for the velocities to recover star clusters. When the observed system deviates from the model, for example, because an environmental effect is not taken into account in the model, important parts of the stellar population may not be evident. This has been demonstrated in the lumpy Galaxy model \textbf{M4}, which deviates from the smooth Galactic model \textbf{M1} in all phase spaces. 
While this can present a certain obstacle in the search for tidal tails, the comparison with models becomes necessary in order to understand the reason for this 
when the search is incomplete (as it appears to be in the case of Hyades). 

In addition, the identified candidate members of the Hyades and its tail need to be further verified in detailed chemical studies. This is especially important for tidal tails with ages of about 500~Myr to 1~Gyr because the  identification based on isochrones becomes very difficult in these cases; see for example the study by \cite{Hawkins2020} using the LAMOST\footnote{LAMOST survey contains two main parts: the LAMOST ExtraGAlactic Survey (LEGAS), and the LAMOST Experiment for Galactic Understanding and Exploration (LEGUE) survey of MW stellar structure.} spectroscopic survey to establish the age for the nearby Pisces-Eridanus stellar stream as 120~Myr and not 1~Gyr, as estimated before through isochrone fitting.

\subsection{Relation of period distribution to age  using \textit{Gaia}-derived periods}

One way to test whether candidate stars are likely past Hyades members is to determine whether their spin  periods are similar to those of the current Hyades members, that is, determine whether the spin speeds of these stars are consistent with the age of the Hyades stars.
\cite{Curtis2019} used data from the Transiting Exoplanet Survey Satellite (TESS) to estimate 
period distributions for the Pisces-Eridanus stellar stream. They compared the period distribution of 
the stream members with period distributions of stars belonging to nearby open clusters of different ages. They acquired spin period distributions for the 120 Myr old Pleiades, for the 670 Myr old Praesepe cluster, and for the 1000 Myr old NGC 6811. Based on this comparison, they concluded that the age of the Pisces-Eridanus stellar stream is comparable to that of the Pleiades, that is, about 120 Myr. 
This method thus significantly increases the precision of age estimates for stellar groups with ages $> 100\,$Myr for which 
low-mass stars (which lie below the turnoff in the colour-magnitude diagram) can no longer be distinguished from the main sequence. 

Together with chemical tagging \citep[see e.g.][]{Hawkins2020}, this method is very important for the purpose of testing the membership of the identified stars in the Hyades tidal tail. 
There is a difference between estimating the age of a star cluster and the age of a large-scale co-moving group of stars:
For the cluster, the membership probability is given by the spatial and kinematic signature, and because of the compactness in phase-space, these constraints usually allow reducing contamination from the field. For large-scale structures such as relic filaments and tidal tails, which are close to becoming the field population, this is not the case. Data in addition to CaMD-age estimates, such as spin-period--age data, are therefore extremely useful. 

\begin{figure}
        \includegraphics[scale=1.0]{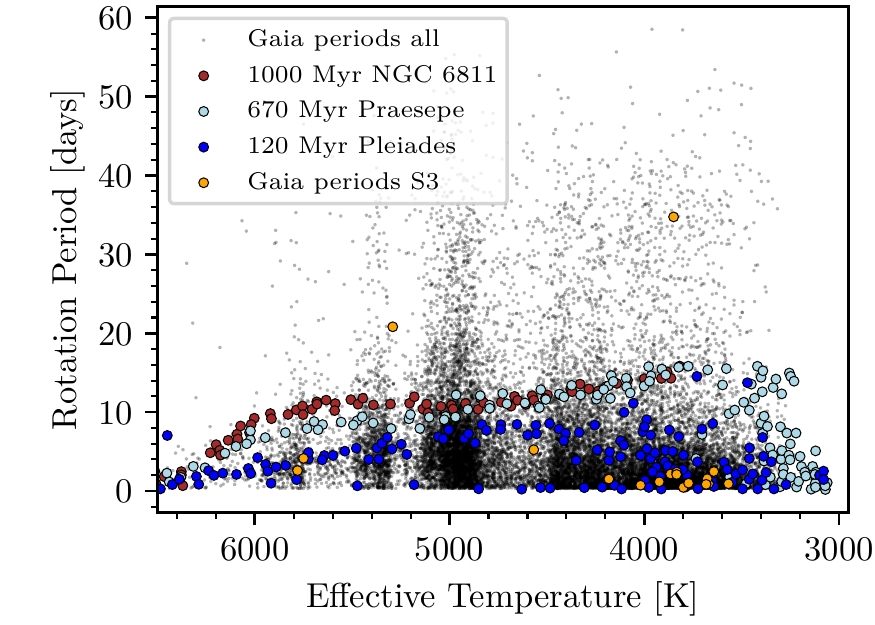}
        \caption{Stellar spin-rotation periods (for tail candidates after selection S3), either estimated from TESS data by \cite{Curtis2019} or taken from \cite{Lanzafame2018}, as a function of the Gaia-estimated effective temperature. }
        \label{fig:periods}
    \end{figure}

A separate dedicated study would be needed to derive the period distribution from the TESS data, as was done by \cite{Curtis2019}. However, rotational periods have been estimated by \cite{Lanzafame2018} for \textit{Gaia} DR2 data. 
We acquired all the periods from the \cite{Lanzafame2018} \texttt{gaiadr2.vari\_rotatio\_modulation} catalogue in our initial data sample, see the black points in Fig.~\ref{fig:periods}. 
The orange points in this figure show the available periods in the catalogue for our final Hyades and tail candidate members: only a handfull of points are available at present. All these points belong to the cluster area and are not from the tidal tail. 
For comparison we plot the periods derived from TESS data by \cite{Curtis2019}.

Unfortunately, the incompleteness in the data does not allow us to reach any conclusions on the tidal tails candidates. However, it is interesting to note that the Hyades cluster members that have Gaia-estimated spin periods appear to be more consistent with the period distribution of the Pleiades than with Praesepe, which is closer in age to the Hyades. The clear conclusion from this exercise is the need for further analysis and more spin-period and chemical composition measurements of Hyades and its tail member candidates.

\subsection{Future prospects and general context}
The \textit{Gaia} early-data release~3
(EDR3), available since December~2020, promises an improved precision in the determination of the astrometric 
parameters. This will significantly improve the confidence in detecting structures 
in proper motion values and will help us to better constrain distances. One key question that might be answered in a follow-up study with more precise astrometric parameters is whether the epicyclic overdensities are truly absent from the data. 

On the other hand, there is no expectation that radial velocity information will 
be available for faint targets ($G>15$ mag) in the future \textit{Gaia} data releases. 
This will always present a limitation of the detection method and thus enforces the use of the CP method and the newly introduced CCP method, neither of which relies on the radial velocity information. 

Future surveys such as the ESO 4MOST (4-metre Multi-Object Spectroscopic Telescope) and the ING WEAVE (William Herschel Telescope Enhanced Area Velocity Explorer), with expected first data approximately in 2025, will change the situation regarding radial velocities. While these surveys will provide radial velocities for more targets than \textit{Gaia,} neither of these instruments will deliver a full-sky RV survey complete down to $G \approx 20$ mag.

\section{Conclusions}\label{sec:concl}
We performed $N-$body simulations of Hyades-like star clusters on a realistic orbit in the Galactic potential within the AMUSE computational environment. 
We explored the effect of initial cluster rotation and presence of Galactic lumps in addition to the smooth Galactic potential on the evolution of the tidal tail. The main results are summarized below.

\begin{enumerate}
    \item The $N-$body simulations show that tidal tails are extended low-density structures. The contamination by Galactic-field stars means that they are difficult to recover in their full extent using standard methods that rely on finding clusters and overdensities in the data. The most naturally clustered tidal tails appear to be in galactocentric cylindrical velocities, but the required RV measurements for computing the stellar velocities in galactocentric coordinates are (and will remain) highly incomplete in the \textit{Gaia} catalogue. 
    \item The most powerful method for detecting co-moving structures is the so-called CP method \citep{vanLeeuwen09, Roeser+19}. It does not rely on the RV measurements a priori and removes the projection effect arising from the large extent on the sky. However, as demonstrated in the top panel of  Fig.~\ref{fig:CCP}, the full extent of the tail (out to a distance of 600~pc from the cluster for the Hyades) is extended (by $\pm20\,$km/s) in the CP plot. 
    Because the contamination  increases with the third power of the distance from the Sun in the case of the Hyades, the CP method will therefore not reliably recover the overdensity for the full extent of the tidal tail. 
    \item We suggest using the model-dependent CCP method. The CCP method uses 
    the velocity versus distance properties of tidal tail models to formulate a transformation such that the extended 5D structure becomes a compact collection of points (see Fig.~\ref{fig:CCP}). Using this method, we were able to recover candidate members of the Hyades tidal tails at its full expected extent of 800 pc. The CCP method is only marginally model dependent because the transformation remains robust and largely invariant in all models we used. Moreover, the CP method  assumes that the structure being sought consists of a group of stars with the same velocity vector direction, which introduces less realistic model assumptions.
    
    \item We confirmed the previously noted asymmetry in the detected tidal tail: The leading tail is significantly more pronounced than the trailing tail, which is almost absent. 
    We showed that a close encounter with a massive Galactic lump (or dark matter substructure,  as suggested for some halo streams, \citealt{Bonaca2019}) could lead to an asymmetric tidal tail. 
    
    \item In the eDR3 data we recovered spatial overdensities in leading and trailing tails that are kinematically consistent with being epicyclic overdensities and thus may present the first such detection for an open star cluster. By using two different MW potentials, \cite{Allen1991} and \cite{Irrgang2013} in our N-body simulations, we showed that the epicyclic overdensities of open star clusters are able to provide constraints not only on the cluster properties, but also on the Galactic potential.
\end{enumerate}

Future \textit{Gaia} data releases will deliver more precise astrometric measurements, and with upcoming spectral surveys (4MOST and WEAVE) a significantly improved  6D phase-space map of the solar neighbourhood is expected.
With this information in hand, we will have the opportunity to explore the full potential of extended Galactic structures. These structures provide not only valuable clues to the detailed physics of star-formation, but are also a link to the large scale of the Galactic potential, its tides, and shears. We showed that detailed modelling is fruitful, if not necessary, for interpreting and finding extended tidal tails in the \textit{Gaia} data and for distinguishing them from the Galactic field star populations that the members of tidal tails are becoming part of.

\begin{acknowledgements}
TJ would like to wholeheartedly acknowledge the encouragement and positiveness of astronomer, observer and friend Rebeca Galera Rosillo ($\star 1988$-$\dagger 2020$) especially during the pandemic lock-down at La Palma: Gracias por todo, tu recuerdo permanecerá en el cielo nocturno para siempre. \\
The authors are thankful to the anonymous referee for useful comments and suggestions that improved the manuscript. 
TJ is grateful to Siegfried R\"{o}ser for useful discussions and clarification of the CP method implementation. TJ acknowledges support through  the European Space Agency fellowship programme,  the Erasmus+ programme of the European Union under grant number 2017-1-CZ01- KA203-035562 on La Palma and the support from the HISPK in Bonn.
This research has made use of the
SIMBAD database and of the VizieR catalogue access tool, operated at CDS, Strasbourg, France. This work has made use of data from the European Space
Agency (ESA) mission \textit{Gaia} (https://www.cosmos.esa.int/gaia), processed by the \textit{Gaia} Data Processing and Analysis Consortium (DPAC, https:
//www.cosmos.esa.int/web/gaia/dpac/consortium). Funding for the
DPAC has been provided by national institutions, in particular the institutions
participating in the \textit{Gaia} Multilateral Agreement.
      \end{acknowledgements}

   \bibliographystyle{aa} % style aa.bst
   \bibliography{AATTreferences} % your references Yourfile.bib

\begin{thebibliography}{100}
\expandafter\ifx\csname natexlab\endcsname\relax\def\natexlab#1{#1}\fi

\bibitem[{{Allen} \& {Santillan}(1991)}]{Allen1991}
{Allen}, C. \& {Santillan}, A. 1991, \rmxaa, 22, 255

\bibitem[{{Andr{\'e}} {et~al.}(2014){Andr{\'e}}, {Di Francesco},
  {Ward-Thompson}, {Inutsuka}, {Pudritz}, \& {Pineda}}]{Andre2014}
{Andr{\'e}}, P., {Di Francesco}, J., {Ward-Thompson}, D., {et~al.} 2014, in
  Protostars and Planets VI, ed. H.~{Beuther}, R.~S. {Klessen}, C.~P.
  {Dullemond}, \& T.~{Henning}, 27

\bibitem[{{Astropy Collaboration} {et~al.}(2018){Astropy Collaboration},
  {Price-Whelan}, {Sip{H{o}}cz}, {G{"u}nther}, {Lim}, {Crawford}, {Conseil},
  {Shupe}, {Craig}, {Dencheva}, {Ginsburg}, {Vand erPlas}, {Bradley},
  {P{'e}rez-Su{'a}rez}, {de Val-Borro}, {Aldcroft}, {Cruz}, {Robitaille},
  {Tollerud}, {Ardelean}, {Babej}, {Bach}, {Bachetti}, {Bakanov}, {Bamford},
  {Barentsen}, {Barmby}, {Baumbach}, {Berry}, {Biscani}, {Boquien}, {Bostroem},
  {Bouma}, {Brammer}, {Bray}, {Breytenbach}, {Buddelmeijer}, {Burke},
  {Calderone}, {Cano Rodr{'i}guez}, {Cara}, {Cardoso}, {Cheedella}, {Copin},
  {Corrales}, {Crichton}, {D'Avella}, {Deil}, {Depagne}, {Dietrich}, {Donath},
  {Droettboom}, {Earl}, {Erben}, {Fabbro}, {Ferreira}, {Finethy}, {Fox},
  {Garrison}, {Gibbons}, {Goldstein}, {Gommers}, {Greco}, {Greenfield},
  {Groener}, {Grollier}, {Hagen}, {Hirst}, {Homeier}, {Horton}, {Hosseinzadeh},
  {Hu}, {Hunkeler}, {Ivezi{'c}}, {Jain}, {Jenness}, {Kanarek}, {Kendrew},
  {Kern}, {Kerzendorf}, {Khvalko}, {King}, {Kirkby}, {Kulkarni}, {Kumar},
  {Lee}, {Lenz}, {Littlefair}, {Ma}, {Macleod}, {Mastropietro}, {McCully},
  {Montagnac}, {Morris}, {Mueller}, {Mumford}, {Muna}, {Murphy}, {Nelson},
  {Nguyen}, {Ninan}, {N{"o}the}, {Ogaz}, {Oh}, {Parejko}, {Parley}, {Pascual},
  {Patil}, {Patil}, {Plunkett}, {Prochaska}, {Rastogi}, {Reddy Janga},
  {Sabater}, {Sakurikar}, {Seifert}, {Sherbert}, {Sherwood-Taylor}, {Shih},
  {Sick}, {Silbiger}, {Singanamalla}, {Singer}, {Sladen}, {Sooley},
  {Sornarajah}, {Streicher}, {Teuben}, {Thomas}, {Tremblay}, {Turner},
  {Terr{'o}n}, {van Kerkwijk}, {de la Vega}, {Watkins}, {Weaver}, {Whitmore},
  {Woillez}, {Zabalza}, \& {Astropy Contributors}}]{astropy}
{Astropy Collaboration}, {Price-Whelan}, A.~M., {Sip{H{o}}cz}, B.~M., {et~al.}
  2018, aj, 156, 123

\bibitem[{{Astropy Collaboration} {et~al.}(2013){Astropy Collaboration},
  {Robitaille}, {Tollerud}, {Greenfield}, {Droettboom}, {Bray}, {Aldcroft},
  {Davis}, {Ginsburg}, {Price-Whelan}, {Kerzendorf}, {Conley}, {Crighton},
  {Barbary}, {Muna}, {Ferguson}, {Grollier}, {Parikh}, {Nair}, {Unther},
  {Deil}, {Woillez}, {Conseil}, {Kramer}, {Turner}, {Singer}, {Fox}, {Weaver},
  {Zabalza}, {Edwards}, {Azalee Bostroem}, {Burke}, {Casey}, {Crawford},
  {Dencheva}, {Ely}, {Jenness}, {Labrie}, {Lim}, {Pierfederici}, {Pontzen},
  {Ptak}, {Refsdal}, {Servillat}, \& {Streicher}}]{astropy2013}
{Astropy Collaboration}, {Robitaille}, T.~P., {Tollerud}, E.~J., {et~al.} 2013,
  \aap, 558, A33

\bibitem[{{Banerjee} \& {Kroupa}(2017)}]{BK17}
{Banerjee}, S. \& {Kroupa}, P. 2017, \aap, 597, A28

\bibitem[{{Baumgardt} \& {Makino}(2003)}]{BM03}
{Baumgardt}, H. \& {Makino}, J. 2003, \mnras, 340, 227

\bibitem[{{Beccari} {et~al.}(2020){Beccari}, {Boffin}, \&
  {Jerabkova}}]{Beccari2020}
{Beccari}, G., {Boffin}, H. M.~J., \& {Jerabkova}, T. 2020, \mnras, 491, 2205

\bibitem[{{Beccari} {et~al.}(2018){Beccari}, {Boffin}, {Jerabkova}, {Wright},
  {Kalari}, {Carraro}, {De Marchi}, \& {de Wit}}]{Beccari2018}
{Beccari}, G., {Boffin}, H. M.~J., {Jerabkova}, T., {et~al.} 2018, \mnras, 481,
  L11

\bibitem[{{Beccari} {et~al.}(2017){Beccari}, {Petr-Gotzens}, {Boffin},
  {Romaniello}, {Fedele}, {Carraro}, {De Marchi}, {de Wit}, {Drew}, {Kalari},
  {Manara}, {Martin}, {Mieske}, {Panagia}, {Testi}, {Vink}, {Walsh}, \&
  {Wright}}]{Beccari2017}
{Beccari}, G., {Petr-Gotzens}, M.~G., {Boffin}, H.~M.~J., {et~al.} 2017, \aap,
  604, A22

\bibitem[{{Binney} \& {Tremaine}(1987)}]{BinneyTremaine87}
{Binney}, J. \& {Tremaine}, S. 1987, {Galactic dynamics} ({Princeton University
  Press})

\bibitem[{{Bland-Hawthorn} \& {Gerhard}(2016)}]{Bland2016}
{Bland-Hawthorn}, J. \& {Gerhard}, O. 2016, \araa, 54, 529

\bibitem[{{Bonaca} {et~al.}(2019){Bonaca}, {Hogg}, {Price-Whelan}, \&
  {Conroy}}]{Bonaca2019}
{Bonaca}, A., {Hogg}, D.~W., {Price-Whelan}, A.~M., \& {Conroy}, C. 2019, \apj,
  880, 38

\bibitem[{{Cantat-Gaudin} {et~al.}(2020){Cantat-Gaudin}, {Anders},
  {Castro-Ginard}, {Jordi}, {Romero-G{\'o}mez}, {Soubiran}, {Casamiquela},
  {Tarricq}, {Moitinho}, {Vallenari}, {Bragaglia}, {Krone-Martins}, \&
  {Kounkel}}]{CantatGaudin2020}
{Cantat-Gaudin}, T., {Anders}, F., {Castro-Ginard}, A., {et~al.} 2020, \aap,
  640, A1

\bibitem[{{Cantat-Gaudin} {et~al.}(2019){Cantat-Gaudin}, {Mapelli},
  {Balaguer-N{\'u}{\~n}ez}, {Jordi}, {Sacco}, \&
  {Vallenari}}]{Cantat-Gaudin2019}
{Cantat-Gaudin}, T., {Mapelli}, M., {Balaguer-N{\'u}{\~n}ez}, L., {et~al.}
  2019, \aap, 621, A115

\bibitem[{{Carballo-Bello} {et~al.}(2020){Carballo-Bello}, {Salinas}, \&
  {Piatti}}]{Carballo+20}
{Carballo-Bello}, J.~A., {Salinas}, R., \& {Piatti}, A.~E. 2020, \mnras
  [\eprint[arXiv]{2009.11320}]

\bibitem[{{Chumak} {et~al.}(2005){Chumak}, {Rastorguev}, \&
  {Aarseth}}]{Chumak+05}
{Chumak}, Y.~O., {Rastorguev}, A.~S., \& {Aarseth}, S.~J. 2005, Astronomy
  Letters, 31, 308

\bibitem[{{Curtis} {et~al.}(2019){Curtis}, {Ag{\"u}eros}, {Mamajek}, {Wright},
  \& {Cummings}}]{Curtis2019}
{Curtis}, J.~L., {Ag{\"u}eros}, M.~A., {Mamajek}, E.~E., {Wright}, J.~T., \&
  {Cummings}, J.~D. 2019, \aj, 158, 77

\bibitem[{{Dabringhausen} {et~al.}(2016){Dabringhausen}, {Kroupa}, {Famaey}, \&
  {Fellhauer}}]{Dabringhausen2016}
{Dabringhausen}, J., {Kroupa}, P., {Famaey}, B., \& {Fellhauer}, M. 2016,
  \mnras, 463, 1865

\bibitem[{{de Bruijne}(1999)}]{Jos1999}
{de Bruijne}, J. H.~J. 1999, \mnras, 306, 381

\bibitem[{{De Gennaro} {et~al.}(2009){De Gennaro}, {von Hippel}, {Jefferys},
  {Stein}, {van Dyk}, \& {Jeffery}}]{DeGennaro2009}
{De Gennaro}, S., {von Hippel}, T., {Jefferys}, W.~H., {et~al.} 2009, \apj,
  696, 12

\bibitem[{{Dinnbier} \& {Kroupa}(2020{\natexlab{a}})}]{DK20a}
{Dinnbier}, F. \& {Kroupa}, P. 2020{\natexlab{a}}, \aap, 640, A84

\bibitem[{{Dinnbier} \& {Kroupa}(2020{\natexlab{b}})}]{DK20b}
{Dinnbier}, F. \& {Kroupa}, P. 2020{\natexlab{b}}, \aap, 640, A85

\bibitem[{{Dinnbier} \& {Walch}(2020)}]{DW20}
{Dinnbier}, F. \& {Walch}, S. 2020, \mnras [\eprint[arXiv]{2008.08602}]

\bibitem[{{Douglas} {et~al.}(2019){Douglas}, {Curtis}, {Ag{\"u}eros},
  {Cargile}, {Brewer}, {Meibom}, \& {Jansen}}]{Hy_age1}
{Douglas}, S.~T., {Curtis}, J.~L., {Ag{\"u}eros}, M.~A., {et~al.} 2019, \apj,
  879, 100

\bibitem[{{Ernst} {et~al.}(2011){Ernst}, {Just}, {Berczik}, \&
  {Olczak}}]{Ernst+11}
{Ernst}, A., {Just}, A., {Berczik}, P., \& {Olczak}, C. 2011, \aap, 536, A64

\bibitem[{{F{\"u}rnkranz} {et~al.}(2019){F{\"u}rnkranz}, {Meingast}, \&
  {Alves}}]{Fuernkranz+19}
{F{\"u}rnkranz}, V., {Meingast}, S., \& {Alves}, J. 2019, \aap, 624, L11

\bibitem[{{Gaburov} {et~al.}(2009){Gaburov}, {Harfst}, \& {Portegies
  Zwart}}]{grape2009}
{Gaburov}, E., {Harfst}, S., \& {Portegies Zwart}, S. 2009, \na, 14, 630

\bibitem[{{Gaia Collaboration} {et~al.}(2018{\natexlab{a}}){Gaia
  Collaboration}, {Babusiaux}, {van Leeuwen}, {Barstow}, {Jordi}, {Vallenari},
  {Bossini}, {Bressan}, {Cantat-Gaudin}, {van Leeuwen}, {Brown}, {Prusti}, {de
  Bruijne}, {Bailer-Jones}, {Biermann}, {Evans}, {Eyer}, {Jansen}, {Klioner},
  {Lammers}, {Lindegren}, {Luri}, {Mignard}, {Panem}, {Pourbaix}, {Randich},
  {Sartoretti}, {Siddiqui}, {Soubiran}, {Walton}, {Arenou}, {Bastian},
  {Cropper}, {Drimmel}, {Katz}, {Lattanzi}, {Bakker}, {Cacciari},
  {Casta{\~n}eda}, {Chaoul}, {Cheek}, {De Angeli}, {Fabricius}, {Guerra},
  {Holl}, {Masana}, {Messineo}, {Mowlavi}, {Nienartowicz}, {Panuzzo},
  {Portell}, {Riello}, {Seabroke}, {Tanga}, {Th{\'e}venin}, {Gracia-Abril},
  {Comoretto}, {Garcia-Reinaldos}, {Teyssier}, {Altmann}, {Andrae}, {Audard},
  {Bellas-Velidis}, {Benson}, {Berthier}, {Blomme}, {Burgess}, {Busso},
  {Carry}, {Cellino}, {Clementini}, {Clotet}, {Creevey}, {Davidson}, {De
  Ridder}, {Delchambre}, {Dell'Oro}, {Ducourant},
  {Fern{\'a}ndez-Hern{\'a}ndez}, {Fouesneau}, {Fr{\'e}mat}, {Galluccio},
  {Garc{\'\i}a-Torres}, {Gonz{\'a}lez-N{\'u}{\~n}ez}, {Gonz{\'a}lez-Vidal},
  {Gosset}, {Guy}, {Halbwachs}, {Hambly}, {Harrison}, {Hern{\'a}ndez},
  {Hestroffer}, {Hodgkin}, {Hutton}, {Jasniewicz}, {Jean-Antoine-Piccolo},
  {Jordan}, {Korn}, {Krone-Martins}, {Lanzafame}, {Lebzelter}, {L{\"o}ffler},
  {Manteiga}, {Marrese}, {Mart{\'\i}n-Fleitas}, {Moitinho}, {Mora}, {Muinonen},
  {Osinde}, {Pancino}, {Pauwels}, {Petit}, {Recio-Blanco}, {Richards},
  {Rimoldini}, {Robin}, {Sarro}, {Siopis}, {Smith}, {Sozzetti}, {S{\"u}veges},
  {Torra}, {van Reeven}, {Abbas}, {Abreu Aramburu}, {Accart}, {Aerts},
  {Altavilla}, {{\'A}lvarez}, {Alvarez}, {Alves}, {Anderson}, {Andrei},
  {Anglada Varela}, {Antiche}, {Antoja}, {Arcay}, {Astraatmadja}, {Bach},
  {Baker}, {Balaguer-N{\'u}{\~n}ez}, {Balm}, {Barache}, {Barata}, {Barbato},
  {Barblan}, {Barklem}, {Barrado}, {Barros}, {Bartholom{\'e} Mu{\~n}oz},
  {Bassilana}, {Becciani}, {Bellazzini}, {Berihuete}, {Bertone}, {Bianchi},
  {Bienaym{\'e}}, {Blanco-Cuaresma}, {Boch}, {Boeche}, {Bombrun}, {Borrachero},
  {Bouquillon}, {Bourda}, {Bragaglia}, {Bramante}, {Breddels}, {Brouillet},
  {Br{\"u}semeister}, {Brugaletta}, {Bucciarelli}, {Burlacu}, {Busonero},
  {Butkevich}, {Buzzi}, {Caffau}, {Cancelliere}, {Cannizzaro}, {Carballo},
  {Carlucci}, {Carrasco}, {Casamiquela}, {Castellani}, {Castro-Ginard},
  {Charlot}, {Chemin}, {Chiavassa}, {Cocozza}, {Costigan}, {Cowell}, {Crifo},
  {Crosta}, {Crowley}, {Cuypers}, {Dafonte}, {Damerdji}, {Dapergolas}, {David},
  {David}, {de Laverny}, {De Luise}, {De March}, {de Martino}, {de Souza}, {de
  Torres}, {Debosscher}, {del Pozo}, {Delbo}, {Delgado}, {Delgado}, {Diakite},
  {Diener}, {Distefano}, {Dolding}, {Drazinos}, {Dur{\'a}n}, {Edvardsson},
  {Enke}, {Eriksson}, {Esquej}, {Eynard Bontemps}, {Fabre}, {Fabrizio},
  {Faigler}, {Falc{\~a}o}, {Farr{\`a}s Casas}, {Federici}, {Fedorets},
  {Fernique}, {Figueras}, {Filippi}, {Findeisen}, {Fonti}, {Fraile}, {Fraser},
  {Fr{\'e}zouls}, {Gai}, {Galleti}, {Garabato}, {Garc{\'\i}a-Sedano},
  {Garofalo}, {Garralda}, {Gavel}, {Gavras}, {Gerssen}, {Geyer}, {Giacobbe},
  {Gilmore}, {Girona}, {Giuffrida}, {Glass}, {Gomes}, {Granvik}, {Gueguen},
  {Guerrier}, {Guiraud}, {Guti{\'e}}, {Haigron}, {Hatzidimitriou}, {Hauser},
  {Haywood}, {Heiter}, {Helmi}, {Heu}, {Hilger}, {Hobbs}, {Hofmann}, {Holland},
  {Huckle}, {Hypki}, {Icardi}, {Jan{\ss}en}, {Jevardat de Fombelle}, {Jonker},
  {Juh{\'a}sz}, {Julbe}, {Karampelas}, {Kewley}, {Klar}, {Kochoska}, {Kohley},
  {Kolenberg}, {Kontizas}, {Kontizas}, {Koposov}, {Kordopatis},
  {Kostrzewa-Rutkowska}, {Koubsky}, {Lambert}, {Lanza}, {Lasne}, {Lavigne}, {Le
  Fustec}, {Le Poncin-Lafitte}, {Lebreton}, {Leccia}, {Leclerc},
  {Lecoeur-Taibi}, {Lenhardt}, {Leroux}, {Liao}, {Licata}, {Lindstr{\o}m},
  {Lister}, {Livanou}, {Lobel}, {L{\'o}pez}, {Managau}, {Mann}, {Mantelet},
  {Marchal}, {Marchant}, {Marconi}, {Marinoni}, {Marschalk{\'o}}, {Marshall},
  {Martino}, {Marton}, {Mary}, {Massari}, {Matijevi{\v{c}}}, {Mazeh},
  {McMillan}, {Messina}, {Michalik}, {Millar}, {Molina}, {Molinaro},
  {Moln{\'a}r}, {Montegriffo}, {Mor}, {Morbidelli}, {Morel}, {Morris},
  {Mulone}, {Muraveva}, {Musella}, {Nelemans}, {Nicastro}, {Noval},
  {O'Mullane}, {Ord{\'e}novic}, {Ord{\'o}{\~n}ez-Blanco}, {Osborne}, {Pagani},
  {Pagano}, {Pailler}, {Palacin}, {Palaversa}, {Panahi}, {Pawlak},
  {Piersimoni}, {Pineau}, {Plachy}, {Plum}, {Poggio}, {Poujoulet},
  {Pr{\v{s}}a}, {Pulone}, {Racero}, {Ragaini}, {Rambaux}, {Ramos-Lerate},
  {Regibo}, {Reyl{\'e}}, {Riclet}, {Ripepi}, {Riva}, {Rivard}, {Rixon},
  {Roegiers}, {Roelens}, {Romero-G{\'o}mez}, {Rowell}, {Royer}, {Ruiz-Dern},
  {Sadowski}, {Sagrist{\`a} Sell{\'e}s}, {Sahlmann}, {Salgado}, {Salguero},
  {Sanna}, {Santana-Ros}, {Sarasso}, {Savietto}, {Schultheis}, {Sciacca},
  {Segol}, {Segovia}, {S{\'e}gransan}, {Shih}, {Siltala}, {Silva}, {Smart},
  {Smith}, {Solano}, {Solitro}, {Sordo}, {Soria Nieto}, {Souchay}, {Spagna},
  {Spoto}, {Stampa}, {Steele}, {Steidelm{\"u}ller}, {Stephenson}, {Stoev},
  {Suess}, {Surdej}, {Szabados}, {Szegedi-Elek}, {Tapiador}, {Taris}, {Tauran},
  {Taylor}, {Teixeira}, {Terrett}, {Teyssand ier}, {Thuillot}, {Titarenko},
  {Torra Clotet}, {Turon}, {Ulla}, {Utrilla}, {Uzzi}, {Vaillant}, {Valentini},
  {Valette}, {van Elteren}, {Van Hemelryck}, {Vaschetto}, {Vecchiato},
  {Veljanoski}, {Viala}, {Vicente}, {Vogt}, {von Essen}, {Voss}, {Votruba},
  {Voutsinas}, {Walmsley}, {Weiler}, {Wertz}, {Wevers}, {Wyrzykowski},
  {Yoldas}, {{\v{Z}}erjal}, {Ziaeepour}, {Zorec}, {Zschocke}, {Zucker},
  {Zurbach}, \& {Zwitter}}]{GDR2_Hyades}
{Gaia Collaboration}, {Babusiaux}, C., {van Leeuwen}, F., {et~al.}
  2018{\natexlab{a}}, \aap, 616, A10

\bibitem[{{Gaia Collaboration} {et~al.}(2018{\natexlab{b}}){Gaia
  Collaboration}, {Brown}, {Vallenari}, {Prusti}, {de Bruijne}, {Babusiaux},
  {Bailer-Jones}, {Biermann}, {Evans}, {Eyer}, {Jansen}, {Jordi}, {Klioner},
  {Lammers}, {Lindegren}, {Luri}, {Mignard}, {Panem}, {Pourbaix}, {Randich},
  {Sartoretti}, {Siddiqui}, {Soubiran}, {van Leeuwen}, {Walton}, {Arenou},
  {Bastian}, {Cropper}, {Drimmel}, {Katz}, {Lattanzi}, {Bakker}, {Cacciari},
  {Casta{\~n}eda}, {Chaoul}, {Cheek}, {De Angeli}, {Fabricius}, {Guerra},
  {Holl}, {Masana}, {Messineo}, {Mowlavi}, {Nienartowicz}, {Panuzzo},
  {Portell}, {Riello}, {Seabroke}, {Tanga}, {Th{\'e}venin}, {Gracia-Abril},
  {Comoretto}, {Garcia-Reinaldos}, {Teyssier}, {Altmann}, {Andrae}, {Audard},
  {Bellas-Velidis}, {Benson}, {Berthier}, {Blomme}, {Burgess}, {Busso},
  {Carry}, {Cellino}, {Clementini}, {Clotet}, {Creevey}, {Davidson}, {De
  Ridder}, {Delchambre}, {Dell'Oro}, {Ducourant},
  {Fern{\'a}ndez-Hern{\'a}ndez}, {Fouesneau}, {Fr{\'e}mat}, {Galluccio},
  {Garc{\'\i}a-Torres}, {Gonz{\'a}lez-N{\'u}{\~n}ez}, {Gonz{\'a}lez-Vidal},
  {Gosset}, {Guy}, {Halbwachs}, {Hambly}, {Harrison}, {Hern{\'a}ndez},
  {Hestroffer}, {Hodgkin}, {Hutton}, {Jasniewicz}, {Jean-Antoine-Piccolo},
  {Jordan}, {Korn}, {Krone-Martins}, {Lanzafame}, {Lebzelter}, {L{\"o}ffler},
  {Manteiga}, {Marrese}, {Mart{\'\i}n-Fleitas}, {Moitinho}, {Mora}, {Muinonen},
  {Osinde}, {Pancino}, {Pauwels}, {Petit}, {Recio-Blanco}, {Richards},
  {Rimoldini}, {Robin}, {Sarro}, {Siopis}, {Smith}, {Sozzetti}, {S{\"u}veges},
  {Torra}, {van Reeven}, {Abbas}, {Abreu Aramburu}, {Accart}, {Aerts},
  {Altavilla}, {{\'A}lvarez}, {Alvarez}, {Alves}, {Anderson}, {Andrei},
  {Anglada Varela}, {Antiche}, {Antoja}, {Arcay}, {Astraatmadja}, {Bach},
  {Baker}, {Balaguer-N{\'u}{\~n}ez}, {Balm}, {Barache}, {Barata}, {Barbato},
  {Barblan}, {Barklem}, {Barrado}, {Barros}, {Barstow}, {Bartholom{\'e}
  Mu{\~n}oz}, {Bassilana}, {Becciani}, {Bellazzini}, {Berihuete}, {Bertone},
  {Bianchi}, {Bienaym{\'e}}, {Blanco-Cuaresma}, {Boch}, {Boeche}, {Bombrun},
  {Borrachero}, {Bossini}, {Bouquillon}, {Bourda}, {Bragaglia}, {Bramante},
  {Breddels}, {Bressan}, {Brouillet}, {Br{\"u}semeister}, {Brugaletta},
  {Bucciarelli}, {Burlacu}, {Busonero}, {Butkevich}, {Buzzi}, {Caffau},
  {Cancelliere}, {Cannizzaro}, {Cantat-Gaudin}, {Carballo}, {Carlucci},
  {Carrasco}, {Casamiquela}, {Castellani}, {Castro-Ginard}, {Charlot},
  {Chemin}, {Chiavassa}, {Cocozza}, {Costigan}, {Cowell}, {Crifo}, {Crosta},
  {Crowley}, {Cuypers}, {Dafonte}, {Damerdji}, {Dapergolas}, {David}, {David},
  {de Laverny}, {De Luise}, {De March}, {de Martino}, {de Souza}, {de Torres},
  {Debosscher}, {del Pozo}, {Delbo}, {Delgado}, {Delgado}, {Di Matteo},
  {Diakite}, {Diener}, {Distefano}, {Dolding}, {Drazinos}, {Dur{\'a}n},
  {Edvardsson}, {Enke}, {Eriksson}, {Esquej}, {Eynard Bontemps}, {Fabre},
  {Fabrizio}, {Faigler}, {Falc{\~a}o}, {Farr{\`a}s Casas}, {Federici},
  {Fedorets}, {Fernique}, {Figueras}, {Filippi}, {Findeisen}, {Fonti},
  {Fraile}, {Fraser}, {Fr{\'e}zouls}, {Gai}, {Galleti}, {Garabato},
  {Garc{\'\i}a-Sedano}, {Garofalo}, {Garralda}, {Gavel}, {Gavras}, {Gerssen},
  {Geyer}, {Giacobbe}, {Gilmore}, {Girona}, {Giuffrida}, {Glass}, {Gomes},
  {Granvik}, {Gueguen}, {Guerrier}, {Guiraud}, {Guti{\'e}rrez-S{\'a}nchez},
  {Haigron}, {Hatzidimitriou}, {Hauser}, {Haywood}, {Heiter}, {Helmi}, {Heu},
  {Hilger}, {Hobbs}, {Hofmann}, {Holland}, {Huckle}, {Hypki}, {Icardi},
  {Jan{\ss}en}, {Jevardat de Fombelle}, {Jonker}, {Juh{\'a}sz}, {Julbe},
  {Karampelas}, {Kewley}, {Klar}, {Kochoska}, {Kohley}, {Kolenberg},
  {Kontizas}, {Kontizas}, {Koposov}, {Kordopatis}, {Kostrzewa-Rutkowska},
  {Koubsky}, {Lambert}, {Lanza}, {Lasne}, {Lavigne}, {Le Fustec}, {Le
  Poncin-Lafitte}, {Lebreton}, {Leccia}, {Leclerc}, {Lecoeur-Taibi},
  {Lenhardt}, {Leroux}, {Liao}, {Licata}, {Lindstr{\o}m}, {Lister}, {Livanou},
  {Lobel}, {L{\'o}pez}, {Managau}, {Mann}, {Mantelet}, {Marchal}, {Marchant},
  {Marconi}, {Marinoni}, {Marschalk{\'o}}, {Marshall}, {Martino}, {Marton},
  {Mary}, {Massari}, {Matijevi{\v{c}}}, {Mazeh}, {McMillan}, {Messina},
  {Michalik}, {Millar}, {Molina}, {Molinaro}, {Moln{\'a}r}, {Montegriffo},
  {Mor}, {Morbidelli}, {Morel}, {Morris}, {Mulone}, {Muraveva}, {Musella},
  {Nelemans}, {Nicastro}, {Noval}, {O'Mullane}, {Ord{\'e}novic},
  {Ord{\'o}{\~n}ez-Blanco}, {Osborne}, {Pagani}, {Pagano}, {Pailler},
  {Palacin}, {Palaversa}, {Panahi}, {Pawlak}, {Piersimoni}, {Pineau}, {Plachy},
  {Plum}, {Poggio}, {Poujoulet}, {Pr{\v{s}}a}, {Pulone}, {Racero}, {Ragaini},
  {Rambaux}, {Ramos-Lerate}, {Regibo}, {Reyl{\'e}}, {Riclet}, {Ripepi}, {Riva},
  {Rivard}, {Rixon}, {Roegiers}, {Roelens}, {Romero-G{\'o}mez}, {Rowell},
  {Royer}, {Ruiz-Dern}, {Sadowski}, {Sagrist{\`a} Sell{\'e}s}, {Sahlmann},
  {Salgado}, {Salguero}, {Sanna}, {Santana-Ros}, {Sarasso}, {Savietto},
  {Schultheis}, {Sciacca}, {Segol}, {Segovia}, {S{\'e}gransan}, {Shih},
  {Siltala}, {Silva}, {Smart}, {Smith}, {Solano}, {Solitro}, {Sordo}, {Soria
  Nieto}, {Souchay}, {Spagna}, {Spoto}, {Stampa}, {Steele},
  {Steidelm{\"u}ller}, {Stephenson}, {Stoev}, {Suess}, {Surdej}, {Szabados},
  {Szegedi-Elek}, {Tapiador}, {Taris}, {Tauran}, {Taylor}, {Teixeira},
  {Terrett}, {Teyssand ier}, {Thuillot}, {Titarenko}, {Torra Clotet}, {Turon},
  {Ulla}, {Utrilla}, {Uzzi}, {Vaillant}, {Valentini}, {Valette}, {van Elteren},
  {Van Hemelryck}, {van Leeuwen}, {Vaschetto}, {Vecchiato}, {Veljanoski},
  {Viala}, {Vicente}, {Vogt}, {von Essen}, {Voss}, {Votruba}, {Voutsinas},
  {Walmsley}, {Weiler}, {Wertz}, {Wevers}, {Wyrzykowski}, {Yoldas},
  {{\v{Z}}erjal}, {Ziaeepour}, {Zorec}, {Zschocke}, {Zucker}, {Zurbach}, \&
  {Zwitter}}]{GaiaDR2_2018}
{Gaia Collaboration}, {Brown}, A.~G.~A., {Vallenari}, A., {et~al.}
  2018{\natexlab{b}}, \aap, 616, A1

\bibitem[{{Gaia Collaboration} {et~al.}(2020){Gaia Collaboration}, {Brown},
  {Vallenari}, {Prusti}, {de Bruijne}, {Babusiaux}, \& {Biermann}}]{eDR3}
{Gaia Collaboration}, {Brown}, A.~G.~A., {Vallenari}, A., {et~al.} 2020, arXiv
  e-prints, arXiv:2012.01533

\bibitem[{{Goldman} {et~al.}(2013){Goldman}, {R{\"o}ser}, {Schilbach},
  {Magnier}, {Olczak}, {Henning}, {Juri{\'c}}, {Schlafly}, {Chen}, {Platais},
  {Burgett}, {Hodapp}, {Heasley}, {Kudritzki}, {Morgan}, {Price}, {Tonry}, \&
  {Wainscoat}}]{Goldman2013}
{Goldman}, B., {R{\"o}ser}, S., {Schilbach}, E., {et~al.} 2013, \aap, 559, A43

\bibitem[{{Gossage} {et~al.}(2018){Gossage}, {Conroy}, {Dotter}, {Choi},
  {Rosenfield}, {Cargile}, \& {Dolphin}}]{Hy_age2}
{Gossage}, S., {Conroy}, C., {Dotter}, A., {et~al.} 2018, \apj, 863, 67

\bibitem[{{Harfst} {et~al.}(2007){Harfst}, {Gualandris}, {Merritt}, {Spurzem},
  {Portegies Zwart}, \& {Berczik}}]{grape2007}
{Harfst}, S., {Gualandris}, A., {Merritt}, D., {et~al.} 2007, \na, 12, 357

\bibitem[{{Hawkins} {et~al.}(2020){Hawkins}, {Lucey}, \&
  {Curtis}}]{Hawkins2020}
{Hawkins}, K., {Lucey}, M., \& {Curtis}, J. 2020, \mnras, 496, 2422

\bibitem[{{Ibata} {et~al.}(2019){Ibata}, {Bellazzini}, {Malhan}, {Martin}, \&
  {Bianchini}}]{Ibata2019}
{Ibata}, R.~A., {Bellazzini}, M., {Malhan}, K., {Martin}, N., \& {Bianchini},
  P. 2019, Nature Astronomy, 3, 667

\bibitem[{{Irrgang} {et~al.}(2013){Irrgang}, {Wilcox}, {Tucker}, \&
  {Schiefelbein}}]{Irrgang2013}
{Irrgang}, A., {Wilcox}, B., {Tucker}, E., \& {Schiefelbein}, L. 2013, \aap,
  549, A137

\bibitem[{{J{\"a}nes} {et~al.}(2014){J{\"a}nes}, {Pelupessy}, \& {Portegies
  Zwart}}]{Huayno2014}
{J{\"a}nes}, J., {Pelupessy}, I., \& {Portegies Zwart}, S. 2014, \aap, 570, A20

\bibitem[{{Jerabkova} {et~al.}(2019{\natexlab{a}}){Jerabkova}, {Beccari},
  {Boffin}, {Petr-Gotzens}, {Manara}, {Prada Moroni}, {Tognelli}, \&
  {Degl'Innocenti}}]{Jerabkova2019_ONC}
{Jerabkova}, T., {Beccari}, G., {Boffin}, H. M.~J., {et~al.}
  2019{\natexlab{a}}, \aap, 627, A57

\bibitem[{{Jerabkova} {et~al.}(2019{\natexlab{b}}){Jerabkova}, {Boffin},
  {Beccari}, \& {Anderson}}]{Jerabkova2019}
{Jerabkova}, T., {Boffin}, H. M.~J., {Beccari}, G., \& {Anderson}, R.~I.
  2019{\natexlab{b}}, \mnras, 489, 4418

\bibitem[{{Je{\v{r}}{\'a}bkov{\'a}} {et~al.}(2018){Je{\v{r}}{\'a}bkov{\'a}},
  {Hasani Zonoozi}, {Kroupa}, {Beccari}, {Yan}, {Vazdekis}, \&
  {Zhang}}]{Jerabkova+18}
{Je{\v{r}}{\'a}bkov{\'a}}, T., {Hasani Zonoozi}, A., {Kroupa}, P., {et~al.}
  2018, \aap, 620, A39

\bibitem[{{Just} {et~al.}(2009){Just}, {Berczik}, {Petrov}, \&
  {Ernst}}]{Just+09}
{Just}, A., {Berczik}, P., {Petrov}, M.~I., \& {Ernst}, A. 2009, \mnras, 392,
  969

\bibitem[{{Karim} \& {Mamajek}(2017)}]{Karim2017}
{Karim}, M.~T. \& {Mamajek}, E.~E. 2017, \mnras, 465, 472

\bibitem[{{Kharchenko} {et~al.}(2009){Kharchenko}, {Berczik}, {Petrov},
  {Piskunov}, {R{\"o}ser}, {Schilbach}, \& {Scholz}}]{Kharchenko2009}
{Kharchenko}, N.~V., {Berczik}, P., {Petrov}, M.~I., {et~al.} 2009, \aap, 495,
  807

\bibitem[{{Kounkel} \& {Covey}(2019{\natexlab{a}})}]{KounkelCovey19}
{Kounkel}, M. \& {Covey}, K. 2019{\natexlab{a}}, \aj, 158, 122

\bibitem[{{Kounkel} \& {Covey}(2019{\natexlab{b}})}]{Kounkel2019}
{Kounkel}, M. \& {Covey}, K. 2019{\natexlab{b}}, \aj, 158, 122

\bibitem[{{Kounkel} {et~al.}(2018){Kounkel}, {Covey}, {Su{\'a}rez},
  {Rom{\'a}n-Z{\'u}{\~n}iga}, {Hernandez}, {Stassun}, {Jaehnig}, {Feigelson},
  {Pe{\~n}a Ram{\'\i}rez}, {Roman-Lopes}, {Da Rio}, {Stringfellow}, {Kim},
  {Borissova}, {Fern{\'a}ndez-Trincado}, {Burgasser},
  {Garc{\'\i}a-Hern{\'a}ndez}, {Zamora}, {Pan}, \& {Nitschelm}}]{Kounkel+18}
{Kounkel}, M., {Covey}, K., {Su{\'a}rez}, G., {et~al.} 2018, \aj, 156, 84

\bibitem[{{Kroupa}(1995{\natexlab{a}})}]{Kroupa95a}
{Kroupa}, P. 1995{\natexlab{a}}, \mnras, 277, 1491

\bibitem[{{Kroupa}(1995{\natexlab{b}})}]{Kroupa95c}
{Kroupa}, P. 1995{\natexlab{b}}, \mnras, 277, 1522

\bibitem[{{Kroupa}(1995{\natexlab{c}})}]{Kroupa95b}
{Kroupa}, P. 1995{\natexlab{c}}, \mnras, 277, 1507

\bibitem[{{Kroupa}(1997)}]{Kroupa97}
{Kroupa}, P. 1997, \na, 2, 139

\bibitem[{{Kroupa}(2005)}]{Kroupa05}
{Kroupa}, P. 2005, in ESA Special Publication, Vol. 576, The Three-Dimensional
  Universe with Gaia, ed. C.~{Turon}, K.~S. {O'Flaherty}, \& M.~A.~C.
  {Perryman}, 629

\bibitem[{{Kroupa}(2011)}]{Kroupa11}
{Kroupa}, P. 2011, in Computational Star Formation, ed. J.~{Alves}, B.~G.
  {Elmegreen}, J.~M. {Girart}, \& V.~{Trimble}, Vol. 270, 141--149

\bibitem[{{Kroupa} {et~al.}(2001){Kroupa}, {Aarseth}, \& {Hurley}}]{Kroupa+01}
{Kroupa}, P., {Aarseth}, S., \& {Hurley}, J. 2001, \mnras, 321, 699

\bibitem[{{Kroupa} {et~al.}(2018){Kroupa}, {Je{\v{r}}{\'a}bkov{\'a}},
  {Dinnbier}, {Beccari}, \& {Yan}}]{Kroupa2018}
{Kroupa}, P., {Je{\v{r}}{\'a}bkov{\'a}}, T., {Dinnbier}, F., {Beccari}, G., \&
  {Yan}, Z. 2018, \aap, 612, A74

\bibitem[{{K{\"u}pper} {et~al.}(2015){K{\"u}pper}, {Balbinot}, {Bonaca},
  {Johnston}, {Hogg}, {Kroupa}, \& {Santiago}}]{Kuepper+15}
{K{\"u}pper}, A. H.~W., {Balbinot}, E., {Bonaca}, A., {et~al.} 2015, \apj, 803,
  80

\bibitem[{{K{\"u}pper} {et~al.}(2010){K{\"u}pper}, {Kroupa}, {Baumgardt}, \&
  {Heggie}}]{Kuepper+10}
{K{\"u}pper}, A. H.~W., {Kroupa}, P., {Baumgardt}, H., \& {Heggie}, D.~C. 2010,
  \mnras, 401, 105

\bibitem[{{K{\"u}pper} {et~al.}(2012){K{\"u}pper}, {Lane}, \&
  {Heggie}}]{Kuepper+12}
{K{\"u}pper}, A. H.~W., {Lane}, R.~R., \& {Heggie}, D.~C. 2012, \mnras, 420,
  2700

\bibitem[{{K{\"u}pper} {et~al.}(2008){K{\"u}pper}, {MacLeod}, \&
  {Heggie}}]{Kuepper+08}
{K{\"u}pper}, A. H.~W., {MacLeod}, A., \& {Heggie}, D.~C. 2008, \mnras, 387,
  1248

\bibitem[{{Lada} \& {Lada}(2003)}]{LL03}
{Lada}, C.~J. \& {Lada}, E.~A. 2003, \araa, 41, 57

\bibitem[{{Lanzafame} {et~al.}(2018){Lanzafame}, {Distefano}, {Messina},
  {Pagano}, {Lanza}, {Eyer}, {Guy}, {Rimoldini}, {Lecoeur-Taibi}, {Holl},
  {Audard}, {de Fombelle}, {Nienartowicz}, {Marchal}, \&
  {Mowlavi}}]{Lanzafame2018}
{Lanzafame}, A.~C., {Distefano}, E., {Messina}, S., {et~al.} 2018, \aap, 616,
  A16

\bibitem[{{Lebreton} {et~al.}(2001){Lebreton}, {Fernandes}, \&
  {Lejeune}}]{Lebreton2001}
{Lebreton}, Y., {Fernandes}, J., \& {Lejeune}, T. 2001, \aap, 374, 540

\bibitem[{{Lodieu}(2020)}]{Hy_age4}
{Lodieu}, N. 2020, \memsai, 91, 84

\bibitem[{{Madsen}(2003)}]{Madsen03}
{Madsen}, S. 2003, \aap, 401, 565

\bibitem[{{Makarov} {et~al.}(2000){Makarov}, {Odenkirchen}, \&
  {Urban}}]{Makarov2000}
{Makarov}, V.~V., {Odenkirchen}, M., \& {Urban}, S. 2000, \aap, 358, 923

\bibitem[{{Marks} \& {Kroupa}(2011)}]{MarksKroupa11}
{Marks}, M. \& {Kroupa}, P. 2011, \mnras, 417, 1702

\bibitem[{{Marks} \& {Kroupa}(2012)}]{MarksKroupa12}
{Marks}, M. \& {Kroupa}, P. 2012, \aap, 543, A8

\bibitem[{{Mart{\'\i}nez-Barbosa} {et~al.}(2016){Mart{\'\i}nez-Barbosa},
  {Brown}, {Boekholt}, {Portegies Zwart}, {Antiche}, \&
  {Antoja}}]{Martinez2016}
{Mart{\'\i}nez-Barbosa}, C.~A., {Brown}, A.~G.~A., {Boekholt}, T., {et~al.}
  2016, \mnras, 457, 1062

\bibitem[{{Mart{\'\i}nez-Barbosa} {et~al.}(2017){Mart{\'\i}nez-Barbosa},
  {J{\'\i}lkov{\'a}}, {Portegies Zwart}, \& {Brown}}]{Martinez2017}
{Mart{\'\i}nez-Barbosa}, C.~A., {J{\'\i}lkov{\'a}}, L., {Portegies Zwart}, S.,
  \& {Brown}, A.~G.~A. 2017, \mnras, 464, 2290

\bibitem[{{Meingast} \& {Alves}(2019)}]{MeingastAlves19}
{Meingast}, S. \& {Alves}, J. 2019, \aap, 621, L3

\bibitem[{{Meingast} {et~al.}(2020){Meingast}, {Alves}, \&
  {Rottensteiner}}]{Meingast+20}
{Meingast}, S., {Alves}, J., \& {Rottensteiner}, A. 2020, arXiv e-prints,
  arXiv:2010.06591

\bibitem[{{Odenkirchen} {et~al.}(2003){Odenkirchen}, {Grebel}, {Dehnen}, {Rix},
  {Yanny}, {Newberg}, {Rockosi}, {Mart{\'\i}nez-Delgado}, {Brinkmann}, \&
  {Pier}}]{Odenkirchen+03}
{Odenkirchen}, M., {Grebel}, E.~K., {Dehnen}, W., {et~al.} 2003, \aj, 126, 2385

\bibitem[{{Oh} \& {Evans}(2020)}]{Oh2020}
{Oh}, S. \& {Evans}, N.~W. 2020, \mnras, 498, 1920

\bibitem[{{Oh} \& {Kroupa}(2016)}]{Oh2016}
{Oh}, S. \& {Kroupa}, P. 2016, \aap, 590, A107

\bibitem[{{Oh} {et~al.}(2015){Oh}, {Kroupa}, \& {Pflamm-Altenburg}}]{Oh2015}
{Oh}, S., {Kroupa}, P., \& {Pflamm-Altenburg}, J. 2015, \apj, 805, 92

\bibitem[{{Pelupessy} {et~al.}(2012){Pelupessy}, {J{\"a}nes}, \& {Portegies
  Zwart}}]{Huayno2012}
{Pelupessy}, F.~I., {J{\"a}nes}, J., \& {Portegies Zwart}, S. 2012, \na, 17,
  711

\bibitem[{{Pelupessy} {et~al.}(2013){Pelupessy}, {van Elteren}, {de Vries},
  {McMillan}, {Drost}, \& {Portegies Zwart}}]{amuse2013b}
{Pelupessy}, F.~I., {van Elteren}, A., {de Vries}, N., {et~al.} 2013, \aap,
  557, A84

\bibitem[{{Perryman} {et~al.}(1998){Perryman}, {Brown}, {Lebreton}, {Gomez},
  {Turon}, {Cayrel de Strobel}, {Mermilliod}, {Robichon}, {Kovalevsky}, \&
  {Crifo}}]{Perryman1998}
{Perryman}, M.~A.~C., {Brown}, A.~G.~A., {Lebreton}, Y., {et~al.} 1998, \aap,
  331, 81

\bibitem[{{Piatek} \& {Pryor}(1995)}]{PiatekPryor95}
{Piatek}, S. \& {Pryor}, C. 1995, \aj, 109, 1071

\bibitem[{{Piatti} \& {Carballo-Bello}(2020)}]{PiattiCarballo20}
{Piatti}, A.~E. \& {Carballo-Bello}, J.~A. 2020, \aap, 637, L2

\bibitem[{{Portegies Zwart} \& {McMillan}(2018)}]{amuse_book}
{Portegies Zwart}, S. \& {McMillan}, S. 2018, {Astrophysical Recipes; The art
  of AMUSE} ({Iop Publishing Ltd})

\bibitem[{{Portegies Zwart} {et~al.}(2009){Portegies Zwart}, {McMillan},
  {Harfst}, {Groen}, {Fujii}, {Nuall{\'a}in}, {Glebbeek}, {Heggie}, {Lombardi},
  {Hut}, {Angelou}, {Banerjee}, {Belkus}, {Fragos}, {Fregeau}, {Gaburov},
  {Izzard}, {Juri{\'c}}, {Justham}, {Sottoriva}, {Teuben}, {van Bever},
  {Yaron}, \& {Zemp}}]{amuse2009}
{Portegies Zwart}, S., {McMillan}, S., {Harfst}, S., {et~al.} 2009, \na, 14,
  369

\bibitem[{{Portegies Zwart} {et~al.}(2013){Portegies Zwart}, {McMillan}, {van
  Elteren}, {Pelupessy}, \& {de Vries}}]{amuse2013}
{Portegies Zwart}, S., {McMillan}, S.~L.~W., {van Elteren}, E., {Pelupessy},
  I., \& {de Vries}, N. 2013, Computer Physics Communications, 184, 456

\bibitem[{{Portegies Zwart} {et~al.}(2001){Portegies Zwart}, {McMillan}, {Hut},
  \& {Makino}}]{Portegies+01}
{Portegies Zwart}, S.~F., {McMillan}, S. L.~W., {Hut}, P., \& {Makino}, J.
  2001, \mnras, 321, 199

\bibitem[{{Portegies Zwart} \& {Verbunt}(1996)}]{Zwart1996}
{Portegies Zwart}, S.~F. \& {Verbunt}, F. 1996, \aap, 309, 179

\bibitem[{{Reid} {et~al.}(2014){Reid}, {Menten}, {Brunthaler}, {Zheng}, {Dame},
  {Xu}, {Wu}, {Zhang}, {Sanna}, {Sato}, {Hachisuka}, {Choi}, {Immer},
  {Moscadelli}, {Rygl}, \& {Bartkiewicz}}]{Reid2014}
{Reid}, M.~J., {Menten}, K.~M., {Brunthaler}, A., {et~al.} 2014, \apj, 783, 130

\bibitem[{{Reino} {et~al.}(2018){Reino}, {de Bruijne}, {Zari}, {d'Antona}, \&
  {Ventura}}]{Hy_age3}
{Reino}, S., {de Bruijne}, J., {Zari}, E., {d'Antona}, F., \& {Ventura}, P.
  2018, \mnras, 477, 3197

\bibitem[{{Roeser} {et~al.}(2010){Roeser}, {Demleitner}, \&
  {Schilbach}}]{roeser2010}
{Roeser}, S., {Demleitner}, M., \& {Schilbach}, E. 2010, \aj, 139, 2440

\bibitem[{{R{\"o}ser} \& {Schilbach}(2019)}]{RS19}
{R{\"o}ser}, S. \& {Schilbach}, E. 2019, \aap, 627, A4

\bibitem[{{R{\"o}ser} \& {Schilbach}(2020)}]{Roser2020}
{R{\"o}ser}, S. \& {Schilbach}, E. 2020, \aap, 638, A9

\bibitem[{{R{\"o}ser} {et~al.}(2019){R{\"o}ser}, {Schilbach}, \&
  {Goldman}}]{Roeser+19}
{R{\"o}ser}, S., {Schilbach}, E., \& {Goldman}, B. 2019, \aap, 621, L2

\bibitem[{{R{\"o}ser} {et~al.}(2011){R{\"o}ser}, {Schilbach}, {Piskunov},
  {Kharchenko}, \& {Scholz}}]{Roeser+11}
{R{\"o}ser}, S., {Schilbach}, E., {Piskunov}, A.~E., {Kharchenko}, N.~V., \&
  {Scholz}, R.~D. 2011, \aap, 531, A92

\bibitem[{{Str{\"o}mberg}(1939)}]{Smart1939}
{Str{\"o}mberg}, G. 1939, Popular Astronomy, 47, 172

\bibitem[{{Tang} {et~al.}(2019){Tang}, {Pang}, {Yuan}, {Chen}, {Hong},
  {Goldman}, {Just}, {Shukirgaliyev}, \& {Lin}}]{Tang+19}
{Tang}, S.-Y., {Pang}, X., {Yuan}, Z., {et~al.} 2019, \apj, 877, 12

\bibitem[{{Thomas} {et~al.}(2018){Thomas}, {Famaey}, {Ibata}, {Renaud},
  {Martin}, \& {Kroupa}}]{Thomas+18}
{Thomas}, G.~F., {Famaey}, B., {Ibata}, R., {et~al.} 2018, \aap, 609, A44

\bibitem[{{Toonen} {et~al.}(2012){Toonen}, {Nelemans}, \& {Portegies
  Zwart}}]{Toonen2012}
{Toonen}, S., {Nelemans}, G., \& {Portegies Zwart}, S. 2012, \aap, 546, A70

\bibitem[{{van Leeuwen}(2009)}]{vanLeeuwen09}
{van Leeuwen}, F. 2009, \aap, 497, 209

\bibitem[{{Yao} {et~al.}(2017){Yao}, {Manchester}, \& {Wang}}]{Yao2017}
{Yao}, J.~M., {Manchester}, R.~N., \& {Wang}, N. 2017, \mnras, 468, 3289

\bibitem[{{Zari} {et~al.}(2019){Zari}, {Brown}, \& {de Zeeuw}}]{Zari+19}
{Zari}, E., {Brown}, A.~G.~A., \& {de Zeeuw}, P.~T. 2019, \aap, 628, A123

\bibitem[{{Zari} {et~al.}(2018){Zari}, {Hashemi}, {Brown}, {Jardine}, \& {de
  Zeeuw}}]{Zari+18}
{Zari}, E., {Hashemi}, H., {Brown}, A.~G.~A., {Jardine}, K., \& {de Zeeuw},
  P.~T. 2018, \aap, 620, A172

\bibitem[{{Zhang} {et~al.}(2020){Zhang}, {Tang}, {Chen}, {Pang}, \&
  {Liu}}]{Zhang+20}
{Zhang}, Y., {Tang}, S.-Y., {Chen}, W.~P., {Pang}, X., \& {Liu}, J.~Z. 2020,
  \apj, 889, 99

\end{thebibliography}
\begin{appendix}

\section{\textit{Gaia} queries} \label{sec:gaia}
To query Gaia data, we used the ESA archive\footnote{https://gea.esac.esa.int/archive/} to acquire \textit{Gaia}  DR2 data in order to search for the Hyades tidal tails using the following \texttt{ADQL} set of commands:

\begin{lstlisting}
SELECT  gs.source_id,gs.ra,gs.ra_error,
gs.dec,gs.dec_error,gs.parallax,
gs.parallax_error,gs.pmra,gs.pmra_error,
gs.pmdec,gs.pmdec_error,gs.phot_g_mean_mag,
gs.bp_rp,gs.radial_velocity, 
gs.radial_velocity_error, r.ruwe
FROM gaiadr2.gaia_source AS gs, 
     gaiadr2.ruwe AS r
WHERE (gs.source_id = r.source_id 
       AND gs.parallax>=2.0 
       AND gs.parallax_over_error>=10 
       AND r.ruwe < 1.4)
\end{lstlisting}
 
\texttt{RUWE} in the download script above notes the renormalised unit weight error that is used to  filter out objects with spurious astrometric solutions (Lindegren, document \texttt{GAIA-C3-TN-LU-LL-124-01}).
We used several parallax bins to query the full range of parallaxes ($\varpi >2 \, \mathrm{mas}$), 
resulting in a total number of queried objects of~11,223,898.

To query Gaia eDR3 data, we used following set of \texttt{ADQL} commands:

\begin{lstlisting}
SELECT  gs.source_id,gs.ra,gs.ra_error,
gs.dec,gs.dec_error,gs.parallax,
gs.parallax_error,gs.pmra,gs.pmra_error,
gs.pmdec,gs.pmdec_error,gs.phot_g_mean_mag,
gs.bp_rp,gs.dr2_radial_velocity, 
gs.dr2_radial_velocity_error, gs.ruwe
FROM gaiaedr3.gaia_source AS gs, 
WHERE (gs.parallax>=2.0 
       AND gs.parallax_over_error>=10 
       AND gs.ruwe < 1.4)
\end{lstlisting}

The changed catalogue name for the radial velocity parameter and the \texttt{RUWE} parameter is part of the official eDR3 table. This query results in 13,896,271 objects.

\subsection{Query the Gaia rotation periods}

\begin{lstlisting}
SELECT  gs.source_id,gs.ra,gs.ra_error,
gs.dec,gs.dec_error,gs.parallax,
gs.parallax_error,gs.pmra,gs.pmra_error,
gs.pmdec,gs.pmdec_error,gs.phot_g_mean_mag,
gs.bp_rp,gs.radial_velocity, 
gs.radial_velocity_error, gs.teff_val, r.ruwe, 
gv.source_id, gv.best_rotation_period, gv.max_activity_index
FROM gaiadr2.gaia_source AS gs, 
     gaiadr2.ruwe AS r, 
     gaiadr2.vari_rotation_modulation as gv
WHERE (gs.source_id = r.source_id 
           AND gs.source_id = gv.source_id
           AND gs.parallax>=2.0 
       AND gs.parallax_over_error>=10 
       AND r.ruwe < 1.4)
\end{lstlisting}

\section{Catalogue tables}

\begin{table*}
\centering                                
\begin{tabular}{l l l l l}          
\hline\hline                       
source ID & R.A. [deg] & Dec [deg] & $\mu_{\alpha^{*}}$ [mas/yr] & $\mu_{\delta}$ [mas/yr] \\
2171661054009007616 & 322.706 & 51.214 & 13.535 & 11.171 \\
2171661054003013504 & 322.705 & 51.214 & 12.889 & 11.188 \\
1978513488498261888 & 322.009 & 47.876 & 14.500 & 13.828 \\
\vdots & \vdots & \vdots & \vdots & \vdots \\

\end{tabular}
\caption{Final selection of tidal tail members in Gaia DR2 data based on model M1. The full table is available at the CDS.}
\label{tab:DR2}
\end{table*}

\begin{table*}
\centering                                
\begin{tabular}{l l l l l}          
\hline\hline                       
source ID & R.A. [deg] & Dec [deg] & $\mu_{\alpha^{*}}$ [mas/yr] & $\mu_{\delta}$ [mas/yr] \\
5305130139316222080 & 135.090 & -55.267 & -16.274 & 19.302 \\
2174289161681454720 & 325.689 & 54.0383 & 10.528 & 9.471 \\
5311109828992983168 & 137.422  & -54.213 & -16.763 & 17.373 \\
\vdots & \vdots & \vdots & \vdots & \vdots \\

\end{tabular}
\caption{Final selection of tidal tail members in Gaia eDR3 data based of model M5. The full table is available at the CDS.}
\label{tab:eDR3}
\end{table*}

\end{appendix}

\end{document}